\documentclass[aip,jcp,amsmath,amssymb,floatfix,reprint,citeautoscript,noeprint,longbibliography]{revtex4-2}
\usepackage[utf8]{inputenc}
\usepackage{graphicx}
\usepackage{wrapfig}
\usepackage[colorlinks,allcolors=black,citecolor=blue,urlcolor=blue]{hyperref}
\usepackage[mathlines,displaymath]{lineno}
\usepackage{chemformula}
\usepackage{placeins}
\usepackage{booktabs}
\DeclareUnicodeCharacter{2212}{-}

\graphicspath{{figures/}}

\begin{document}
\def\mytitle{%
{To Pair or not to Pair? Machine-Learned Explicitly-Correlated Electronic Structure for NaCl in Water}}

\title{\mytitle}

\author{Niamh O'Neill}%
\affiliation{%
Yusuf Hamied Department of Chemistry, University of Cambridge, Lensfield Road, Cambridge, CB2 1EW, UK
}
\affiliation{%
Cavendish Laboratory, Department of Physics, University of Cambridge, Cambridge, CB3 0HE, UK
}
\affiliation{%
Lennard-Jones Centre, University of Cambridge, Trinity Ln, Cambridge, CB2 1TN, UK
}
\author{Benjamin X. Shi}%
\affiliation{%
Yusuf Hamied Department of Chemistry, University of Cambridge, Lensfield Road, Cambridge, CB2 1EW, UK
}
\affiliation{%
Lennard-Jones Centre, University of Cambridge, Trinity Ln, Cambridge, CB2 1TN, UK
}
\author{Kara Fong}%
\affiliation{%
Yusuf Hamied Department of Chemistry, University of Cambridge, Lensfield Road, Cambridge, CB2 1EW, UK
}
\affiliation{%
Lennard-Jones Centre, University of Cambridge, Trinity Ln, Cambridge, CB2 1TN, UK
}
\author{Angelos Michaelides}%
\email{am452@cam.ac.uk}
\affiliation{%
Yusuf Hamied Department of Chemistry, University of Cambridge, Lensfield Road, Cambridge, CB2 1EW, UK
}
\affiliation{%
Lennard-Jones Centre, University of Cambridge, Trinity Ln, Cambridge, CB2 1TN, UK
}

\author{Christoph Schran}%
\email{cs2121@cam.ac.uk}
\affiliation{%
Cavendish Laboratory, Department of Physics, University of Cambridge, Cambridge, CB3 0HE, UK
}
\affiliation{%
Lennard-Jones Centre, University of Cambridge, Trinity Ln, Cambridge, CB2 1TN, UK
}


\begin{abstract}
\setlength\intextsep{0pt}
\begin{wrapfigure}{r}{0.25\textwidth}
  \hspace{-1.5cm}
  \includegraphics[width=0.25\textwidth]{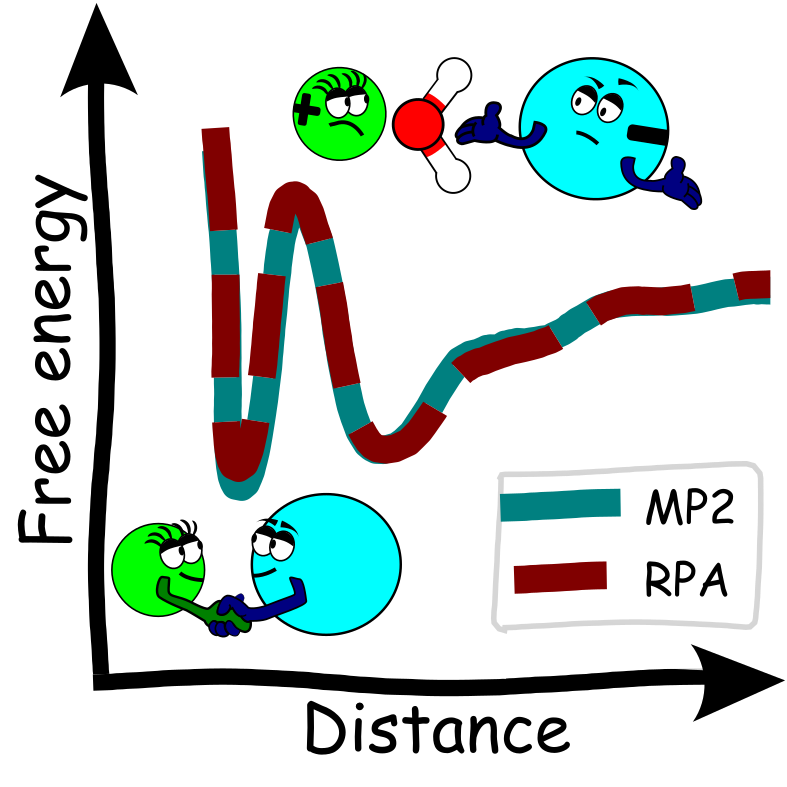}
\end{wrapfigure}

The extent of ion pairing in solution is an important phenomenon to rationalise transport and thermodynamic properties of electrolytes. A fundamental measure of this pairing is the potential of mean force (PMF) between solvated ions. The relative stabilities of the paired and solvent shared states in the PMF and the barrier between them are highly sensitive to the
underlying potential energy surface. However direct application of accurate electronic structure methods is challenging, since long simulations are required. We develop wavefunction based machine learning potentials with the Random Phase Approximation (RPA) and second order Møller–Plesset (MP2) perturbation theory for the prototypical system of Na and Cl ions in water. We show both methods in agreement, predicting the paired and solvent shared states to have similar energies (within 0.2 kcal/mol). We also provide the same benchmarks for different DFT functionals as well as insight into the PMF based on simple analyses of the interactions in the system. 
\end{abstract}

\maketitle

Understanding the nature of ion pairing and the solvation structure of electrolyte solutions is a fundamental challenge in the quest to design efficient next generation energy storage devices \cite{Holoubek2022/10.1039/D1EE03422G}.
For example, ionic conductivity and redox stability are predominantly influenced by the electrolyte solvation structure \cite{Yao2022/10.1021/ACS.CHEMREV.1C00904, Chen2022/10.1039/D2EE00004K}.
Moreover, an understanding of the solvation behaviour of ions is crucial to ensure a uniform and stable solid-liquid interphase to limit dendrite formation, which currently presents a significant challenge with respect to efficiency and safety of electrochemical devices \cite{Zhang2019/10.1021/ACSENERGYLETT.8B02376}.
More generally for transport properties, Peng et al. showed that the diffusion of sodium ions at interfaces is significantly affected by their hydration number \cite{Peng2018/10.1038/s41586-018-0122-2}.

The potential of mean force (PMF) between an ion pair in solution provides a direct window into the ion pairing behaviour and solution structure of electrolytes.
However, for the prototypical electrolyte solution of NaCl in water, no experimental benchmark exists.
Moreover, from a computational modelling perspective, it is highly sensitive to the underlying potential energy surface \cite{Duignan2016/10.1016/j.cocis.2016.05.009, Baer2016/10.1021/acs.jpcb.5b09579}
and there is currently no quantitative or qualitative consensus on the PMF of NaCl in water.
For example, two of the best-established classical force-fields for NaCl in water -- the Joung-Cheatham (JC) \cite{Joung2008/10.1021/jp8001614} and Smith-Dang (SD) \cite{Dang1993/10.1063/1.465441} models -- both disagree significantly in their PMFs, with the JC model predicting thermodynamically unfavourable ion pairing \cite{Zhang2020/10.1038/s41467-020-16704-9,Smith1994/10.1063/1.466363}. 
In general electrolytes are very challenging to model, where on top of the already arduous situation for bulk water \cite{Gillan2016/10.1063/1.4944633/152221}, they have the additional complication of ion-water and ion-ion interactions that need to be accurately described \cite{Duignan2021/10.1021/ACS.ACCOUNTS.1C00107}.
Therefore, it is clear that to rationalise the PMF, it is imperative to first have a model that accurately describes the structure of NaCl in water.

The past decades have seen pioneering work in the development of classical force field models to describe ions in water, with the most recent Madrid scaled charge models for ions giving excellent agreement with experiment for a range of structural and dynamical properties \cite{Blazquez2023/10.1063/5.0136498/2871744}.
However in general, such models can typically lack accuracy beyond the property or phase space to which they have been parameterised.
It has been suggested that to simultaneously and faithfully capture dynamical and structural properties of electrolytes, a model that explicitly treats its electronic structure is desirable\cite{Ding2014/10.1073/PNAS.1400675111, Panagiotopoulos2023/10.1021/ACS.JPCB.2C07477, Panagiotopoulos2023/10.1021/ACS.JPCB.2C07477}.

Until now, the workhorse method of \textit{ab initio} computational chemistry - density functional theory (DFT) - has been the method of choice for studying these systems, offering an acceptable balance of accuracy and computational overhead.
While the solvation structure of water around ions has been the subject of numerous previous \textit{ab initio} simulation studies \cite{Bankura2014/10.1080/00268976.2014.905721, Gaiduk2017/10.1021/ACS.JPCLETT.7B00239, Gaiduk2017/10.1021/ACS.JPCLETT.7B00239, Galib2017/10.1063/1.4975608, Duignan2020/10.1039/C9CP06161D, Rozsa2020/10.1063/1.5143317/15575136, Dellostritto2020/10.1039/C9CP06821J}, there is currently no consensus on a sufficiently accurate model to describe NaCl in water.
In fact both Duignan et al. and Panagiotopoulos et al. have recently highlighted the urgent need for accurate \textit{ab initio} models for electrolytes to capture their dynamics \cite{Panagiotopoulos2023/10.1021/ACS.JPCB.2C07477} and collective properties \cite{Duignan2016/10.1016/j.cocis.2016.05.009}.

For example, tried and tested exchange-correlation (XC) functionals for liquid water, such as the generalised gradient approximation (GGA) revPBE-D3 are not guaranteed to perform well when ions are added \cite{Galib2017/10.1063/1.4975608, Galib2017/10.1063/1.4986284, Duignan2020/10.1039/C9CP06161D}.
This is because standard GGAs tend to perform poorly for interactions involving charges, overestimating electrostatic contributions to binding energies \cite{Roza2020/10.1021/ACS.JPCA.9B10257} along with their well-known delocalisation error \cite{Bryenton2023/10.1002/WCMS.1631, Cheng2014/10.1021/AR500268Y}, both arising from the self-interaction error.
Moreover, the interplay of the electronic structure method and nuclear quantum effects has been shown to be highly sensitive for liquid water \cite{Marsalek2017/10.1021/ACS.JPCLETT.7B00391, Ruiz2017/10.1039/C6SC04711D, Ceriotti2016/10.1021/ACS.CHEMREV.5B00674}, but their impact on electrolyte solutions has not been exhaustively explored so far \cite{Xu2021/10.1103/PhysRevMaterials.5.L012801, Wilkins2015/10.1063/1.4907554}.
Inclusion of more complex ingredients into the density functional approximation following Perdew's Jacob's Ladder \cite{Perdew2003/10.1063/1.1390175} may improve the description of water-water \cite{Pestana2018/10.1021/ACS.JPCLETT.8B02400, Dasgupta2022/10.1021/acs.jctc.2c00313} or ion-water \cite{Riera2017/10.1063/1.4993213} interactions, however a model that faithfully captures the behaviour of ions in solution for the right reasons remains elusive.

Going beyond DFT, correlated wavefunction-based methods such as the Random Phase Approximation (RPA) and second-order M{\o}ller Plesset Perturbation Theory (MP2) are expected to perform well for electrolyte systems. 
These methods naturally incorporate van der Waals interactions and do not suffer from delocalisation error \cite{AlHamdani2019/10.1063/1.5075487/152312}. 
They have shown initial promise for liquid water, accurately predicting the correct relative densities for ice and water, which is governed sensitively by a balance of van der Waals and and hydrogen bonding interactions \cite{Ben2015/10.1063/1.4927325/193869, Ben2015/10.1016/J.CPC.2014.10.021, Ben2013/10.1021/JZ401931F}. %
While Duignan et al. hint that RPA could outperform lower rungs of Jacobs Ladder for the case of the solvated sodium ion \cite{Duignan2020/10.1039/C9CP06161D}, routine application of these high-level methods in condensed phase simulations has been sporadic. This is primarily because of their high computational cost to implement, with canonical scaling behaviour between $\mathcal{O}(N^4)$ and $\mathcal{O}(N^5)$, albeit reduced-scaling variants also exist \cite{Seewald2016/10.1021/ACS.JCTC.6B00840}.
Meanwhile, to obtain statistically converged properties to probe the structure of electrolyte solutions including radial distribution functions, densities and solvation free energies, even DFT becomes extremely computationally challenging.
While valuable developments in electronic structure code algorithms and computer hardware \cite{Ben2015/10.1063/1.4919238/898068, Ben2015/10.1016/J.CPC.2014.10.021} have significantly increased the accessibility of these methods, it would be highly desirable to confine them to a small number of single-point energy (and force) calculations rather than finite temperature simulations which typically require hundreds of thousands of such computations.

Fortunately machine learning potentials provide a gateway to perform simulations at \textit{ab initio} levels of theory with a decrease in several orders of magnitude in their computational cost \cite{Unke2021/10.1021/ACS.CHEMREV.0C01111, Behler2021/10.1140/EPJB/S10051-021-00156-1}.
Machine learning models have been successfully trained and applied to study bulk water at various levels of theory \cite{Morawietz2016/10.1073/pnas.160237511, Zhang2021/10.1021/ACS.JPCB.1C03884, Yao2020/10.1063/5.0012815/14721885, Schran2020/10.1063/5.0016004/199713, Cheng2019/10.1073/PNAS.1815117116}, including recent neural network models for bulk water with MP2 \cite{Lan2021/10.26434/CHEMRXIV-2021-N32Q8-V2, Liu2022/10.1021/acs.jpca.2c00601} and RPA \cite{Yao2021/10.1021/acs.jpclett.1c01566}.
While the natural next step is to add ions to the water \cite{Zhang2023/2310.12535, pagotto2022predicting, Baker2023/10.1021/ACS.JPCLETT.3C01783}, this brings additional complexity to the configuration space to be explored and so the training set must be judiciously chosen to reflect this.
To this end we use a previously developed automated active learning framework \cite{Schran2021/10.1073/pnas.2110077118} and an initial training set describing NaCl dissolution in water \cite{ONeill2022/2211.04345} to generate MLPs at various levels of electronic structure theory for Na and Cl ions in water.
\begin{figure*}[!ht]
    \centering
    \includegraphics[width=\textwidth]{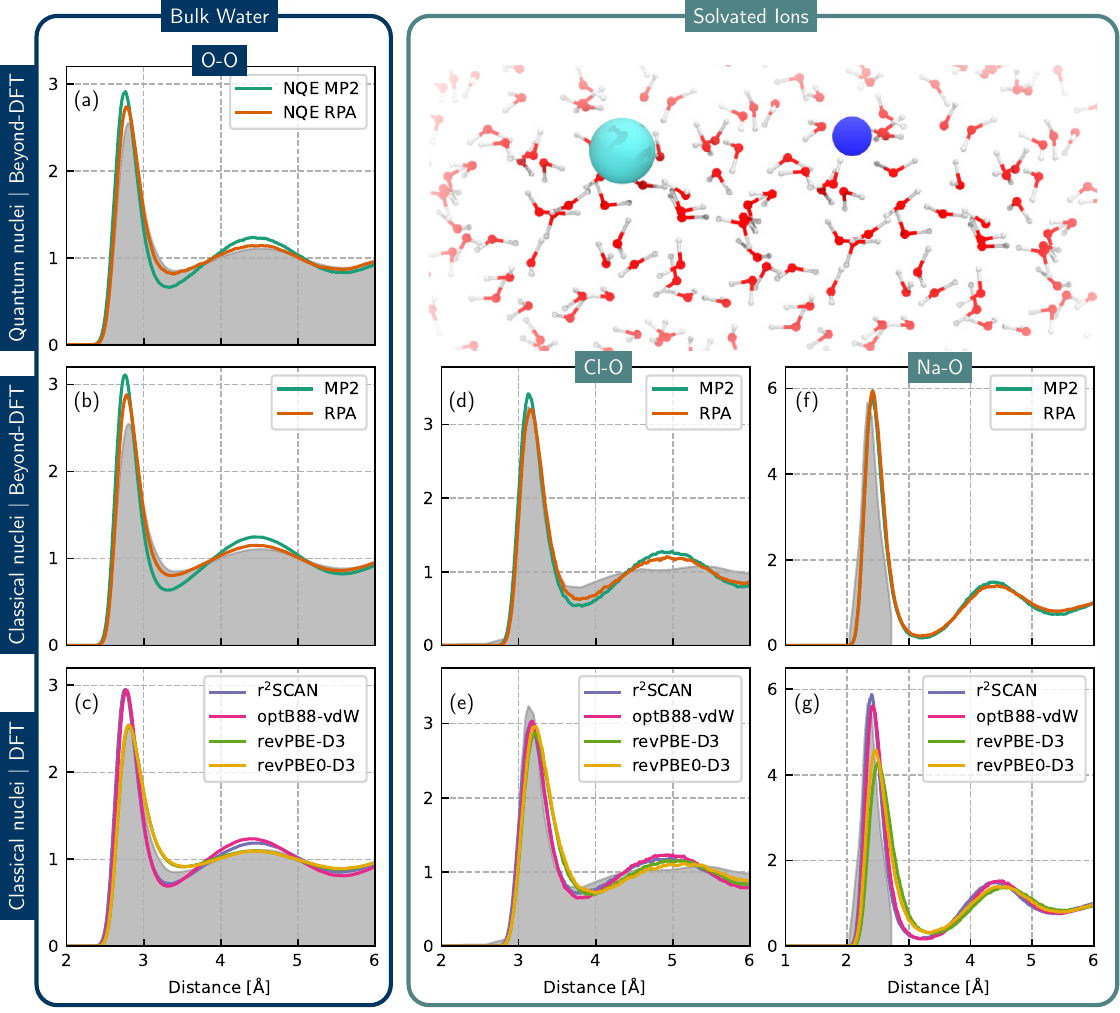}
	
	\caption{Comparison of radial distribution functions with experiment for correlated wavefunction methods and various DFT XC functionals. The first column shows the O-O RDF for bulk water, considering nuclear quantum effects with RPA and MP2 (a), classical nuclei for RPA and MP2 (b) and classical nuclei for DFT (c). Columns 2 and 3 show the O-Cl and O-Na RDFs respectively of a solvated NaCl ion pair in water for RPA and MP2 with classical nuclei (d and f) and DFT with classical nuclei (e and g). Experimental references in each case are shaded in grey. The bulk water O-O experimental RDF is taken from Ref.~\citenum{Skinner2013/10.1063/1.4790861}. Cl-O is neutron scattering data from Ref.~\citenum{Soper2006/10.1016/j.bpc.2006.04.009} for KCl. The Na-O reference from Ref.~\citenum{Galib2017/10.1063/1.4975608} is the rescaled peak from X-ray diffraction data.
    \label{fig:rdf}}
\end{figure*}
The forthcoming discussion will simply refer to all of these models by the name of the reference method to which they have been trained.
We believe this is valid since we have rigorously benchmarked the capability of the MLPs to reproduce their underlying reference method (see Supplementary Information (SI)). %
We first explore the capabilities of different DFT XC functionals and correlated wave-function methods to accurately describe the structure of ions in water, showing that both RPA and MP2 generally reproduce experimental densities and radial distribution functions for both bulk water and solvated ions.
We then use these wavefunction based methods as a benchmark against which to compare the predictions of DFT and classical force-fields for the potential of mean force of a NaCl ion pair in water. 
We provide simple metrics based on the constituent interactions to rationalise the overall shape of the PMF, while also providing insight into the ability of DFT to capture these interactions.
In all cases, the simulations performed are well-converged with respect to statistical sampling and are not significantly impacted by finite-size effects, highlighting a major advantage of the MLP approach over \textit{ab initio} methods.
We finish with an outlook on the potential of these models for describing more complex situations such as confined electrolytes and for computing dynamical properties.

\begin{figure}[ht]
    \centering
    \includegraphics[width=0.5\textwidth]{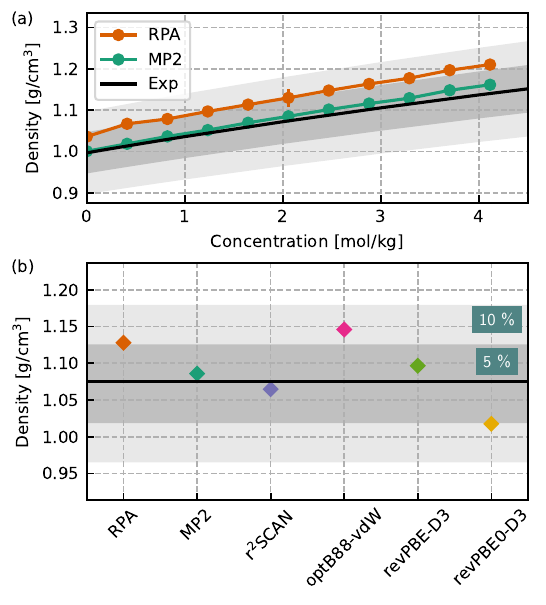}
	
	\caption{Comparison of densities with experiment \cite{Pitzer1984/10.1063/1.555709} for different DFT XC functionals and beyond DFT methods. Panel (a) shows the NaCl concentration-dependent density for RPA and MP2 at 300 K.  Panel (b) compares the experimental prediction at 300 K for a 2 M NaCl solution. In both cases a 5 and 10 \% boundary around experiment is shown shaded in dark and light grey respectively to facilitate comparisons.
    \label{fig:density}}
\end{figure}
In Figure \ref{fig:rdf}, we compare with experiment the performance of various DFT XC functionals, along with two wave function based methods MP2 and RPA, in predicting the radial distribution functions (RDFs) of both bulk water and Na and Cl ions in water. 
The DFT functionals have been chosen to span the various levels of Jacob's Ladder, with each level incorporating increased complexity into the XC functional description.
Specifically we have chosen the GGA revPBE-D3 \cite{Zhang1998/10.1103/PhysRevLett.80.890, Grimme2010/10.1063/1.3382344}, the van der Waals inclusive optB88-vdW \cite{Klimes2009/10.1088/0953-8984/22/2/022201}, the meta-GGA r$^2$SCAN \cite{Furness2020/10.1021/ACS.JPCLETT.0C02405} and the hybrid revPBE0-D3 \cite{Grimme2010/10.1063/1.3382344}.
Beyond DFT, we consider RPA and MP2, with both classical and quantum nuclei.
Inclusion of nuclear quantum effects have been shown to be necessary to obtain accurate agreement with experiment for liquid water structural and dynamical properties, including RDFs, diffusion coefficients and spectroscopy \cite{Pestana2018/10.1021/ACS.JPCLETT.8B02400,Marsalek2017/10.1021/ACS.JPCLETT.7B00391}.

We first consider the performance of the RPA and MP2 wavefunction-based methods.
Overall, out of all methods tested, RPA performs best in describing the structure of both bulk water and Cl and Na ions in water (Figure \ref{fig:rdf} (a), (d) and (f) respectively), accurately reproducing experimental RDFs.
For bulk water, the first peak height of the O-O RDF with RPA and classical nuclei (b) is overestimated by approximately 16 \%.
Inclusion of nuclear quantum effects (a) reduces the height of the first peak compared to classical nuclei (b), yielding excellent agreement with experiment, with just a slight overstructuring of the first peak, consistant with previous literature \cite{Yao2021/10.1021/acs.jpclett.1c01566}.
Meanwhile MP2, even with quantum nuclei shows some deficiencies for bulk water, predicting a more structured liquid by overestimating the height of the first peak by approximately 11\%, underestimating the first minimum also by approximately 11 \% and also predicting greater long-range order than experiment.
The poorer performance of MP2 for bulk water is consistent with previous literature.
Lan et al. showed that MP2 predicts overstructured water and a lower diffusion coefficient compared to experiment \cite{Lan2021/10.26434/CHEMRXIV-2021-N32Q8-V2}, attributing some of the shortcomings in that study to an incomplete basis set, while Willow et al. in Ref.~\citenum{Willow2016/10.1021/ACS.JPCLETT.5B02430} also show MP2 predicts denser water than experiment and DFT at ambient conditions. 

Upon comparing correlated wavefunction simulations with classical and quantum nuclei (see SI) it seems that the ion-water structure is more forgiving with respect to NQEs than bulk water for both RPA and MP2, with no significant improvement to the structural description when NQEs are included.
Both RPA and MP2 with classical nuclei are in excellent agreement with experiment for the Na-O and Cl-O RDFs, accurately reproducing the position and height of the first peak.
Simulations that include NQEs result in essentially identical RDFs for these cases, as shown in the SI.
It should be noted that only the first peak of the Na-O RDF was quoted from experimental X-ray diffraction (XRD) measurements \cite{Galib2017/10.1063/1.4975608}.
Also, the experimental Cl-O RDF is only available from a KCl solution \cite{{Soper2006/10.1016/j.bpc.2006.04.009}}.
While there are discrepancies beyond the first peak, the cation should not significantly influence the first solvation peak in the large separation limit. 

The challenges of accurately modelling bulk water \textit{and} ions in solution for DFT is apparent when comparing the different DFT XC functional predictions for the RDFs with experiment.
None of the functionals tested can simultaneously describe water and ions with the same accuracy as the best performing correlated wavefunction method, RPA.
As has been previously observed, revPBE-D3 performs very well for liquid water \cite{Skinner2016/10.1063/1.4944935/13698109, Marsalek2017/10.1021/ACS.JPCLETT.7B00391}, however it has been shown that this is in part due to a fortuitous cancellation of errors and this breaks down upon inclusion of NQEs \cite{Marsalek2017/10.1021/ACS.JPCLETT.7B00391}. 
Moreover, while it also performs well for the Cl-O RDF, it significantly underestimates the first peak of the Na-O RDF.
Inclusion of a fraction of exact exchange via the hybrid revPBE0-D3 functional shows similar good performance for bulk water and Cl-O, but does not improve the performance for Na-O. 
In contrast, r$^2$SCAN significantly overstructures bulk water, predicting a first RDF peak for O-O approximately 20\% greater than experiment, yet shows good agreement with experiment for both ion types in water. 
Various density corrections to the r$^2$SCAN functional have shown to improve its description of liquid water, \cite{Dasgupta2021/10.1038/s41467-021-26618-9} which warrant further tests on their suitability for aqueous electrolytes.
Similar to r$^2$SCAN, the van der Waals inclusive optB88-vdW functional also overstructures bulk water, but accurately predicts the ion-water RDF for both Na-O and Cl-O.
The poor performance for both revPBE-D3 and revPBE0-D3 for Na-O can be potentially ascribed to the fact that the D3 correction does not account for changes in dispersion due to charge-transfer effects (i.e., the formation of ions).
Cations have been shown to have a significantly different polarizability (i.e.\ dispersion) than their neutral atom counterpart compared to anions \cite{Ehrlich2011/10.1002/CPHC.201100521}, thus explaining the greater impact on sodium.
More sophisticated treatment of these cases such as the D4 correction \cite{Caldeweyher2020/10.1039/D0CP00502A}, methods incorporating iterative Hirschfeld partitioning \cite{Bucko2014/10.1063/1.4890003/194229}, and using van der Waals inclusive methods (as shown here with optB88-vdW) may alleviate this problem.
It has also been suggested that simply neglecting the D3 correction for the problematic cation interactions can improve agreement with experiment \cite{Kostal2023/10.1021/ACS.JPCLETT.3C00856}.

Accurately predicting the density is another important benchmark of the ML models and their underlying electronic structure reference method.
Figure \ref{fig:density} (a) compares the concentration dependent density prediction of RPA and MP2 with experiment.
This is a stern test of the electronic structure method but also the MLP quality, covering a significant concentration range.
Both MP2 and RPA show a similar qualitative increase in density with concentration of NaCl as observed in experiment. 
MP2 is within 1 \% of the experimental values, while RPA more significantly overestimates the density by approximately 5 \%.
Both RPA and MP2 MLP predictions for the bulk water density are also in close agreement to previous \textit{ab initio} predictions of 0.994 g/cm$^3$ \cite{Ben2015/10.1063/1.4927325/193869} and 1.020 g/cm$^3$ \cite{Ben2013/10.1021/JZ401931F} for RPA and MP2 respectively, and are within 5\% of the experimental value.
Figure \ref{fig:density} (b) compares the DFT and the wave-function MLP predictions of the density of a 2 M NaCl solution with the experimental value.
There is again a wide spread in DFT predictions, with most functionals tested overestimating the density of the 2 M NaCl solution apart from revPBE0-D3 which underestimates the density by 5\% compared to experiment, while r$^2$SCAN is in good agreement with experiment.
In comparison, RPA slightly overestimates the density, while MP2 is in very good agreement with experiment.
It should also be noted that while classical force-field models can accurately reproduce experimental density predictions, particularly in low concentration regimes \cite{Benavides2016/10.1063/1.4943780}, its agreement arises because the density is a property to which the water model has been explicitly fitted \cite{Berendsen1987/10.1021/J100308A038}.
At higher salt concentrations, away from the regions explicitly used in the parameterisation, force-field predictions also deviate from experiment \cite{Benavides2016/10.1063/1.4943780}.
To summarise, although RPA gives excellent structural properties for NaCl in water, it is slightly outperformed by MP2 in terms of the density response with increasing NaCl concentration. 
However, MP2 yields poorer agreement with experiment for structural properties than RPA. 
Nevertheless, the overall commendable all-round performance of RPA and MP2 for all cases of ions and water suggests them as reliable methods with which to study NaCl in water.
Of course the enhanced computational cost for initial training set computations for both RPA and MP2 over DFT should be mentioned, however this is a one-off investment during training.
Once a model is obtained, it can then be applied in simulations with a reduction of several orders of magnitude in computational cost - where the saving is greater for models based on more expensive methods.

\begin{figure*}[ht]
    \centering
    \includegraphics[width=\textwidth]{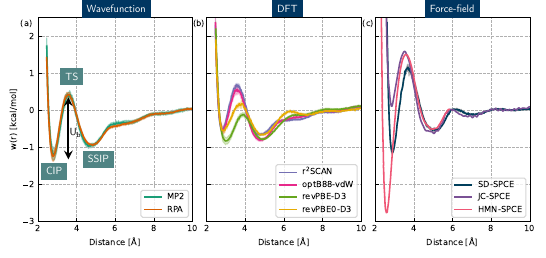}
        
	\caption{Potential of mean force for Na Cl ion pair in water with associated statistical error bars for various MLPs and classical forcefields: (a) Wavefunction methods RPA and MP2, with contact ion pair (CIP), transition state (TS), solvent-shared ion pair (SSIP) and barrier height ($\mathrm{U_b}$) annotated. (b) DFT, (c) Classical force-fields. HMN-SPCE is taken from Ref.~\citenum{Yao2018/10.1021/ACS.JCTC.7B00846} and JC-SPCE from Ref.~\citenum{Zhang2020/10.1038/s41467-020-16704-9}.
 \label{fig:pmf}}
\end{figure*}

Going beyond experimentally accessible properties, the potential of mean force (PMF), denoted here as w(r) is a fundamental property of the electrolyte governing the extent of ion pairing in solution and capturing collective solvent motion.
The key features of this quantity for \ch{Na^+} and \ch{Cl^-} in water are two local minima corresponding to the so-called contact-ion pair (CIP) and solvent-shared ion pair (SSIP), with free energies $\mathrm{U_{CIP}}$ and $\mathrm{U_{SSIP}}$ respectively.
These are separated by a barrier (TS) along the inter-ionic separation coordinate as illustrated in Figure \ref{fig:pmf} (a), where $\mathrm{U_b}$ is defined as $\mathrm{U_{TS} - U_{CIP}}$.
The relative stabilities of these minima ($\Delta \mathrm{U_{SSIP-CIP} } = \mathrm{U_{SSIP} - U_{CIP}}$), along with the barrier height ($\mathrm{U_b}$) are crucial to understanding the kinetics of ion-pair association and dissociation \cite{Geissler1999}.
However as previously mentioned, the PMF is highly sensitive to its underlying potential energy surface \cite{Duignan2016/10.1016/j.cocis.2016.05.009, Baer2016/10.1021/acs.jpcb.5b09579}.
Among both DFT and classical models, there is a major unresolved question regarding the relative stabilities of the CIP and SSIP \cite{Timko2010/10.1063/1.3360310/939701, Yao2018/10.1021/ACS.JCTC.7B00846, Zhang2020/10.1038/s41467-020-16704-9, Finney2022/10.1039/D1FD00089F}, with some classical models predicting a more stable CIP by up to 4 kcal/mol compared to DFT, which generally predicts almost degenerate CIP and SSIP states.
It should also be noted that the error bars on the DFT predictions are typically much greater, due to the significant computational effort required to converge the PMF.
Despite commendable efforts to resolve the PMF using \textit{ab initio} methods \cite{Yao2018/10.1021/ACS.JCTC.7B00846, Timko2010/10.1063/1.3360310/939701, Baer2016/10.1021/acs.jpcb.5b09579, Pluharova2013/10.1021/JZ402177Q}, it is computationally intensive to  statistically converge, requiring long simulation times ($\sim$ 2-3 ns) and several replicates to obtain statistical error bars.
Therefore, in absence of experimental benchmarks a reference PMF based on high-level electronic structure method is particularly valuable to the community.
It can provide clear atomistic insight to explain experimental observations of ion-pairing, and allow comparison among theoretical models -- both \textit{ab initio} and classical.

The MLPs developed in this work are ideal for efficient and statistically converged calculation of the NaCl PMF at a given level of theory.
We compute the PMF via the Na-Cl RDF of a 1 M NaCl solution comprising six ion-pairs in a box of 332 waters. 
This highlights another advantage of the MLP approach, where the potential energy surface along the inter-ionic separation can be sufficiently sampled through standard molecular dynamics simulations, without the requirement for enhanced sampling methods such as thermodynamic integration (we show in the SI the equivalence of both methods).
Moreover, these models have enabled sampling of the PMF out to the large-separation limit, providing valuable insights into the ion-pair binding energies.

Overall, over 200 ns of MD simulations were performed, which is significantly beyond the capabilities of \textit{ab initio} methods.
Figure \ref{fig:pmf} compares the PMF predictions
for MLPs at the same levels of electronic structure theory described above, as well as some commonly used classical force fields.
Statistical convergence and finite size effects have been carefully investigated and details are given in the SI.
We first note that the correlated wavefunction methods MP2 and RPA are in very good agreement, predicting that the CIP and SSIP have very similar stabilities.
They both predict that the two states are within $\sim$ 0.2 kcal/mol of each other.
This resolves the so far unanswered question regarding the relative stabilities of the CIP and SSIP states.
Moreover, of all the electronic structure methods tested, MP2 and RPA predict the largest barrier between the CIP and SSIP states of roughly 1.6 kcal/mol.
The excellent agreement between both high-level methods suggests this PMF can now be used as a benchmark to compare other methods.

\begin{figure*}[ht]
    \centering
    \includegraphics[width=\textwidth]{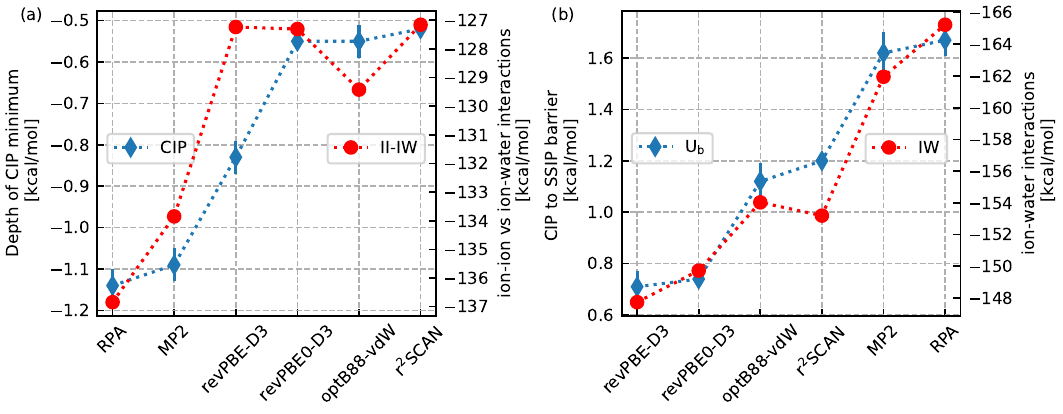}
        
	\caption{Relationship between constituent interactions in the \ch{Na^+} - \ch{Cl^-} solution and PMF observables. (a) The depth of the CIP minimum (CIP) is strongly correlated to balance of the ion-ion vs ion-water interactions whereby we have subtracted the ion-water interactions from ion-ion interactions (II-IW). (b) The barrier height between the CIP and SSIP (U$_\mathrm{b}$) is strongly correlated with the strength of the ion-water interactions (I-W).}.
 \label{fig:interactions}
\end{figure*}

Similar to the RDFs and densities shown above, DFT predictions also vary significantly (although not to the same extent as classical force-fields), and reinforce the considerations for selecting a suitable functional to study electrolyte systems.
While all methods are in relatively good agreement (within 0.2 kcal/mol) for the relative stabilities of the CIP and SSIP states, there is a larger range in the predictions of the barrier from SSIP and CIP and the depth of the CIP state.
The free energy barrier between the CIP and SSIP for both r$^2$SCAN and optB88-vdW is similar at approximately 1.25 kcal/mol. 
Meanwhile, although revPBE-D3 and revPBE0-D3 predict similar barrier heights for the CIP to SSIP transition, revPBE-D3 shows a shift to slightly larger distances for the locations of the CIP, SSIP and transition state.
Given the wide-ranging performance of the DFT methods tested here with respect to various aspects of the electrolyte structure (shown in Figures \ref{fig:rdf} and \ref{fig:density}), it is perhaps unsurprising that they also struggle to concur on the PMF, a property that depends sensitively on the collective interactions of sodium and chlorine with the surrounding water.

We now rationalise these trends in the PMFs by a heuristic analysis of the subtle balance of interactions that qualitatively match our findings. 
We focus on the two key interactions, namely the ion-ion and ion-water interactions.
We compute the binding energy of an ion pair in vacuum up to 3 \AA\space (before the transition to SSIP), to understand ion-ion (II) effects.
For ion-water (IW) effects, we compute the energy of interaction for an ion pair in water from equilibrium periodic snapshots taken from the points of interest along the PMF (CIP, SSIP, TS).
Details of these calculations as well as further comparisons are given in the SI.
Figure \ref{fig:interactions} shows the relationships between these constituent interactions and the PMF observables.
First, in Figure \ref{fig:interactions} (a), there is a strong correlation between the depth of the CIP minimum in the PMF and the difference of the ion-water and ion-ion interactions.
We show in the SI that considering the ion-ion interactions alone is not the best descriptor for the CIP well depth.
However we find that subtracting the ion-water interactions from the ion-ion interactions as shown here effectively captures the trend of CIP well depth across the functionals. 
All functionals except optB88-vdW show the same ordering for both properties (within error bars), suggesting the CIP region is primarily governed by a delicate interplay between ion-ion and ion-water interactions, whereby stronger ion-ion (ion-water) interactions strengthen (weaken) the CIP well depth.
However, the deviation of optB88-vdW from this trend suggests that there still may be important yet subtle effects for example from water-water interactions in the close contact region. 
This analysis builds on work by Duignan et al. \cite{Duignan2020/10.1039/C9CP06161D, Duignan2020/10.1021/ACS.JCTC.0C00300} who compute interaction energies from clusters extracted from simulations of cations in water. 
In Figure \ref{fig:interactions} (b) there is again a clear correlation between the barrier going from the CIP to SSIP states and the strength of the ion(-pair)-water interaction energy.
Here, within error bars, the largest transition barrier corresponds to the largest ion-water interaction energy.
We rationalise this observation by the fact that going from the SSIP to CIP state requires a rearrangement of the ions' solvation shell.
A lower ion-water interaction energy will thus result in a more facile transition from one minimum to the other, thereby lowering the barrier between the two states.

These qualitative metrics can also be further used to understand the performance of the various electronic structure methods employed here.
Firstly, for the ion-ion interactions, MP2 and RPA both shows the strongest binding compared to DFT. \cite{Riley2012/10.1021/JP211997B, Ren2011/10.1103/PHYSREVLETT.106.153003}
With respect to the ion-water interactions, MP2 and RPA also exhibit the strongest interactions, with the accuracy of the former having been verified by previous hydrated sodium cluster analysis. \cite{Riera2016/10.1039/C6CP02553F}
Meanwhile, revPBE-D3 and revPBE0-D3 demonstrate the weakest ion-water interactions.
We show in the SI that while the \ch{Na^+} -- \ch{H2O} interaction is overbound due to the D3 (as discussed previously), there is an underbinding in the \ch{Cl-} -- \ch{H2O} interaction.
This arises from the revPBE exchange enhancement factor causing stronger repulsion, tending to underbind \cite{Otero2012/10.1063/1.4705760/191680}, which then dominates the overall ion-water interaction.
In agreement with analyses on hydrated sodium ion clusters by Duignan et al. \cite{Duignan2020/10.1039/C9CP06161D}, we find that r$^2$SCAN (or SCAN in their work) gives better agreement to RPA than revPBE-D3.

Finally, the largest discrepancies in PMFs from all the classes of methods tested are in the classical force-field predictions.
The Smith-Dang/SPCE model \cite{Dang1993/10.1063/1.465441} is similar to the MP2 and RPA prediction, with the position of the SSIP slightly shifted to larger distances and a larger barrier by roughly 0.4 kcal/mol.
However, the Joung-Cheatham/SPC/E and HMN/SPC/E models significantly differ from all the other tested methods.
JC predicts a more stable SSIP by almost $1\,$kcal/mol, while HMN/SPCE predicts a more stable CIP by over $2\,$kcal/mol, and a barrier of over 4kcal/mol between CIP and SSIP.
As shown in the analysis of interactions above, given the sensitive dependance of the PMF on the interplay of interactions, this perhaps explains why classical force fields, which are typically fitted to observables rather than the interactions may struggle to converge on the PMF.

The machine learning approach used in this work offers a high level of accuracy and efficiency for studying electrolyte solutions, yielding high-quality, well-converged structural properties.
We compute the disputed potential of mean force of the NaCl ion pair in water using correlated wavefunction methods.
We find that correlated wavefunction methods are in very good agreement for the description of structural properties and densities.
Importantly, they also agree that the CIP and SSIP are essentially degenerate (within 0.2 kcal/mol).
DFT can reproduce these results with larger variations depending on the chosen functional, where we do not identify a functional that delivers convincing performance throughout the chosen properties.
Compared to the scale of force field predictions, in particular for ion-pairing, these differences remain small and candidates such as revPBE-D3 or r$^2$SCAN are expected to deliver a good compromise when simulating extended system sizes.
Going forward, the remaining differences among DFT XC functionals and even the discrepancies that remain among the wavefunction based methods point towards a need for an unambiguous benchmark quality model for these systems.
Given the recent successes by Chen et al. \cite{Chen2023/10.1021/ACS.JCTC.2C01203} and Daru et al. \cite{Daru2022/10.1103/PHYSREVLETT.129.226001} in performing CCSD(T) --considered the `gold-standard' of quantum chemistry-- based molecular dynamics for bulk water, there is an exciting opportunity to build on this work for the case of ions in water.
Their machine learning models accurately predict structural and dynamical properties of liquid water compared to experiment.
Using these approaches to train a model for water including ions is therefore a promising prospect to obtain a high quality potential energy surface for this system that can be used to further benchmark the lower level methods, and also to be directly used for simulations.

Furthermore, we have explored the effect of the various interactions of the system on the PMF, providing insight into the delicate interplay of ion-ion and ion-water interactions.
The simple metrics can now be used to benchmark other electronic structure or forcefield methods, using for example future CCSD(T)-based models as references.
From a more general perspective, our work highlights how ML potentials can now be used to efficiently screen various levels of theory directly on properties of interest, as has also been recently shown for the case of perovskite phase transitions \cite{Fransson2023/10.1021/ACS.JPCC.3C01542}.

Beyond simply comparing levels of theory, the PMFs and models generated in this work can next be used to provide valuable insights into experiment and computationally measured dynamical properties of electrolytes.
Obtaining accurate dynamical properties remains a frontier of research in the field of electrolyte simulations.
There has been significant work using \textit{ab initio} data to parameterise continuum scale models of electrolytes to access quantities such as osmotic coefficients and activities, along with transport properties such as diffusion coefficients and conductivity \cite{Duignan2016/10.1016/j.cocis.2016.05.009}.
This work shows that using MLPs offers an alternative yet complementary approach in the quest to obtain a fully \textit{ab initio} description of electrolytes.
In Figure S.10 of the SI, we report the water self-diffusion coefficients as a function of concentration relative to pure water for RPA and MP2 in good agreement with experiment, in particular for MP2.
We note that at the accurate electronic structure level employed in this work other factors such as the quantum nature of the nuclei~\cite{Ceriotti2016/10.1021/ACS.CHEMREV.5B00674} start to become a dominating factor to reach quantitative agreement with experiment, as shown recently for pure water~\cite{Marsalek2017/10.1021/ACS.JPCLETT.7B00391,Zhang2021/10.1021/ACS.JPCB.1C03884}.
The here developed machine learning potentials combined with path integral molecular dynamics enable to address this issue in future studies via efficient sampling of the potential energy surface.
From a fundamental perspective, building on the work of Geissler et al., a high level reference model is imperative to establish a quantitative understanding of the kinetics of ion pair dissociation \cite{Geissler1999}.
Similarly, a reliable model capable of accurately describing the bulk structure of electrolytes is primed to target dynamical properties of interest, including activity and osmotic coefficients, along with transport properties such as diffusion coefficients and conductivity.
Relating the PMF to Onsager transport coefficients, which quantify correlations in ion motion \cite{fong2020transport,Fong2021/10.1021/ACS.MACROMOL.0C02545} would be highly insightful to rationalise various transport phenomena, and an interesting first step would be to compute standard ion pair association energies \cite{Raiteri2020/10.1021/ACS.JPCB.0C01582}.
In particular, the ionic conductivity as a measure of the quantity of current an electrolyte can transport is vital for the development of next generation energy storage devices \cite{Chao2020/10.1126/SCIADV.ABA4098}.
Another important consideration and open question for these problems is the influence of long-range effects, given the large simulation cells required to compute these properties and the long-range electrostatic interactions between ions.
More sophisticated long-range schemes than that used in this work to name but a few include Refs. ~\citenum{Gao2022/10.1038/s41467-022-29243-2, grisafi2019/10.1063/1.5128375/197247, Finkler2023/10.1021/ACS.JCTC.2C01146} and a comparison of their performance would be highly insightful.
Moreover, recent equivarient graph-based architectures including MACE \cite{Batatia2022/2205.06643}, NEQUIP \cite{Batzner2021/10.1038/s41467-022-29939-5} and GRACE \cite{Bochkarev2022/10.1103/PHYSREVRESEARCH.4.L042019} are by construction longer range and are again highly promising to address the question of long-range interactions in an efficient manner.
Beyond the bulk, confinement of electrolytes leads to interesting physics and unexpected phenomena that are highly relevant to a range of applications, including blue energy harvesting \cite{Siria201710.1038/s41570-017-0091} and desalination \cite{Marbach2019/10.1039/C8CS00420J, Tunuguntla2017/10.1126/SCIENCE.AAN2438}.
In particular, extension of our current models to explore the intriguing phenomenon of confinement-induced ion pairing is very attractive.
Recent work has shown the assembly of confined ions in solution into long chains under an electric field, resulting in the so-called `memristor' effect \cite{Robin2021/10.1126/SCIENCE.ABF7923}, for which accurate atomistic insights are urgently needed \cite{Fong2024/10.26434/CHEMRXIV-2024-R67MX}.
\section*{Methods}
\noindent\textbf{Machine learning potential}
MLPs provide a direct mapping between a structure and its energy and forces, bypassing the computationally costly step of solving the Schrödinger equation for each timestep of a simulation. 
All of the models in this work are based on the seminal work of Behler and Parrinello \cite{Behler2007/10.1103/PHYSREVLETT.98.146401}, where we train a committee of eight neural network models \cite{Schran2020/10.1063/5.0016004/199713} for Na and Cl ions in water on forces and energies computed at various levels of electronic structure theory.
The model is systematically trained over three generations. The first comprises a general training set common to all models obtained from previous work \cite{ONeill2022/2211.04345} for NaCl ions in water. 
We then use an active learning procedure \cite{Schran2021/10.1073/pnas.2110077118} screening different solution concentrations to augment the training set for each model to ensure relevant configurations for a given level of theory are included to give the second generation model used in production simulations.
The third generation ensures the large ion-ion separation limit is accessible, by performing another active learning scheme on non-isotropic simulations cells or transfer learning depending on the specifc model. 
To address the question of long-range forces in our models, the final model is trained only on short range energies and forces, where the long-range electrostatic term (based on fixed point charges assigned to each atom, and computed by particle mesh Ewald summation) has been subtracted. This Coulombic baseline is then added on during simulations to give the full forces and energies.
Complete technical details of the models' training and validation are given in the SI.

\noindent\textbf{Molecular dynamics simulations}
All classical simulations were carried out using the CP2K/Quickstep code at a constant temperature of 300 K maintained using the Canonical Sampling Through Velocity Rescaling (CSVR) thermostat \cite{Bussi2007/10.1063/1.2408420} and with a 1 fs timestep.

\textbf{PMF}: The PMF calculations were carried out with a 1.0 mol/kg NaCl solution comprising 6 NaCl ion pairs in 332 waters.
This was first equillibrated in the NpT ensemble to obtain the equillibrium density for a given level of theory. 
To ensure sufficient statistical sampling, between five to ten uncorrelated configurations were then sampled from this trajectory and used as independent initial conditions.
Production simulations were then performed for at least 2 ns in the NVT ensemble at the previously computed equilibrium density, after which the radial distribution functions were obtained.
The potential of mean force $w({r})$ between Na and Cl was computed using the relation $w(
{r}) = -k_BT\,\mathrm{ln}\,g({r_{\mathrm{Na-Cl}}})$, where ${k_B}$ and ${T}$ are the Boltzman constant and temperature, with $g_{\mathrm{Na-Cl}}(r)$ being the radial distribution function between Na and Cl ion pairs. 
The final result was given by the average of these 10 runs, with the standard deviation providing an error estimate. 
We show in the SI that this approach is fully consistant to performing constrained molecular dynamics with thermodynamic integration and does not suffer from finite size effects.

\textbf{RDFs:}
The RDF calculations for bulk water were obtained from simulations of 126 water molecules in the NVT ensemble at the experimental density. RDFs for ions were obtained from simulations of a single ion pair and 95 water molecules. 
Path integral simulations for RPA and MP2 were performed using ring polymer molecular dynamics with 16 replicas, using the PILE thermostat \cite{Ceriotti2010/10.1063/1.3489925}.
\textbf{Electronic structure}
Electronic structure calculations were all carried out using the CP2K/Quickstep code. A plane-wave cutoff of 1200 Ry was required to converge the PMF (see SI for detailed convergence tests), and a TZ quality basis set was used for each model. Further details of specific electronic structure settings for various levels of theory are given in the SI, including convergence tests for the various electronic structure settings.
\section*{Data \& Code Availability}
All data required to reproduce the findings of this study is available at GitHub (\url{https://github.com/niamhon/nacl-water}).
All simulations were performed with publicly available simulation software (\texttt{n2p2}, \texttt{CP2K}), while the active learning package is available at GitHub (\url{https://github.com/MarsalekGroup/aml-dev}).
 \section*{Supporting Information}
Supporting Information (SI) contains electronic structure settings and convergence tests, details on machine learning potential development, simulation details and additional results including diffusion coefficients, interaction analysis and rdfs.
\section*{Acknowledgments}
We are grateful for insightful discussions with Erin Johnson and Julian Gale.
N.O.N acknowledges financial support from the Gates Cambridge Trust.
B.X.S. acknowledges support from the EPSRC Doctoral Training Partnership (EP/T517847/1).
KF acknowledges financial support from Schmidt Science Fellows.
AM acknowledges support from the European Union under the “n-AQUA” European Research Council project (Grant No. 101071937).
C.S acknowledges financial support from the Alexander von Humboldt Stiftung and the Deutsche Forschungsgemeinschaft (DFG, German Research Foundation) project number 500244608.
We are grateful for computational support and resources from the UK Materials and Molecular Modeling Hub which is partially funded by EPSRC (Grant Nos. EP/P020194/1 and EP/T022213/1).
We are also grateful for computational support and resources from the UK national high-performance computing service, Advanced Research Computing High End Resource (ARCHER2).
Access for both the UK Materials and Molecular Modeling Hub and ARCHER2 were obtained via the UK Car-Parrinello consortium, funded by EPSRC grant reference EP/P022561/1.
Access to CSD3 was obtained through a University of Cambridge EPSRC Core Equipment Award (EP/X034712/1).
%\bibliography{references.bib}

\begin{thebibliography}{103}%
\makeatletter
\providecommand \@ifxundefined [1]{%
 \@ifx{#1\undefined}
}%
\providecommand \@ifnum [1]{%
 \ifnum #1\expandafter \@firstoftwo
 \else \expandafter \@secondoftwo
 \fi
}%
\providecommand \@ifx [1]{%
 \ifx #1\expandafter \@firstoftwo
 \else \expandafter \@secondoftwo
 \fi
}%
\providecommand \natexlab [1]{#1}%
\providecommand \enquote  [1]{``#1''}%
\providecommand \bibnamefont  [1]{#1}%
\providecommand \bibfnamefont [1]{#1}%
\providecommand \citenamefont [1]{#1}%
\providecommand \href@noop [0]{\@secondoftwo}%
\providecommand \href [0]{\begingroup \@sanitize@url \@href}%
\providecommand \@href[1]{\@@startlink{#1}\@@href}%
\providecommand \@@href[1]{\endgroup#1\@@endlink}%
\providecommand \@sanitize@url [0]{\catcode `\\12\catcode `\$12\catcode
  `\&12\catcode `\#12\catcode `\^12\catcode `\_12\catcode `\%12\relax}%
\providecommand \@@startlink[1]{}%
\providecommand \@@endlink[0]{}%
\providecommand \url  [0]{\begingroup\@sanitize@url \@url }%
\providecommand \@url [1]{\endgroup\@href {#1}{\urlprefix }}%
\providecommand \urlprefix  [0]{URL }%
\providecommand \Eprint [0]{\href }%
\providecommand \doibase [0]{https://doi.org/}%
\providecommand \selectlanguage [0]{\@gobble}%
\providecommand \bibinfo  [0]{\@secondoftwo}%
\providecommand \bibfield  [0]{\@secondoftwo}%
\providecommand \translation [1]{[#1]}%
\providecommand \BibitemOpen [0]{}%
\providecommand \bibitemStop [0]{}%
\providecommand \bibitemNoStop [0]{.\EOS\space}%
\providecommand \EOS [0]{\spacefactor3000\relax}%
\providecommand \BibitemShut  [1]{\csname bibitem#1\endcsname}%
\let\auto@bib@innerbib\@empty
%</preamble>
\bibitem [{\citenamefont {Holoubek}\ \emph {et~al.}(2022)\citenamefont
  {Holoubek}, \citenamefont {Kim}, \citenamefont {Yin}, \citenamefont {Wu},
  \citenamefont {Liu}, \citenamefont {Li}, \citenamefont {Chen}, \citenamefont
  {Gao}, \citenamefont {Cai}, \citenamefont {Pascal}, \citenamefont {Liu},\
  and\ \citenamefont {Chen}}]{Holoubek2022/10.1039/D1EE03422G}%
  \BibitemOpen
  \bibfield  {author} {\bibinfo {author} {\bibfnamefont {J.}~\bibnamefont
  {Holoubek}}, \bibinfo {author} {\bibfnamefont {K.}~\bibnamefont {Kim}},
  \bibinfo {author} {\bibfnamefont {Y.}~\bibnamefont {Yin}}, \bibinfo {author}
  {\bibfnamefont {Z.}~\bibnamefont {Wu}}, \bibinfo {author} {\bibfnamefont
  {H.}~\bibnamefont {Liu}}, \bibinfo {author} {\bibfnamefont {M.}~\bibnamefont
  {Li}}, \bibinfo {author} {\bibfnamefont {A.}~\bibnamefont {Chen}}, \bibinfo
  {author} {\bibfnamefont {H.}~\bibnamefont {Gao}}, \bibinfo {author}
  {\bibfnamefont {G.}~\bibnamefont {Cai}}, \bibinfo {author} {\bibfnamefont
  {T.~A.}\ \bibnamefont {Pascal}}, \bibinfo {author} {\bibfnamefont
  {P.}~\bibnamefont {Liu}},\ and\ \bibinfo {author} {\bibfnamefont
  {Z.}~\bibnamefont {Chen}},\ }\bibfield  {title} {\enquote {\bibinfo {title}
  {Electrolyte design implications of ion-pairing in low-temperature li metal
  batteries},}\ }\href {https://doi.org/10.1039/D1EE03422G} {\bibfield
  {journal} {\bibinfo  {journal} {Energy \& Environmental Science}\ }\textbf
  {\bibinfo {volume} {15}},\ \bibinfo {pages} {1647--1658} (\bibinfo {year}
  {2022})}\BibitemShut {NoStop}%
\bibitem [{\citenamefont {Yao}\ \emph {et~al.}(2022)\citenamefont {Yao},
  \citenamefont {Chen}, \citenamefont {Fu},\ and\ \citenamefont
  {Zhang}}]{Yao2022/10.1021/ACS.CHEMREV.1C00904}%
  \BibitemOpen
  \bibfield  {author} {\bibinfo {author} {\bibfnamefont {N.}~\bibnamefont
  {Yao}}, \bibinfo {author} {\bibfnamefont {X.}~\bibnamefont {Chen}}, \bibinfo
  {author} {\bibfnamefont {Z.~H.}\ \bibnamefont {Fu}},\ and\ \bibinfo {author}
  {\bibfnamefont {Q.}~\bibnamefont {Zhang}},\ }\bibfield  {title} {\enquote
  {\bibinfo {title} {Applying classical, ab initio, and machine-learning
  molecular dynamics simulations to the liquid electrolyte for rechargeable
  batteries},}\ }\href {https://doi.org/10.1021/ACS.CHEMREV.1C00904} {\bibfield
   {journal} {\bibinfo  {journal} {Chemical Reviews}\ }\textbf {\bibinfo
  {volume} {122}},\ \bibinfo {pages} {10970--11021} (\bibinfo {year}
  {2022})}\BibitemShut {NoStop}%
\bibitem [{\citenamefont {Chen}\ \emph {et~al.}(2022)\citenamefont {Chen},
  \citenamefont {Zhang}, \citenamefont {Zou}, \citenamefont {Sun},\ and\
  \citenamefont {Tao}}]{Chen2022/10.1039/D2EE00004K}%
  \BibitemOpen
  \bibfield  {author} {\bibinfo {author} {\bibfnamefont {S.}~\bibnamefont
  {Chen}}, \bibinfo {author} {\bibfnamefont {M.}~\bibnamefont {Zhang}},
  \bibinfo {author} {\bibfnamefont {P.}~\bibnamefont {Zou}}, \bibinfo {author}
  {\bibfnamefont {B.}~\bibnamefont {Sun}},\ and\ \bibinfo {author}
  {\bibfnamefont {S.}~\bibnamefont {Tao}},\ }\bibfield  {title} {\enquote
  {\bibinfo {title} {Historical development and novel concepts on electrolytes
  for aqueous rechargeable batteries},}\ }\href
  {https://doi.org/10.1039/D2EE00004K} {\bibfield  {journal} {\bibinfo
  {journal} {Energy \& Environmental Science}\ }\textbf {\bibinfo {volume}
  {15}},\ \bibinfo {pages} {1805--1839} (\bibinfo {year} {2022})}\BibitemShut
  {NoStop}%
\bibitem [{\citenamefont {Zhang}\ \emph {et~al.}(2019)\citenamefont {Zhang},
  \citenamefont {Chen}, \citenamefont {Hou}, \citenamefont {Li}, \citenamefont
  {Cheng}, \citenamefont {Huang},\ and\ \citenamefont
  {Zhang}}]{Zhang2019/10.1021/ACSENERGYLETT.8B02376}%
  \BibitemOpen
  \bibfield  {author} {\bibinfo {author} {\bibfnamefont {X.~Q.}\ \bibnamefont
  {Zhang}}, \bibinfo {author} {\bibfnamefont {X.}~\bibnamefont {Chen}},
  \bibinfo {author} {\bibfnamefont {L.~P.}\ \bibnamefont {Hou}}, \bibinfo
  {author} {\bibfnamefont {B.~Q.}\ \bibnamefont {Li}}, \bibinfo {author}
  {\bibfnamefont {X.~B.}\ \bibnamefont {Cheng}}, \bibinfo {author}
  {\bibfnamefont {J.~Q.}\ \bibnamefont {Huang}},\ and\ \bibinfo {author}
  {\bibfnamefont {Q.}~\bibnamefont {Zhang}},\ }\bibfield  {title} {\enquote
  {\bibinfo {title} {Regulating anions in the solvation sheath of lithium ions
  for stable lithium metal batteries},}\ }\href
  {https://doi.org/10.1021/ACSENERGYLETT.8B02376} {\bibfield  {journal}
  {\bibinfo  {journal} {ACS Energy Letters}\ }\textbf {\bibinfo {volume} {4}},\
  \bibinfo {pages} {411--416} (\bibinfo {year} {2019})}\BibitemShut {NoStop}%
\bibitem [{\citenamefont {Peng}\ \emph {et~al.}(2018)\citenamefont {Peng},
  \citenamefont {Cao}, \citenamefont {He}, \citenamefont {Guo}, \citenamefont
  {Hapala}, \citenamefont {Ma}, \citenamefont {Cheng}, \citenamefont {Chen},
  \citenamefont {Xie}, \citenamefont {Li}, \citenamefont {Jelínek},
  \citenamefont {Xu}, \citenamefont {Gao}, \citenamefont {Wang},\ and\
  \citenamefont {Jiang}}]{Peng2018/10.1038/s41586-018-0122-2}%
  \BibitemOpen
  \bibfield  {author} {\bibinfo {author} {\bibfnamefont {J.}~\bibnamefont
  {Peng}}, \bibinfo {author} {\bibfnamefont {D.}~\bibnamefont {Cao}}, \bibinfo
  {author} {\bibfnamefont {Z.}~\bibnamefont {He}}, \bibinfo {author}
  {\bibfnamefont {J.}~\bibnamefont {Guo}}, \bibinfo {author} {\bibfnamefont
  {P.}~\bibnamefont {Hapala}}, \bibinfo {author} {\bibfnamefont
  {R.}~\bibnamefont {Ma}}, \bibinfo {author} {\bibfnamefont {B.}~\bibnamefont
  {Cheng}}, \bibinfo {author} {\bibfnamefont {J.}~\bibnamefont {Chen}},
  \bibinfo {author} {\bibfnamefont {W.~J.}\ \bibnamefont {Xie}}, \bibinfo
  {author} {\bibfnamefont {X.~Z.}\ \bibnamefont {Li}}, \bibinfo {author}
  {\bibfnamefont {P.}~\bibnamefont {Jelínek}}, \bibinfo {author}
  {\bibfnamefont {L.~M.}\ \bibnamefont {Xu}}, \bibinfo {author} {\bibfnamefont
  {Y.~Q.}\ \bibnamefont {Gao}}, \bibinfo {author} {\bibfnamefont {E.~G.}\
  \bibnamefont {Wang}},\ and\ \bibinfo {author} {\bibfnamefont
  {Y.}~\bibnamefont {Jiang}},\ }\bibfield  {title} {\enquote {\bibinfo {title}
  {The effect of hydration number on the interfacial transport of sodium
  ions},}\ }\href {https://doi.org/10.1038/s41586-018-0122-2} {\bibfield
  {journal} {\bibinfo  {journal} {Nature 2018 557:7707}\ }\textbf {\bibinfo
  {volume} {557}},\ \bibinfo {pages} {701--705} (\bibinfo {year}
  {2018})}\BibitemShut {NoStop}%
\bibitem [{\citenamefont {Duignan}, \citenamefont {Baer},\ and\ \citenamefont
  {Mundy}(2016)}]{Duignan2016/10.1016/j.cocis.2016.05.009}%
  \BibitemOpen
  \bibfield  {author} {\bibinfo {author} {\bibfnamefont {T.~T.}\ \bibnamefont
  {Duignan}}, \bibinfo {author} {\bibfnamefont {M.~D.}\ \bibnamefont {Baer}},\
  and\ \bibinfo {author} {\bibfnamefont {C.~J.}\ \bibnamefont {Mundy}},\ }\href
  {https://doi.org/10.1016/j.cocis.2016.05.009} {\enquote {\bibinfo {title}
  {Ions interacting in solution: Moving from intrinsic to collective
  properties},}\ } (\bibinfo {year} {2016})\BibitemShut {NoStop}%
\bibitem [{\citenamefont {Baer}\ and\ \citenamefont
  {Mundy}(2016)}]{Baer2016/10.1021/acs.jpcb.5b09579}%
  \BibitemOpen
  \bibfield  {author} {\bibinfo {author} {\bibfnamefont {M.~D.}\ \bibnamefont
  {Baer}}\ and\ \bibinfo {author} {\bibfnamefont {C.~J.}\ \bibnamefont
  {Mundy}},\ }\bibfield  {title} {\enquote {\bibinfo {title} {Local aqueous
  solvation structure around ca2+ during ca2+···cl- pair formation},}\
  }\href {https://doi.org/10.1021/acs.jpcb.5b09579} {\bibfield  {journal}
  {\bibinfo  {journal} {Journal of Physical Chemistry B}\ }\textbf {\bibinfo
  {volume} {120}},\ \bibinfo {pages} {1885--1893} (\bibinfo {year}
  {2016})}\BibitemShut {NoStop}%
\bibitem [{\citenamefont {Joung}\ and\ \citenamefont
  {Cheatham}(2008)}]{Joung2008/10.1021/jp8001614}%
  \BibitemOpen
  \bibfield  {author} {\bibinfo {author} {\bibfnamefont {I.~S.}\ \bibnamefont
  {Joung}}\ and\ \bibinfo {author} {\bibfnamefont {T.~E.}\ \bibnamefont
  {Cheatham}},\ }\bibfield  {title} {\enquote {\bibinfo {title} {Determination
  of alkali and halide monovalent ion parameters for use in explicitly solvated
  biomolecular simulations},}\ }\href {https://doi.org/10.1021/jp8001614}
  {\bibfield  {journal} {\bibinfo  {journal} {Journal of Physical Chemistry B}\
  }\textbf {\bibinfo {volume} {112}},\ \bibinfo {pages} {9020--9041} (\bibinfo
  {year} {2008})}\BibitemShut {NoStop}%
\bibitem [{\citenamefont {Dang}\ and\ \citenamefont
  {Smith}(1993)}]{Dang1993/10.1063/1.465441}%
  \BibitemOpen
  \bibfield  {author} {\bibinfo {author} {\bibfnamefont {L.~X.}\ \bibnamefont
  {Dang}}\ and\ \bibinfo {author} {\bibfnamefont {D.~E.}\ \bibnamefont
  {Smith}},\ }\bibfield  {title} {\enquote {\bibinfo {title} {Molecular
  dynamics simulations of aqueous ionic clusters using polarizable water},}\
  }\href {https://doi.org/10.1063/1.465441} {\bibfield  {journal} {\bibinfo
  {journal} {The Journal of Chemical Physics}\ }\textbf {\bibinfo {volume}
  {99}},\ \bibinfo {pages} {6950--6956} (\bibinfo {year} {1993})}\BibitemShut
  {NoStop}%
\bibitem [{\citenamefont {Zhang}\ \emph {et~al.}(2020)\citenamefont {Zhang},
  \citenamefont {Giberti}, \citenamefont {Sevgen}, \citenamefont {de~Pablo},
  \citenamefont {Gygi},\ and\ \citenamefont
  {Galli}}]{Zhang2020/10.1038/s41467-020-16704-9}%
  \BibitemOpen
  \bibfield  {author} {\bibinfo {author} {\bibfnamefont {C.}~\bibnamefont
  {Zhang}}, \bibinfo {author} {\bibfnamefont {F.}~\bibnamefont {Giberti}},
  \bibinfo {author} {\bibfnamefont {E.}~\bibnamefont {Sevgen}}, \bibinfo
  {author} {\bibfnamefont {J.~J.}\ \bibnamefont {de~Pablo}}, \bibinfo {author}
  {\bibfnamefont {F.}~\bibnamefont {Gygi}},\ and\ \bibinfo {author}
  {\bibfnamefont {G.}~\bibnamefont {Galli}},\ }\bibfield  {title} {\enquote
  {\bibinfo {title} {Dissociation of salts in water under pressure},}\ }\href
  {https://doi.org/10.1038/s41467-020-16704-9} {\bibfield  {journal} {\bibinfo
  {journal} {Nature Communications 2020 11:1}\ }\textbf {\bibinfo {volume}
  {11}},\ \bibinfo {pages} {1--9} (\bibinfo {year} {2020})}\BibitemShut
  {NoStop}%
\bibitem [{\citenamefont {Smith}\ and\ \citenamefont
  {Dang}(1994)}]{Smith1994/10.1063/1.466363}%
  \BibitemOpen
  \bibfield  {author} {\bibinfo {author} {\bibfnamefont {D.~E.}\ \bibnamefont
  {Smith}}\ and\ \bibinfo {author} {\bibfnamefont {L.~X.}\ \bibnamefont
  {Dang}},\ }\bibfield  {title} {\enquote {\bibinfo {title} {Computer
  simulations of nacl association in polarizable water},}\ }\href
  {https://doi.org/10.1063/1.466363} {\bibfield  {journal} {\bibinfo  {journal}
  {The Journal of Chemical Physics}\ }\textbf {\bibinfo {volume} {100}},\
  \bibinfo {pages} {3757--3766} (\bibinfo {year} {1994})}\BibitemShut {NoStop}%
\bibitem [{\citenamefont {Gillan}, \citenamefont {Alfè},\ and\ \citenamefont
  {Michaelides}(2016)}]{Gillan2016/10.1063/1.4944633/152221}%
  \BibitemOpen
  \bibfield  {author} {\bibinfo {author} {\bibfnamefont {M.~J.}\ \bibnamefont
  {Gillan}}, \bibinfo {author} {\bibfnamefont {D.}~\bibnamefont {Alfè}},\ and\
  \bibinfo {author} {\bibfnamefont {A.}~\bibnamefont {Michaelides}},\
  }\bibfield  {title} {\enquote {\bibinfo {title} {Perspective: How good is dft
  for water?}}\ }\href {https://doi.org/10.1063/1.4944633} {\bibfield
  {journal} {\bibinfo  {journal} {Journal of Chemical Physics}\ }\textbf
  {\bibinfo {volume} {144}},\ \bibinfo {pages} {130901} (\bibinfo {year}
  {2016})}\BibitemShut {NoStop}%
\bibitem [{\citenamefont {Duignan}\ \emph {et~al.}(2021)\citenamefont
  {Duignan}, \citenamefont {Kathmann}, \citenamefont {Schenter},\ and\
  \citenamefont {Mundy}}]{Duignan2021/10.1021/ACS.ACCOUNTS.1C00107}%
  \BibitemOpen
  \bibfield  {author} {\bibinfo {author} {\bibfnamefont {T.~T.}\ \bibnamefont
  {Duignan}}, \bibinfo {author} {\bibfnamefont {S.~M.}\ \bibnamefont
  {Kathmann}}, \bibinfo {author} {\bibfnamefont {G.~K.}\ \bibnamefont
  {Schenter}},\ and\ \bibinfo {author} {\bibfnamefont {C.~J.}\ \bibnamefont
  {Mundy}},\ }\bibfield  {title} {\enquote {\bibinfo {title} {Toward a
  first-principles framework for predicting collective properties of
  electrolytes},}\ }\href {https://doi.org/10.1021/ACS.ACCOUNTS.1C00107}
  {\bibfield  {journal} {\bibinfo  {journal} {Accounts of Chemical Research}\
  }\textbf {\bibinfo {volume} {54}},\ \bibinfo {pages} {2833--2843} (\bibinfo
  {year} {2021})}\BibitemShut {NoStop}%
\bibitem [{\citenamefont {Blazquez}, \citenamefont {Conde},\ and\ \citenamefont
  {Vega}(2023)}]{Blazquez2023/10.1063/5.0136498/2871744}%
  \BibitemOpen
  \bibfield  {author} {\bibinfo {author} {\bibfnamefont {S.}~\bibnamefont
  {Blazquez}}, \bibinfo {author} {\bibfnamefont {M.~M.}\ \bibnamefont
  {Conde}},\ and\ \bibinfo {author} {\bibfnamefont {C.}~\bibnamefont {Vega}},\
  }\bibfield  {title} {\enquote {\bibinfo {title} {Scaled charges for ions: An
  improvement but not the final word for modeling electrolytes in water},}\
  }\href {https://doi.org/10.1063/5.0136498/2871744} {\bibfield  {journal}
  {\bibinfo  {journal} {Journal of Chemical Physics}\ }\textbf {\bibinfo
  {volume} {158}} (\bibinfo {year} {2023}),\
  10.1063/5.0136498/2871744}\BibitemShut {NoStop}%
\bibitem [{\citenamefont {Ding}, \citenamefont {Hassanali},\ and\ \citenamefont
  {Parrinello}(2014)}]{Ding2014/10.1073/PNAS.1400675111}%
  \BibitemOpen
  \bibfield  {author} {\bibinfo {author} {\bibfnamefont {Y.}~\bibnamefont
  {Ding}}, \bibinfo {author} {\bibfnamefont {A.~A.}\ \bibnamefont
  {Hassanali}},\ and\ \bibinfo {author} {\bibfnamefont {M.}~\bibnamefont
  {Parrinello}},\ }\bibfield  {title} {\enquote {\bibinfo {title} {Anomalous
  water diffusion in salt solutions},}\ }\href
  {https://doi.org/10.1073/PNAS.1400675111} {\bibfield  {journal} {\bibinfo
  {journal} {Proceedings of the National Academy of Sciences of the United
  States of America}\ }\textbf {\bibinfo {volume} {111}},\ \bibinfo {pages}
  {3310--3315} (\bibinfo {year} {2014})}\BibitemShut {NoStop}%
\bibitem [{\citenamefont {Panagiotopoulos}\ and\ \citenamefont
  {Yue}(2023)}]{Panagiotopoulos2023/10.1021/ACS.JPCB.2C07477}%
  \BibitemOpen
  \bibfield  {author} {\bibinfo {author} {\bibfnamefont {A.~Z.}\ \bibnamefont
  {Panagiotopoulos}}\ and\ \bibinfo {author} {\bibfnamefont {S.}~\bibnamefont
  {Yue}},\ }\bibfield  {title} {\enquote {\bibinfo {title} {Dynamics of aqueous
  electrolyte solutions: Challenges for simulations},}\ }\href
  {https://doi.org/10.1021/ACS.JPCB.2C07477} {\bibfield  {journal} {\bibinfo
  {journal} {Journal of Physical Chemistry B}\ }\textbf {\bibinfo {volume}
  {127}},\ \bibinfo {pages} {430--437} (\bibinfo {year} {2023})}\BibitemShut
  {NoStop}%
\bibitem [{\citenamefont {Bankura}, \citenamefont {Carnevale},\ and\
  \citenamefont {Klein}(2014)}]{Bankura2014/10.1080/00268976.2014.905721}%
  \BibitemOpen
  \bibfield  {author} {\bibinfo {author} {\bibfnamefont {A.}~\bibnamefont
  {Bankura}}, \bibinfo {author} {\bibfnamefont {V.}~\bibnamefont {Carnevale}},\
  and\ \bibinfo {author} {\bibfnamefont {M.~L.}\ \bibnamefont {Klein}},\
  }\bibfield  {title} {\enquote {\bibinfo {title} {Hydration structure of na+
  and k+ from ab initio molecular dynamics based on modern density functional
  theory},}\ }\href {https://doi.org/10.1080/00268976.2014.905721} {\bibfield
  {journal} {\bibinfo  {journal} {Molecular Physics}\ }\textbf {\bibinfo
  {volume} {112}},\ \bibinfo {pages} {1448--1456} (\bibinfo {year}
  {2014})}\BibitemShut {NoStop}%
\bibitem [{\citenamefont {Gaiduk}\ and\ \citenamefont
  {Galli}(2017)}]{Gaiduk2017/10.1021/ACS.JPCLETT.7B00239}%
  \BibitemOpen
  \bibfield  {author} {\bibinfo {author} {\bibfnamefont {A.~P.}\ \bibnamefont
  {Gaiduk}}\ and\ \bibinfo {author} {\bibfnamefont {G.}~\bibnamefont {Galli}},\
  }\bibfield  {title} {\enquote {\bibinfo {title} {Local and global effects of
  dissolved sodium chloride on the structure of water},}\ }\href
  {https://doi.org/10.1021/ACS.JPCLETT.7B00239} {\bibfield  {journal} {\bibinfo
   {journal} {Journal of Physical Chemistry Letters}\ }\textbf {\bibinfo
  {volume} {8}},\ \bibinfo {pages} {1496--1502} (\bibinfo {year}
  {2017})}\BibitemShut {NoStop}%
\bibitem [{\citenamefont {Galib}\ \emph
  {et~al.}(2017{\natexlab{a}})\citenamefont {Galib}, \citenamefont {Baer},
  \citenamefont {Skinner}, \citenamefont {Mundy}, \citenamefont {Huthwelker},
  \citenamefont {Schenter}, \citenamefont {Benmore}, \citenamefont {Govind},\
  and\ \citenamefont {Fulton}}]{Galib2017/10.1063/1.4975608}%
  \BibitemOpen
  \bibfield  {author} {\bibinfo {author} {\bibfnamefont {M.}~\bibnamefont
  {Galib}}, \bibinfo {author} {\bibfnamefont {M.~D.}\ \bibnamefont {Baer}},
  \bibinfo {author} {\bibfnamefont {L.~B.}\ \bibnamefont {Skinner}}, \bibinfo
  {author} {\bibfnamefont {C.~J.}\ \bibnamefont {Mundy}}, \bibinfo {author}
  {\bibfnamefont {T.}~\bibnamefont {Huthwelker}}, \bibinfo {author}
  {\bibfnamefont {G.~K.}\ \bibnamefont {Schenter}}, \bibinfo {author}
  {\bibfnamefont {C.~J.}\ \bibnamefont {Benmore}}, \bibinfo {author}
  {\bibfnamefont {N.}~\bibnamefont {Govind}},\ and\ \bibinfo {author}
  {\bibfnamefont {J.~L.}\ \bibnamefont {Fulton}},\ }\bibfield  {title}
  {\enquote {\bibinfo {title} {Revisiting the hydration structure of aqueous
  na+},}\ }\href {https://doi.org/10.1063/1.4975608} {\bibfield  {journal}
  {\bibinfo  {journal} {Journal of Chemical Physics}\ }\textbf {\bibinfo
  {volume} {146}} (\bibinfo {year} {2017}{\natexlab{a}}),\
  10.1063/1.4975608}\BibitemShut {NoStop}%
\bibitem [{\citenamefont {Duignan}\ \emph
  {et~al.}(2020{\natexlab{a}})\citenamefont {Duignan}, \citenamefont
  {Schenter}, \citenamefont {Fulton}, \citenamefont {Huthwelker}, \citenamefont
  {Balasubramanian}, \citenamefont {Galib}, \citenamefont {Baer}, \citenamefont
  {Wilhelm}, \citenamefont {Hutter}, \citenamefont {Ben}, \citenamefont
  {Zhao},\ and\ \citenamefont {Mundy}}]{Duignan2020/10.1039/C9CP06161D}%
  \BibitemOpen
  \bibfield  {author} {\bibinfo {author} {\bibfnamefont {T.~T.}\ \bibnamefont
  {Duignan}}, \bibinfo {author} {\bibfnamefont {G.~K.}\ \bibnamefont
  {Schenter}}, \bibinfo {author} {\bibfnamefont {J.~L.}\ \bibnamefont
  {Fulton}}, \bibinfo {author} {\bibfnamefont {T.}~\bibnamefont {Huthwelker}},
  \bibinfo {author} {\bibfnamefont {M.}~\bibnamefont {Balasubramanian}},
  \bibinfo {author} {\bibfnamefont {M.}~\bibnamefont {Galib}}, \bibinfo
  {author} {\bibfnamefont {M.~D.}\ \bibnamefont {Baer}}, \bibinfo {author}
  {\bibfnamefont {J.}~\bibnamefont {Wilhelm}}, \bibinfo {author} {\bibfnamefont
  {J.}~\bibnamefont {Hutter}}, \bibinfo {author} {\bibfnamefont {M.~D.}\
  \bibnamefont {Ben}}, \bibinfo {author} {\bibfnamefont {X.~S.}\ \bibnamefont
  {Zhao}},\ and\ \bibinfo {author} {\bibfnamefont {C.~J.}\ \bibnamefont
  {Mundy}},\ }\bibfield  {title} {\enquote {\bibinfo {title} {Quantifying the
  hydration structure of sodium and potassium ions: Taking additional steps on
  jacob's ladder},}\ }\href {https://doi.org/10.1039/c9cp06161d} {\bibfield
  {journal} {\bibinfo  {journal} {Physical Chemistry Chemical Physics}\
  }\textbf {\bibinfo {volume} {22}},\ \bibinfo {pages} {10641--10652} (\bibinfo
  {year} {2020}{\natexlab{a}})}\BibitemShut {NoStop}%
\bibitem [{\citenamefont {Rozsa}, \citenamefont {Pham},\ and\ \citenamefont
  {Galli}(2020)}]{Rozsa2020/10.1063/1.5143317/15575136}%
  \BibitemOpen
  \bibfield  {author} {\bibinfo {author} {\bibfnamefont {V.}~\bibnamefont
  {Rozsa}}, \bibinfo {author} {\bibfnamefont {T.~A.}\ \bibnamefont {Pham}},\
  and\ \bibinfo {author} {\bibfnamefont {G.}~\bibnamefont {Galli}},\ }\bibfield
   {title} {\enquote {\bibinfo {title} {Molecular polarizabilities as
  fingerprints of perturbations to water by ions and confinement},}\ }\href
  {https://doi.org/10.1063/1.5143317} {\bibfield  {journal} {\bibinfo
  {journal} {Journal of Chemical Physics}\ }\textbf {\bibinfo {volume} {152}},\
  \bibinfo {pages} {124501} (\bibinfo {year} {2020})}\BibitemShut {NoStop}%
\bibitem [{\citenamefont {Dellostritto}\ \emph {et~al.}(2020)\citenamefont
  {Dellostritto}, \citenamefont {Xu}, \citenamefont {Wu},\ and\ \citenamefont
  {Klein}}]{Dellostritto2020/10.1039/C9CP06821J}%
  \BibitemOpen
  \bibfield  {author} {\bibinfo {author} {\bibfnamefont {M.}~\bibnamefont
  {Dellostritto}}, \bibinfo {author} {\bibfnamefont {J.}~\bibnamefont {Xu}},
  \bibinfo {author} {\bibfnamefont {X.}~\bibnamefont {Wu}},\ and\ \bibinfo
  {author} {\bibfnamefont {M.~L.}\ \bibnamefont {Klein}},\ }\bibfield  {title}
  {\enquote {\bibinfo {title} {Aqueous solvation of the chloride ion revisited
  with density functional theory: impact of correlation and exchange
  approximations},}\ }\href {https://doi.org/10.1039/C9CP06821J} {\bibfield
  {journal} {\bibinfo  {journal} {Physical Chemistry Chemical Physics}\
  }\textbf {\bibinfo {volume} {22}},\ \bibinfo {pages} {10666--10675} (\bibinfo
  {year} {2020})}\BibitemShut {NoStop}%
\bibitem [{\citenamefont {Galib}\ \emph
  {et~al.}(2017{\natexlab{b}})\citenamefont {Galib}, \citenamefont {Duignan},
  \citenamefont {Misteli}, \citenamefont {Baer}, \citenamefont {Schenter},
  \citenamefont {Hutter},\ and\ \citenamefont
  {Mundy}}]{Galib2017/10.1063/1.4986284}%
  \BibitemOpen
  \bibfield  {author} {\bibinfo {author} {\bibfnamefont {M.}~\bibnamefont
  {Galib}}, \bibinfo {author} {\bibfnamefont {T.~T.}\ \bibnamefont {Duignan}},
  \bibinfo {author} {\bibfnamefont {Y.}~\bibnamefont {Misteli}}, \bibinfo
  {author} {\bibfnamefont {M.~D.}\ \bibnamefont {Baer}}, \bibinfo {author}
  {\bibfnamefont {G.~K.}\ \bibnamefont {Schenter}}, \bibinfo {author}
  {\bibfnamefont {J.}~\bibnamefont {Hutter}},\ and\ \bibinfo {author}
  {\bibfnamefont {C.~J.}\ \bibnamefont {Mundy}},\ }\bibfield  {title} {\enquote
  {\bibinfo {title} {Mass density fluctuations in quantum and classical
  descriptions of liquid water},}\ }\href {https://doi.org/10.1063/1.4986284}
  {\bibfield  {journal} {\bibinfo  {journal} {Journal of Chemical Physics}\
  }\textbf {\bibinfo {volume} {146}},\ \bibinfo {pages} {244501} (\bibinfo
  {year} {2017}{\natexlab{b}})}\BibitemShut {NoStop}%
\bibitem [{\citenamefont {Otero-De-La-Roza}\ and\ \citenamefont
  {Johnson}(2020)}]{Roza2020/10.1021/ACS.JPCA.9B10257}%
  \BibitemOpen
  \bibfield  {author} {\bibinfo {author} {\bibfnamefont {A.}~\bibnamefont
  {Otero-De-La-Roza}}\ and\ \bibinfo {author} {\bibfnamefont {E.~R.}\
  \bibnamefont {Johnson}},\ }\bibfield  {title} {\enquote {\bibinfo {title}
  {Analysis of density-functional errors for noncovalent interactions between
  charged molecules},}\ }\href {https://doi.org/10.1021/ACS.JPCA.9B10257}
  {\bibfield  {journal} {\bibinfo  {journal} {Journal of Physical Chemistry A}\
  }\textbf {\bibinfo {volume} {124}},\ \bibinfo {pages} {353--361} (\bibinfo
  {year} {2020})}\BibitemShut {NoStop}%
\bibitem [{\citenamefont {Bryenton}\ \emph {et~al.}(2023)\citenamefont
  {Bryenton}, \citenamefont {Adeleke}, \citenamefont {Dale},\ and\
  \citenamefont {Johnson}}]{Bryenton2023/10.1002/WCMS.1631}%
  \BibitemOpen
  \bibfield  {author} {\bibinfo {author} {\bibfnamefont {K.~R.}\ \bibnamefont
  {Bryenton}}, \bibinfo {author} {\bibfnamefont {A.~A.}\ \bibnamefont
  {Adeleke}}, \bibinfo {author} {\bibfnamefont {S.~G.}\ \bibnamefont {Dale}},\
  and\ \bibinfo {author} {\bibfnamefont {E.~R.}\ \bibnamefont {Johnson}},\
  }\bibfield  {title} {\enquote {\bibinfo {title} {Delocalization error: The
  greatest outstanding challenge in density-functional theory},}\ }\href
  {https://doi.org/10.1002/WCMS.1631} {\bibfield  {journal} {\bibinfo
  {journal} {Wiley Interdisciplinary Reviews: Computational Molecular Science}\
  }\textbf {\bibinfo {volume} {13}},\ \bibinfo {pages} {e1631} (\bibinfo {year}
  {2023})}\BibitemShut {NoStop}%
\bibitem [{\citenamefont {Cheng}\ \emph {et~al.}(2014)\citenamefont {Cheng},
  \citenamefont {Liu}, \citenamefont {VandeVondele}, \citenamefont {Sulpizi},\
  and\ \citenamefont {Sprik}}]{Cheng2014/10.1021/AR500268Y}%
  \BibitemOpen
  \bibfield  {author} {\bibinfo {author} {\bibfnamefont {J.}~\bibnamefont
  {Cheng}}, \bibinfo {author} {\bibfnamefont {X.}~\bibnamefont {Liu}}, \bibinfo
  {author} {\bibfnamefont {J.}~\bibnamefont {VandeVondele}}, \bibinfo {author}
  {\bibfnamefont {M.}~\bibnamefont {Sulpizi}},\ and\ \bibinfo {author}
  {\bibfnamefont {M.}~\bibnamefont {Sprik}},\ }\bibfield  {title} {\enquote
  {\bibinfo {title} {Redox potentials and acidity constants from density
  functional theory based molecular dynamics},}\ }\href
  {https://doi.org/10.1021/AR500268Y} {\bibfield  {journal} {\bibinfo
  {journal} {Accounts of Chemical Research}\ }\textbf {\bibinfo {volume}
  {47}},\ \bibinfo {pages} {3522--3529} (\bibinfo {year} {2014})}\BibitemShut
  {NoStop}%
\bibitem [{\citenamefont {Marsalek}\ and\ \citenamefont
  {Markland}(2017)}]{Marsalek2017/10.1021/ACS.JPCLETT.7B00391}%
  \BibitemOpen
  \bibfield  {author} {\bibinfo {author} {\bibfnamefont {O.}~\bibnamefont
  {Marsalek}}\ and\ \bibinfo {author} {\bibfnamefont {T.~E.}\ \bibnamefont
  {Markland}},\ }\bibfield  {title} {\enquote {\bibinfo {title} {Quantum
  dynamics and spectroscopy of ab initio liquid water: The interplay of nuclear
  and electronic quantum effects},}\ }\href
  {https://doi.org/10.1021/ACS.JPCLETT.7B00391} {\bibfield  {journal} {\bibinfo
   {journal} {Journal of Physical Chemistry Letters}\ }\textbf {\bibinfo
  {volume} {8}},\ \bibinfo {pages} {1545--1551} (\bibinfo {year}
  {2017})}\BibitemShut {NoStop}%
\bibitem [{\citenamefont {Pestana}\ \emph {et~al.}(2017)\citenamefont
  {Pestana}, \citenamefont {Mardirossian}, \citenamefont {Head-Gordon},\ and\
  \citenamefont {Head-Gordon}}]{Ruiz2017/10.1039/C6SC04711D}%
  \BibitemOpen
  \bibfield  {author} {\bibinfo {author} {\bibfnamefont {L.~R.}\ \bibnamefont
  {Pestana}}, \bibinfo {author} {\bibfnamefont {N.}~\bibnamefont
  {Mardirossian}}, \bibinfo {author} {\bibfnamefont {M.}~\bibnamefont
  {Head-Gordon}},\ and\ \bibinfo {author} {\bibfnamefont {T.}~\bibnamefont
  {Head-Gordon}},\ }\bibfield  {title} {\enquote {\bibinfo {title} {Ab initio
  molecular dynamics simulations of liquid water using high quality meta-gga
  functionals},}\ }\href {https://doi.org/10.1039/C6SC04711D} {\bibfield
  {journal} {\bibinfo  {journal} {Chemical Science}\ }\textbf {\bibinfo
  {volume} {8}},\ \bibinfo {pages} {3554--3565} (\bibinfo {year}
  {2017})}\BibitemShut {NoStop}%
\bibitem [{\citenamefont {Ceriotti}\ \emph {et~al.}(2016)\citenamefont
  {Ceriotti}, \citenamefont {Fang}, \citenamefont {Kusalik}, \citenamefont
  {McKenzie}, \citenamefont {Michaelides}, \citenamefont {Morales},\ and\
  \citenamefont {Markland}}]{Ceriotti2016/10.1021/ACS.CHEMREV.5B00674}%
  \BibitemOpen
  \bibfield  {author} {\bibinfo {author} {\bibfnamefont {M.}~\bibnamefont
  {Ceriotti}}, \bibinfo {author} {\bibfnamefont {W.}~\bibnamefont {Fang}},
  \bibinfo {author} {\bibfnamefont {P.~G.}\ \bibnamefont {Kusalik}}, \bibinfo
  {author} {\bibfnamefont {R.~H.}\ \bibnamefont {McKenzie}}, \bibinfo {author}
  {\bibfnamefont {A.}~\bibnamefont {Michaelides}}, \bibinfo {author}
  {\bibfnamefont {M.~A.}\ \bibnamefont {Morales}},\ and\ \bibinfo {author}
  {\bibfnamefont {T.~E.}\ \bibnamefont {Markland}},\ }\bibfield  {title}
  {\enquote {\bibinfo {title} {Nuclear quantum effects in water and aqueous
  systems: Experiment, theory, and current challenges},}\ }\href
  {https://doi.org/10.1021/ACS.CHEMREV.5B00674} {\bibfield  {journal} {\bibinfo
   {journal} {Chemical Reviews}\ }\textbf {\bibinfo {volume} {116}},\ \bibinfo
  {pages} {7529--7550} (\bibinfo {year} {2016})}\BibitemShut {NoStop}%
\bibitem [{\citenamefont {Xu}\ \emph {et~al.}(2021)\citenamefont {Xu},
  \citenamefont {Sun}, \citenamefont {Zhang}, \citenamefont {Dellostritto},
  \citenamefont {Lu}, \citenamefont {Klein},\ and\ \citenamefont
  {Wu}}]{Xu2021/10.1103/PhysRevMaterials.5.L012801}%
  \BibitemOpen
  \bibfield  {author} {\bibinfo {author} {\bibfnamefont {J.}~\bibnamefont
  {Xu}}, \bibinfo {author} {\bibfnamefont {Z.}~\bibnamefont {Sun}}, \bibinfo
  {author} {\bibfnamefont {C.}~\bibnamefont {Zhang}}, \bibinfo {author}
  {\bibfnamefont {M.}~\bibnamefont {Dellostritto}}, \bibinfo {author}
  {\bibfnamefont {D.}~\bibnamefont {Lu}}, \bibinfo {author} {\bibfnamefont
  {M.~L.}\ \bibnamefont {Klein}},\ and\ \bibinfo {author} {\bibfnamefont
  {X.}~\bibnamefont {Wu}},\ }\bibfield  {title} {\enquote {\bibinfo {title}
  {Importance of nuclear quantum effects on the hydration of chloride ion},}\
  }\href {https://doi.org/10.1103/PhysRevMaterials.5.L012801} {\bibfield
  {journal} {\bibinfo  {journal} {Physical Review Materials}\ }\textbf
  {\bibinfo {volume} {5}},\ \bibinfo {pages} {L012801} (\bibinfo {year}
  {2021})}\BibitemShut {NoStop}%
\bibitem [{\citenamefont {Wilkins}, \citenamefont {Manolopoulos},\ and\
  \citenamefont {Dang}(2015)}]{Wilkins2015/10.1063/1.4907554}%
  \BibitemOpen
  \bibfield  {author} {\bibinfo {author} {\bibfnamefont {D.~M.}\ \bibnamefont
  {Wilkins}}, \bibinfo {author} {\bibfnamefont {D.~E.}\ \bibnamefont
  {Manolopoulos}},\ and\ \bibinfo {author} {\bibfnamefont {L.~X.}\ \bibnamefont
  {Dang}},\ }\bibfield  {title} {\enquote {\bibinfo {title} {Nuclear quantum
  effects in water exchange around lithium and fluoride ions},}\ }\href
  {https://doi.org/10.1063/1.4907554} {\bibfield  {journal} {\bibinfo
  {journal} {Journal of Chemical Physics}\ }\textbf {\bibinfo {volume} {142}},\
  \bibinfo {pages} {064509} (\bibinfo {year} {2015})}\BibitemShut {NoStop}%
\bibitem [{\citenamefont {Perdew}\ and\ \citenamefont
  {Schmidt}(2001)}]{Perdew2003/10.1063/1.1390175}%
  \BibitemOpen
  \bibfield  {author} {\bibinfo {author} {\bibfnamefont {J.~P.}\ \bibnamefont
  {Perdew}}\ and\ \bibinfo {author} {\bibfnamefont {K.}~\bibnamefont
  {Schmidt}},\ }\bibfield  {title} {\enquote {\bibinfo {title} {Jacob’s
  ladder of density functional approximations for the exchange-correlation
  energy},}\ \ }(\bibinfo  {publisher} {AIP Publishing},\ \bibinfo {year}
  {2001})\ pp.\ \bibinfo {pages} {1--20}\BibitemShut {NoStop}%
\bibitem [{\citenamefont {Pestana}\ \emph {et~al.}(2018)\citenamefont
  {Pestana}, \citenamefont {Marsalek}, \citenamefont {Markland},\ and\
  \citenamefont {Head-Gordon}}]{Pestana2018/10.1021/ACS.JPCLETT.8B02400}%
  \BibitemOpen
  \bibfield  {author} {\bibinfo {author} {\bibfnamefont {L.~R.}\ \bibnamefont
  {Pestana}}, \bibinfo {author} {\bibfnamefont {O.}~\bibnamefont {Marsalek}},
  \bibinfo {author} {\bibfnamefont {T.~E.}\ \bibnamefont {Markland}},\ and\
  \bibinfo {author} {\bibfnamefont {T.}~\bibnamefont {Head-Gordon}},\
  }\bibfield  {title} {\enquote {\bibinfo {title} {The quest for accurate
  liquid water properties from first principles},}\ }\href
  {https://doi.org/10.1021/acs.jpclett.8b02400} {\bibfield  {journal} {\bibinfo
   {journal} {Journal of Physical Chemistry Letters}\ }\textbf {\bibinfo
  {volume} {9}},\ \bibinfo {pages} {5009--5016} (\bibinfo {year}
  {2018})}\BibitemShut {NoStop}%
\bibitem [{\citenamefont {Dasgupta}\ \emph {et~al.}(2022)\citenamefont
  {Dasgupta}, \citenamefont {Shahi}, \citenamefont {Bhetwal}, \citenamefont
  {Perdew},\ and\ \citenamefont
  {Paesani}}]{Dasgupta2022/10.1021/acs.jctc.2c00313}%
  \BibitemOpen
  \bibfield  {author} {\bibinfo {author} {\bibfnamefont {S.}~\bibnamefont
  {Dasgupta}}, \bibinfo {author} {\bibfnamefont {C.}~\bibnamefont {Shahi}},
  \bibinfo {author} {\bibfnamefont {P.}~\bibnamefont {Bhetwal}}, \bibinfo
  {author} {\bibfnamefont {J.~P.}\ \bibnamefont {Perdew}},\ and\ \bibinfo
  {author} {\bibfnamefont {F.}~\bibnamefont {Paesani}},\ }\bibfield  {title}
  {\enquote {\bibinfo {title} {How good is the density-corrected scan
  functional for neutral and ionic aqueous systems, and what is so right about
  the hartree-fock density?}}\ }\href
  {https://doi.org/10.1021/acs.jctc.2c00313} {\bibfield  {journal} {\bibinfo
  {journal} {Journal of Chemical Theory and Computation}\ }\textbf {\bibinfo
  {volume} {18}},\ \bibinfo {pages} {4745--4761} (\bibinfo {year}
  {2022})}\BibitemShut {NoStop}%
\bibitem [{\citenamefont {Riera}\ \emph {et~al.}(2017)\citenamefont {Riera},
  \citenamefont {Mardirossian}, \citenamefont {Bajaj}, \citenamefont {Götz},\
  and\ \citenamefont {Paesani}}]{Riera2017/10.1063/1.4993213}%
  \BibitemOpen
  \bibfield  {author} {\bibinfo {author} {\bibfnamefont {M.}~\bibnamefont
  {Riera}}, \bibinfo {author} {\bibfnamefont {N.}~\bibnamefont {Mardirossian}},
  \bibinfo {author} {\bibfnamefont {P.}~\bibnamefont {Bajaj}}, \bibinfo
  {author} {\bibfnamefont {A.~W.}\ \bibnamefont {Götz}},\ and\ \bibinfo
  {author} {\bibfnamefont {F.}~\bibnamefont {Paesani}},\ }\bibfield  {title}
  {\enquote {\bibinfo {title} {Toward chemical accuracy in the description of
  ion-water interactions through many-body representations. alkali-water dimer
  potential energy surfaces},}\ }\href {https://doi.org/10.1063/1.4993213}
  {\bibfield  {journal} {\bibinfo  {journal} {Journal of Chemical Physics}\
  }\textbf {\bibinfo {volume} {147}},\ \bibinfo {pages} {2698--2705} (\bibinfo
  {year} {2017})}\BibitemShut {NoStop}%
\bibitem [{\citenamefont {Al-Hamdani}\ and\ \citenamefont
  {Tkatchenko}(2019)}]{AlHamdani2019/10.1063/1.5075487/152312}%
  \BibitemOpen
  \bibfield  {author} {\bibinfo {author} {\bibfnamefont {Y.~S.}\ \bibnamefont
  {Al-Hamdani}}\ and\ \bibinfo {author} {\bibfnamefont {A.}~\bibnamefont
  {Tkatchenko}},\ }\bibfield  {title} {\enquote {\bibinfo {title}
  {Understanding non-covalent interactions in larger molecular complexes from
  first principles},}\ }\href {https://doi.org/10.1063/1.5075487/152312}
  {\bibfield  {journal} {\bibinfo  {journal} {Journal of Chemical Physics}\
  }\textbf {\bibinfo {volume} {150}},\ \bibinfo {pages} {10901} (\bibinfo
  {year} {2019})}\BibitemShut {NoStop}%
\bibitem [{\citenamefont {Ben}, \citenamefont {Hutter},\ and\ \citenamefont
  {VandeVondele}(2015{\natexlab{a}})}]{Ben2015/10.1063/1.4927325/193869}%
  \BibitemOpen
  \bibfield  {author} {\bibinfo {author} {\bibfnamefont {M.~D.}\ \bibnamefont
  {Ben}}, \bibinfo {author} {\bibfnamefont {J.}~\bibnamefont {Hutter}},\ and\
  \bibinfo {author} {\bibfnamefont {J.}~\bibnamefont {VandeVondele}},\
  }\bibfield  {title} {\enquote {\bibinfo {title} {Probing the structural and
  dynamical properties of liquid water with models including non-local electron
  correlation},}\ }\href {https://doi.org/10.1063/1.4927325} {\bibfield
  {journal} {\bibinfo  {journal} {Journal of Chemical Physics}\ }\textbf
  {\bibinfo {volume} {143}},\ \bibinfo {pages} {54506} (\bibinfo {year}
  {2015}{\natexlab{a}})}\BibitemShut {NoStop}%
\bibitem [{\citenamefont {Ben}\ \emph {et~al.}(2015)\citenamefont {Ben},
  \citenamefont {Schütt}, \citenamefont {Wentz}, \citenamefont {Messmer},
  \citenamefont {Hutter},\ and\ \citenamefont
  {VandeVondele}}]{Ben2015/10.1016/J.CPC.2014.10.021}%
  \BibitemOpen
  \bibfield  {author} {\bibinfo {author} {\bibfnamefont {M.~D.}\ \bibnamefont
  {Ben}}, \bibinfo {author} {\bibfnamefont {O.}~\bibnamefont {Schütt}},
  \bibinfo {author} {\bibfnamefont {T.}~\bibnamefont {Wentz}}, \bibinfo
  {author} {\bibfnamefont {P.}~\bibnamefont {Messmer}}, \bibinfo {author}
  {\bibfnamefont {J.}~\bibnamefont {Hutter}},\ and\ \bibinfo {author}
  {\bibfnamefont {J.}~\bibnamefont {VandeVondele}},\ }\bibfield  {title}
  {\enquote {\bibinfo {title} {Enabling simulation at the fifth rung of dft:
  Large scale rpa calculations with excellent time to solution},}\ }\href
  {https://doi.org/10.1016/j.cpc.2014.10.021} {\bibfield  {journal} {\bibinfo
  {journal} {Computer Physics Communications}\ }\textbf {\bibinfo {volume}
  {187}},\ \bibinfo {pages} {120--129} (\bibinfo {year} {2015})}\BibitemShut
  {NoStop}%
\bibitem [{\citenamefont {Ben}\ \emph {et~al.}(2013)\citenamefont {Ben},
  \citenamefont {Schönherr}, \citenamefont {Hutter},\ and\ \citenamefont
  {VandeVondele}}]{Ben2013/10.1021/JZ401931F}%
  \BibitemOpen
  \bibfield  {author} {\bibinfo {author} {\bibfnamefont {M.~D.}\ \bibnamefont
  {Ben}}, \bibinfo {author} {\bibfnamefont {M.}~\bibnamefont {Schönherr}},
  \bibinfo {author} {\bibfnamefont {J.}~\bibnamefont {Hutter}},\ and\ \bibinfo
  {author} {\bibfnamefont {J.}~\bibnamefont {VandeVondele}},\ }\bibfield
  {title} {\enquote {\bibinfo {title} {Bulk liquid water at ambient temperature
  and pressure from mp2 theory},}\ }\href {https://doi.org/10.1021/JZ401931F}
  {\bibfield  {journal} {\bibinfo  {journal} {Journal of Physical Chemistry
  Letters}\ }\textbf {\bibinfo {volume} {4}},\ \bibinfo {pages} {3753--3759}
  (\bibinfo {year} {2013})}\BibitemShut {NoStop}%
\bibitem [{\citenamefont {Wilhelm}\ \emph {et~al.}(2016)\citenamefont
  {Wilhelm}, \citenamefont {Seewald}, \citenamefont {Ben},\ and\ \citenamefont
  {Hutter}}]{Seewald2016/10.1021/ACS.JCTC.6B00840}%
  \BibitemOpen
  \bibfield  {author} {\bibinfo {author} {\bibfnamefont {J.}~\bibnamefont
  {Wilhelm}}, \bibinfo {author} {\bibfnamefont {P.}~\bibnamefont {Seewald}},
  \bibinfo {author} {\bibfnamefont {M.~D.}\ \bibnamefont {Ben}},\ and\ \bibinfo
  {author} {\bibfnamefont {J.}~\bibnamefont {Hutter}},\ }\bibfield  {title}
  {\enquote {\bibinfo {title} {Large-scale cubic-scaling random phase
  approximation correlation energy calculations using a gaussian basis},}\
  }\href {https://doi.org/10.1021/ACS.JCTC.6B00840} {\bibfield  {journal}
  {\bibinfo  {journal} {Journal of Chemical Theory and Computation}\ }\textbf
  {\bibinfo {volume} {12}},\ \bibinfo {pages} {5851--5859} (\bibinfo {year}
  {2016})}\BibitemShut {NoStop}%
\bibitem [{\citenamefont {Ben}, \citenamefont {Hutter},\ and\ \citenamefont
  {VandeVondele}(2015{\natexlab{b}})}]{Ben2015/10.1063/1.4919238/898068}%
  \BibitemOpen
  \bibfield  {author} {\bibinfo {author} {\bibfnamefont {M.~D.}\ \bibnamefont
  {Ben}}, \bibinfo {author} {\bibfnamefont {J.}~\bibnamefont {Hutter}},\ and\
  \bibinfo {author} {\bibfnamefont {J.}~\bibnamefont {VandeVondele}},\
  }\bibfield  {title} {\enquote {\bibinfo {title} {Forces and stress in second
  order møller-plesset perturbation theory for condensed phase systems within
  the resolution-of-identity gaussian and plane waves approach},}\ }\href
  {https://doi.org/10.1063/1.4919238/898068} {\bibfield  {journal} {\bibinfo
  {journal} {Journal of Chemical Physics}\ }\textbf {\bibinfo {volume} {143}},\
  \bibinfo {pages} {102803} (\bibinfo {year} {2015}{\natexlab{b}})}\BibitemShut
  {NoStop}%
\bibitem [{\citenamefont {Unke}\ \emph {et~al.}(2021)\citenamefont {Unke},
  \citenamefont {Chmiela}, \citenamefont {Sauceda}, \citenamefont {Gastegger},
  \citenamefont {Poltavsky}, \citenamefont {Schütt}, \citenamefont
  {Tkatchenko},\ and\ \citenamefont
  {Müller}}]{Unke2021/10.1021/ACS.CHEMREV.0C01111}%
  \BibitemOpen
  \bibfield  {author} {\bibinfo {author} {\bibfnamefont {O.~T.}\ \bibnamefont
  {Unke}}, \bibinfo {author} {\bibfnamefont {S.}~\bibnamefont {Chmiela}},
  \bibinfo {author} {\bibfnamefont {H.~E.}\ \bibnamefont {Sauceda}}, \bibinfo
  {author} {\bibfnamefont {M.}~\bibnamefont {Gastegger}}, \bibinfo {author}
  {\bibfnamefont {I.}~\bibnamefont {Poltavsky}}, \bibinfo {author}
  {\bibfnamefont {K.~T.}\ \bibnamefont {Schütt}}, \bibinfo {author}
  {\bibfnamefont {A.}~\bibnamefont {Tkatchenko}},\ and\ \bibinfo {author}
  {\bibfnamefont {K.~R.}\ \bibnamefont {Müller}},\ }\href
  {https://doi.org/10.1021/acs.chemrev.0c01111} {\enquote {\bibinfo {title}
  {Machine learning force fields},}\ } (\bibinfo {year} {2021})\BibitemShut
  {NoStop}%
\bibitem [{\citenamefont {Behler}\ and\ \citenamefont
  {Csányi}(2021)}]{Behler2021/10.1140/EPJB/S10051-021-00156-1}%
  \BibitemOpen
  \bibfield  {author} {\bibinfo {author} {\bibfnamefont {J.}~\bibnamefont
  {Behler}}\ and\ \bibinfo {author} {\bibfnamefont {G.}~\bibnamefont
  {Csányi}},\ }\bibfield  {title} {\enquote {\bibinfo {title} {Machine
  learning potentials for extended systems: a perspective},}\ }\href
  {https://doi.org/10.1140/epjb/s10051-021-00156-1} {\bibfield  {journal}
  {\bibinfo  {journal} {European Physical Journal B}\ }\textbf {\bibinfo
  {volume} {94}},\ \bibinfo {pages} {1--11} (\bibinfo {year}
  {2021})}\BibitemShut {NoStop}%
\bibitem [{\citenamefont {Morawietz}\ \emph {et~al.}(2016)\citenamefont
  {Morawietz}, \citenamefont {Singraber}, \citenamefont {Dellago},\ and\
  \citenamefont {Behler}}]{Morawietz2016/10.1073/pnas.160237511}%
  \BibitemOpen
  \bibfield  {author} {\bibinfo {author} {\bibfnamefont {T.}~\bibnamefont
  {Morawietz}}, \bibinfo {author} {\bibfnamefont {A.}~\bibnamefont
  {Singraber}}, \bibinfo {author} {\bibfnamefont {C.}~\bibnamefont {Dellago}},\
  and\ \bibinfo {author} {\bibfnamefont {J.}~\bibnamefont {Behler}},\
  }\bibfield  {title} {\enquote {\bibinfo {title} {How van der waals
  interactions determine the unique properties of water},}\ }\href
  {https://doi.org/10.1073/PNAS.1602375113/SUPPL_FILE/PNAS.1602375113.SAPP.PDF}
  {\bibfield  {journal} {\bibinfo  {journal} {Proceedings of the National
  Academy of Sciences of the United States of America}\ }\textbf {\bibinfo
  {volume} {113}},\ \bibinfo {pages} {8368--8373} (\bibinfo {year}
  {2016})}\BibitemShut {NoStop}%
\bibitem [{\citenamefont {Zhang}\ \emph {et~al.}(2021)\citenamefont {Zhang},
  \citenamefont {Tang}, \citenamefont {Chen}, \citenamefont {Xu}, \citenamefont
  {Zhang}, \citenamefont {Qiu}, \citenamefont {Perdew}, \citenamefont {Klein},\
  and\ \citenamefont {Wu}}]{Zhang2021/10.1021/ACS.JPCB.1C03884}%
  \BibitemOpen
  \bibfield  {author} {\bibinfo {author} {\bibfnamefont {C.}~\bibnamefont
  {Zhang}}, \bibinfo {author} {\bibfnamefont {F.}~\bibnamefont {Tang}},
  \bibinfo {author} {\bibfnamefont {M.}~\bibnamefont {Chen}}, \bibinfo {author}
  {\bibfnamefont {J.}~\bibnamefont {Xu}}, \bibinfo {author} {\bibfnamefont
  {L.}~\bibnamefont {Zhang}}, \bibinfo {author} {\bibfnamefont {D.~Y.}\
  \bibnamefont {Qiu}}, \bibinfo {author} {\bibfnamefont {J.~P.}\ \bibnamefont
  {Perdew}}, \bibinfo {author} {\bibfnamefont {M.~L.}\ \bibnamefont {Klein}},\
  and\ \bibinfo {author} {\bibfnamefont {X.}~\bibnamefont {Wu}},\ }\bibfield
  {title} {\enquote {\bibinfo {title} {Modeling liquid water by climbing up
  jacob's ladder in density functional theory facilitated by using deep neural
  network potentials},}\ }\href
  {https://doi.org/10.1021/ACS.JPCB.1C03884/ASSET/IMAGES/LARGE/JP1C03884_0006.JPEG}
  {\bibfield  {journal} {\bibinfo  {journal} {Journal of Physical Chemistry B}\
  }\textbf {\bibinfo {volume} {125}},\ \bibinfo {pages} {11444--11456}
  (\bibinfo {year} {2021})}\BibitemShut {NoStop}%
\bibitem [{\citenamefont {Yao}\ and\ \citenamefont
  {Kanai}(2020)}]{Yao2020/10.1063/5.0012815/14721885}%
  \BibitemOpen
  \bibfield  {author} {\bibinfo {author} {\bibfnamefont {Y.}~\bibnamefont
  {Yao}}\ and\ \bibinfo {author} {\bibfnamefont {Y.}~\bibnamefont {Kanai}},\
  }\bibfield  {title} {\enquote {\bibinfo {title} {Temperature dependence of
  nuclear quantum effects on liquid water via artificial neural network model
  based on scan meta-gga functional},}\ }\href
  {https://doi.org/10.1063/5.0012815/14721885/044114_1_ACCEPTED_MANUSCRIPT.PDF}
  {\bibfield  {journal} {\bibinfo  {journal} {Journal of Chemical Physics}\
  }\textbf {\bibinfo {volume} {153}},\ \bibinfo {pages} {44114} (\bibinfo
  {year} {2020})}\BibitemShut {NoStop}%
\bibitem [{\citenamefont {Schran}, \citenamefont {Brezina},\ and\ \citenamefont
  {Marsalek}(2020)}]{Schran2020/10.1063/5.0016004/199713}%
  \BibitemOpen
  \bibfield  {author} {\bibinfo {author} {\bibfnamefont {C.}~\bibnamefont
  {Schran}}, \bibinfo {author} {\bibfnamefont {K.}~\bibnamefont {Brezina}},\
  and\ \bibinfo {author} {\bibfnamefont {O.}~\bibnamefont {Marsalek}},\
  }\bibfield  {title} {\enquote {\bibinfo {title} {Committee neural network
  potentials control generalization errors and enable active learning},}\
  }\href {https://doi.org/10.1063/5.0016004/199713} {\bibfield  {journal}
  {\bibinfo  {journal} {Journal of Chemical Physics}\ }\textbf {\bibinfo
  {volume} {153}},\ \bibinfo {pages} {104105} (\bibinfo {year}
  {2020})}\BibitemShut {NoStop}%
\bibitem [{\citenamefont {Cheng}\ \emph {et~al.}(2019)\citenamefont {Cheng},
  \citenamefont {Engel}, \citenamefont {Behler}, \citenamefont {Dellago},\ and\
  \citenamefont {Ceriotti}}]{Cheng2019/10.1073/PNAS.1815117116}%
  \BibitemOpen
  \bibfield  {author} {\bibinfo {author} {\bibfnamefont {B.}~\bibnamefont
  {Cheng}}, \bibinfo {author} {\bibfnamefont {E.~A.}\ \bibnamefont {Engel}},
  \bibinfo {author} {\bibfnamefont {J.}~\bibnamefont {Behler}}, \bibinfo
  {author} {\bibfnamefont {C.}~\bibnamefont {Dellago}},\ and\ \bibinfo {author}
  {\bibfnamefont {M.}~\bibnamefont {Ceriotti}},\ }\bibfield  {title} {\enquote
  {\bibinfo {title} {Ab initio thermodynamics of liquid and solid water},}\
  }\href {https://doi.org/10.1073/PNAS.1815117116} {\bibfield  {journal}
  {\bibinfo  {journal} {Proceedings of the National Academy of Sciences of the
  United States of America}\ }\textbf {\bibinfo {volume} {116}},\ \bibinfo
  {pages} {1110--1115} (\bibinfo {year} {2019})}\BibitemShut {NoStop}%
\bibitem [{\citenamefont {Lan}\ \emph {et~al.}(2021)\citenamefont {Lan},
  \citenamefont {Wilkins}, \citenamefont {Rybkin}, \citenamefont {Iannuzzi},\
  and\ \citenamefont {Hutter}}]{Lan2021/10.26434/CHEMRXIV-2021-N32Q8-V2}%
  \BibitemOpen
  \bibfield  {author} {\bibinfo {author} {\bibfnamefont {J.}~\bibnamefont
  {Lan}}, \bibinfo {author} {\bibfnamefont {D.~M.}\ \bibnamefont {Wilkins}},
  \bibinfo {author} {\bibfnamefont {V.~V.}\ \bibnamefont {Rybkin}}, \bibinfo
  {author} {\bibfnamefont {M.}~\bibnamefont {Iannuzzi}},\ and\ \bibinfo
  {author} {\bibfnamefont {J.}~\bibnamefont {Hutter}},\ }\bibfield  {title}
  {\enquote {\bibinfo {title} {Quantum dynamics of water from møller-plesset
  perturbation theory via a neural network potential},}\ }\href
  {https://doi.org/10.26434/chemrxiv-2021-n32q8-v2} {\bibfield  {journal}
  {\bibinfo  {journal} {ChemRxiv}\ } (\bibinfo {year} {2021}),\
  10.26434/chemrxiv-2021-n32q8-v2}\BibitemShut {NoStop}%
\bibitem [{\citenamefont {Liu}, \citenamefont {Lan},\ and\ \citenamefont
  {He}(2022)}]{Liu2022/10.1021/acs.jpca.2c00601}%
  \BibitemOpen
  \bibfield  {author} {\bibinfo {author} {\bibfnamefont {J.}~\bibnamefont
  {Liu}}, \bibinfo {author} {\bibfnamefont {J.}~\bibnamefont {Lan}},\ and\
  \bibinfo {author} {\bibfnamefont {X.}~\bibnamefont {He}},\ }\bibfield
  {title} {\enquote {\bibinfo {title} {Toward high-level machine learning
  potential for water based on quantum fragmentation and neural networks},}\
  }\href {https://doi.org/10.1021/acs.jpca.2c00601} {\bibfield  {journal}
  {\bibinfo  {journal} {Journal of Physical Chemistry A}\ }\textbf {\bibinfo
  {volume} {126}},\ \bibinfo {pages} {3926--3936} (\bibinfo {year}
  {2022})}\BibitemShut {NoStop}%
\bibitem [{\citenamefont {Yao}\ and\ \citenamefont
  {Kanai}(2021)}]{Yao2021/10.1021/acs.jpclett.1c01566}%
  \BibitemOpen
  \bibfield  {author} {\bibinfo {author} {\bibfnamefont {Y.}~\bibnamefont
  {Yao}}\ and\ \bibinfo {author} {\bibfnamefont {Y.}~\bibnamefont {Kanai}},\
  }\bibfield  {title} {\enquote {\bibinfo {title} {Nuclear quantum effect and
  its temperature dependence in liquid water from random phase approximation
  via artificial neural network},}\ }\href
  {https://doi.org/10.1021/acs.jpclett.1c01566} {\bibfield  {journal} {\bibinfo
   {journal} {Journal of Physical Chemistry Letters}\ }\textbf {\bibinfo
  {volume} {12}},\ \bibinfo {pages} {6354--6362} (\bibinfo {year}
  {2021})}\BibitemShut {NoStop}%
\bibitem [{\citenamefont {Zhang}\ \emph {et~al.}(2023)\citenamefont {Zhang},
  \citenamefont {Pagotto}, \citenamefont {Gould},\ and\ \citenamefont
  {Duignan}}]{Zhang2023/2310.12535}%
  \BibitemOpen
  \bibfield  {author} {\bibinfo {author} {\bibfnamefont {J.}~\bibnamefont
  {Zhang}}, \bibinfo {author} {\bibfnamefont {J.}~\bibnamefont {Pagotto}},
  \bibinfo {author} {\bibfnamefont {T.}~\bibnamefont {Gould}},\ and\ \bibinfo
  {author} {\bibfnamefont {T.~T.}\ \bibnamefont {Duignan}},\ }\bibfield
  {title} {\enquote {\bibinfo {title} {Accurate, fast and generalisable first
  principles simulation of aqueous lithium chloride},}\ }\href
  {https://doi.org/10.48550/arXiv.2310.12535} {\bibfield  {journal} {\bibinfo
  {journal} {arXiv}\ } (\bibinfo {year} {2023}),\
  10.48550/arXiv.2310.12535}\BibitemShut {NoStop}%
\bibitem [{\citenamefont {Pagotto}, \citenamefont {Zhang},\ and\ \citenamefont
  {Duignan}(2022)}]{pagotto2022predicting}%
  \BibitemOpen
  \bibfield  {author} {\bibinfo {author} {\bibfnamefont {J.}~\bibnamefont
  {Pagotto}}, \bibinfo {author} {\bibfnamefont {J.}~\bibnamefont {Zhang}},\
  and\ \bibinfo {author} {\bibfnamefont {T.~T.}\ \bibnamefont {Duignan}},\
  }\bibfield  {title} {\enquote {\bibinfo {title} {Predicting electrolyte
  solution properties by combining neural network accelerated molecular
  dynamics and continuum solvent theory.}}\ }in\ \href
  {https://openreview.net/forum?id=nIYyCVP3X6e} {\emph {\bibinfo {booktitle}
  {NeurIPS 2022 AI for Science: Progress and Promises}}}\ (\bibinfo {year}
  {2022})\BibitemShut {NoStop}%
\bibitem [{\citenamefont {Baker}\ \emph {et~al.}(2023)\citenamefont {Baker},
  \citenamefont {Pagotto}, \citenamefont {Duignan},\ and\ \citenamefont
  {Page}}]{Baker2023/10.1021/ACS.JPCLETT.3C01783}%
  \BibitemOpen
  \bibfield  {author} {\bibinfo {author} {\bibfnamefont {S.}~\bibnamefont
  {Baker}}, \bibinfo {author} {\bibfnamefont {J.}~\bibnamefont {Pagotto}},
  \bibinfo {author} {\bibfnamefont {T.~T.}\ \bibnamefont {Duignan}},\ and\
  \bibinfo {author} {\bibfnamefont {A.~J.}\ \bibnamefont {Page}},\ }\bibfield
  {title} {\enquote {\bibinfo {title} {High-throughput aqueous electrolyte
  structure prediction using ionsolvr and equivariant graph neural network
  potentials},}\ }\href
  {https://doi.org/10.1021/ACS.JPCLETT.3C01783/ASSET/IMAGES/LARGE/JZ3C01783_0003.JPEG}
  {\bibfield  {journal} {\bibinfo  {journal} {Journal of Physical Chemistry
  Letters}\ }\textbf {\bibinfo {volume} {14}},\ \bibinfo {pages} {9508--9515}
  (\bibinfo {year} {2023})}\BibitemShut {NoStop}%
\bibitem [{\citenamefont {Schran}\ \emph {et~al.}(2021)\citenamefont {Schran},
  \citenamefont {Thiemann}, \citenamefont {Rowe}, \citenamefont {Müller},
  \citenamefont {Marsalek},\ and\ \citenamefont
  {Michaelides}}]{Schran2021/10.1073/pnas.2110077118}%
  \BibitemOpen
  \bibfield  {author} {\bibinfo {author} {\bibfnamefont {C.}~\bibnamefont
  {Schran}}, \bibinfo {author} {\bibfnamefont {F.~L.}\ \bibnamefont
  {Thiemann}}, \bibinfo {author} {\bibfnamefont {P.}~\bibnamefont {Rowe}},
  \bibinfo {author} {\bibfnamefont {E.~A.}\ \bibnamefont {Müller}}, \bibinfo
  {author} {\bibfnamefont {O.}~\bibnamefont {Marsalek}},\ and\ \bibinfo
  {author} {\bibfnamefont {A.}~\bibnamefont {Michaelides}},\ }\bibfield
  {title} {\enquote {\bibinfo {title} {Machine learning potentials for complex
  aqueous systems made simple},}\ }\href
  {https://doi.org/10.1073/pnas.2110077118} {\bibfield  {journal} {\bibinfo
  {journal} {Proceedings of the National Academy of Sciences of the United
  States of America}\ }\textbf {\bibinfo {volume} {118}},\ \bibinfo {pages}
  {e2110077118} (\bibinfo {year} {2021})}\BibitemShut {NoStop}%
\bibitem [{\citenamefont {O'Neill}\ \emph {et~al.}(2022)\citenamefont
  {O'Neill}, \citenamefont {Schran}, \citenamefont {Cox},\ and\ \citenamefont
  {Michaelides}}]{ONeill2022/2211.04345}%
  \BibitemOpen
  \bibfield  {author} {\bibinfo {author} {\bibfnamefont {N.}~\bibnamefont
  {O'Neill}}, \bibinfo {author} {\bibfnamefont {C.}~\bibnamefont {Schran}},
  \bibinfo {author} {\bibfnamefont {S.~J.}\ \bibnamefont {Cox}},\ and\ \bibinfo
  {author} {\bibfnamefont {A.}~\bibnamefont {Michaelides}},\ }\bibfield
  {title} {\enquote {\bibinfo {title} {Crumbling crystals: On the dissolution
  mechanism of nacl in water},}\ }\href
  {https://doi.org/10.48550/arXiv.2211.04345} {\bibfield  {journal} {\bibinfo
  {journal} {arXiv}\ } (\bibinfo {year} {2022}),\
  10.48550/arXiv.2211.04345}\BibitemShut {NoStop}%
\bibitem [{\citenamefont {Skinner}\ \emph {et~al.}(2013)\citenamefont
  {Skinner}, \citenamefont {Huang}, \citenamefont {Schlesinger}, \citenamefont
  {Pettersson}, \citenamefont {Nilsson},\ and\ \citenamefont
  {Benmore}}]{Skinner2013/10.1063/1.4790861}%
  \BibitemOpen
  \bibfield  {author} {\bibinfo {author} {\bibfnamefont {L.~B.}\ \bibnamefont
  {Skinner}}, \bibinfo {author} {\bibfnamefont {C.}~\bibnamefont {Huang}},
  \bibinfo {author} {\bibfnamefont {D.}~\bibnamefont {Schlesinger}}, \bibinfo
  {author} {\bibfnamefont {L.~G.}\ \bibnamefont {Pettersson}}, \bibinfo
  {author} {\bibfnamefont {A.}~\bibnamefont {Nilsson}},\ and\ \bibinfo {author}
  {\bibfnamefont {C.~J.}\ \bibnamefont {Benmore}},\ }\bibfield  {title}
  {\enquote {\bibinfo {title} {Benchmark oxygen-oxygen pair-distribution
  function of ambient water from x-ray diffraction measurements with a wide
  q-range},}\ }\href {https://doi.org/10.1063/1.4790861} {\bibfield  {journal}
  {\bibinfo  {journal} {Journal of Chemical Physics}\ }\textbf {\bibinfo
  {volume} {138}},\ \bibinfo {pages} {074506} (\bibinfo {year}
  {2013})}\BibitemShut {NoStop}%
\bibitem [{\citenamefont {Soper}\ and\ \citenamefont
  {Weckström}(2006)}]{Soper2006/10.1016/j.bpc.2006.04.009}%
  \BibitemOpen
  \bibfield  {author} {\bibinfo {author} {\bibfnamefont {A.~K.}\ \bibnamefont
  {Soper}}\ and\ \bibinfo {author} {\bibfnamefont {K.}~\bibnamefont
  {Weckström}},\ }\bibfield  {title} {\enquote {\bibinfo {title} {Ion
  solvation and water structure in potassium halide aqueous solutions},}\
  }\href {https://doi.org/10.1016/j.bpc.2006.04.009} {\bibfield  {journal}
  {\bibinfo  {journal} {Biophysical Chemistry}\ }\textbf {\bibinfo {volume}
  {124}},\ \bibinfo {pages} {180--191} (\bibinfo {year} {2006})}\BibitemShut
  {NoStop}%
\bibitem [{\citenamefont {Pitzer}, \citenamefont {Peiper},\ and\ \citenamefont
  {Busey}(1984)}]{Pitzer1984/10.1063/1.555709}%
  \BibitemOpen
  \bibfield  {author} {\bibinfo {author} {\bibfnamefont {K.~S.}\ \bibnamefont
  {Pitzer}}, \bibinfo {author} {\bibfnamefont {J.~C.}\ \bibnamefont {Peiper}},\
  and\ \bibinfo {author} {\bibfnamefont {R.~H.}\ \bibnamefont {Busey}},\
  }\bibfield  {title} {\enquote {\bibinfo {title} {Thermodynamic properties of
  aqueous sodium chloride solutions},}\ }\href
  {https://doi.org/10.1063/1.555709} {\bibfield  {journal} {\bibinfo  {journal}
  {Journal of Physical and Chemical Reference Data}\ }\textbf {\bibinfo
  {volume} {13}},\ \bibinfo {pages} {1--102} (\bibinfo {year}
  {1984})}\BibitemShut {NoStop}%
\bibitem [{\citenamefont {Zhang}\ and\ \citenamefont
  {Yang}(1998)}]{Zhang1998/10.1103/PhysRevLett.80.890}%
  \BibitemOpen
  \bibfield  {author} {\bibinfo {author} {\bibfnamefont {Y.}~\bibnamefont
  {Zhang}}\ and\ \bibinfo {author} {\bibfnamefont {W.}~\bibnamefont {Yang}},\
  }\bibfield  {title} {\enquote {\bibinfo {title} {Comment on “generalized
  gradient approximation made simple”},}\ }\href
  {https://doi.org/10.1103/PhysRevLett.80.890} {\bibfield  {journal} {\bibinfo
  {journal} {Physical Review Letters}\ }\textbf {\bibinfo {volume} {80}},\
  \bibinfo {pages} {890} (\bibinfo {year} {1998})}\BibitemShut {NoStop}%
\bibitem [{\citenamefont {Grimme}\ \emph {et~al.}(2010)\citenamefont {Grimme},
  \citenamefont {Antony}, \citenamefont {Ehrlich},\ and\ \citenamefont
  {Krieg}}]{Grimme2010/10.1063/1.3382344}%
  \BibitemOpen
  \bibfield  {author} {\bibinfo {author} {\bibfnamefont {S.}~\bibnamefont
  {Grimme}}, \bibinfo {author} {\bibfnamefont {J.}~\bibnamefont {Antony}},
  \bibinfo {author} {\bibfnamefont {S.}~\bibnamefont {Ehrlich}},\ and\ \bibinfo
  {author} {\bibfnamefont {H.}~\bibnamefont {Krieg}},\ }\bibfield  {title}
  {\enquote {\bibinfo {title} {A consistent and accurate ab initio
  parametrization of density functional dispersion correction (dft-d) for the
  94 elements h-pu},}\ }\href {https://doi.org/10.1063/1.3382344} {\bibfield
  {journal} {\bibinfo  {journal} {The Journal of Chemical Physics}\ }\textbf
  {\bibinfo {volume} {132}},\ \bibinfo {pages} {154104} (\bibinfo {year}
  {2010})}\BibitemShut {NoStop}%
\bibitem [{\citenamefont {Klimeš}, \citenamefont {Bowler},\ and\ \citenamefont
  {Michaelides}(2009)}]{Klimes2009/10.1088/0953-8984/22/2/022201}%
  \BibitemOpen
  \bibfield  {author} {\bibinfo {author} {\bibfnamefont {J.}~\bibnamefont
  {Klimeš}}, \bibinfo {author} {\bibfnamefont {D.~R.}\ \bibnamefont
  {Bowler}},\ and\ \bibinfo {author} {\bibfnamefont {A.}~\bibnamefont
  {Michaelides}},\ }\bibfield  {title} {\enquote {\bibinfo {title} {Chemical
  accuracy for the van der waals density functional},}\ }\href
  {https://doi.org/10.1088/0953-8984/22/2/022201} {\bibfield  {journal}
  {\bibinfo  {journal} {Journal of Physics: Condensed Matter}\ }\textbf
  {\bibinfo {volume} {22}},\ \bibinfo {pages} {022201} (\bibinfo {year}
  {2009})}\BibitemShut {NoStop}%
\bibitem [{\citenamefont {Furness}\ \emph {et~al.}(2020)\citenamefont
  {Furness}, \citenamefont {Kaplan}, \citenamefont {Ning}, \citenamefont
  {Perdew},\ and\ \citenamefont
  {Sun}}]{Furness2020/10.1021/ACS.JPCLETT.0C02405}%
  \BibitemOpen
  \bibfield  {author} {\bibinfo {author} {\bibfnamefont {J.~W.}\ \bibnamefont
  {Furness}}, \bibinfo {author} {\bibfnamefont {A.~D.}\ \bibnamefont {Kaplan}},
  \bibinfo {author} {\bibfnamefont {J.}~\bibnamefont {Ning}}, \bibinfo {author}
  {\bibfnamefont {J.~P.}\ \bibnamefont {Perdew}},\ and\ \bibinfo {author}
  {\bibfnamefont {J.}~\bibnamefont {Sun}},\ }\bibfield  {title} {\enquote
  {\bibinfo {title} {Accurate and numerically efficient r2scan meta-generalized
  gradient approximation},}\ }\href
  {https://doi.org/10.1021/acs.jpclett.0c02405} {\bibfield  {journal} {\bibinfo
   {journal} {Journal of Physical Chemistry Letters}\ }\textbf {\bibinfo
  {volume} {11}},\ \bibinfo {pages} {8208--8215} (\bibinfo {year}
  {2020})}\BibitemShut {NoStop}%
\bibitem [{\citenamefont {Willow}\ \emph {et~al.}(2016)\citenamefont {Willow},
  \citenamefont {Zeng}, \citenamefont {Xantheas}, \citenamefont {Kim},\ and\
  \citenamefont {Hirata}}]{Willow2016/10.1021/ACS.JPCLETT.5B02430}%
  \BibitemOpen
  \bibfield  {author} {\bibinfo {author} {\bibfnamefont {S.~Y.}\ \bibnamefont
  {Willow}}, \bibinfo {author} {\bibfnamefont {X.~C.}\ \bibnamefont {Zeng}},
  \bibinfo {author} {\bibfnamefont {S.~S.}\ \bibnamefont {Xantheas}}, \bibinfo
  {author} {\bibfnamefont {K.~S.}\ \bibnamefont {Kim}},\ and\ \bibinfo {author}
  {\bibfnamefont {S.}~\bibnamefont {Hirata}},\ }\bibfield  {title} {\enquote
  {\bibinfo {title} {Why is mp2-water ''cooler'' and ''denser'' than
  dft-water?}}\ }\href {https://doi.org/10.1021/ACS.JPCLETT.5B02430} {\bibfield
   {journal} {\bibinfo  {journal} {Journal of Physical Chemistry Letters}\
  }\textbf {\bibinfo {volume} {7}},\ \bibinfo {pages} {680--684} (\bibinfo
  {year} {2016})}\BibitemShut {NoStop}%
\bibitem [{\citenamefont {Skinner}\ \emph {et~al.}(2016)\citenamefont
  {Skinner}, \citenamefont {Galib}, \citenamefont {Fulton}, \citenamefont
  {Mundy}, \citenamefont {Parise}, \citenamefont {Pham}, \citenamefont
  {Schenter},\ and\ \citenamefont
  {Benmore}}]{Skinner2016/10.1063/1.4944935/13698109}%
  \BibitemOpen
  \bibfield  {author} {\bibinfo {author} {\bibfnamefont {L.~B.}\ \bibnamefont
  {Skinner}}, \bibinfo {author} {\bibfnamefont {M.}~\bibnamefont {Galib}},
  \bibinfo {author} {\bibfnamefont {J.~L.}\ \bibnamefont {Fulton}}, \bibinfo
  {author} {\bibfnamefont {C.~J.}\ \bibnamefont {Mundy}}, \bibinfo {author}
  {\bibfnamefont {J.~B.}\ \bibnamefont {Parise}}, \bibinfo {author}
  {\bibfnamefont {V.~T.}\ \bibnamefont {Pham}}, \bibinfo {author}
  {\bibfnamefont {G.~K.}\ \bibnamefont {Schenter}},\ and\ \bibinfo {author}
  {\bibfnamefont {C.~J.}\ \bibnamefont {Benmore}},\ }\bibfield  {title}
  {\enquote {\bibinfo {title} {The structure of liquid water up to 360 mpa from
  x-ray diffraction measurements using a high q-range and from molecular
  simulation},}\ }\href {https://doi.org/10.1063/1.4944935} {\bibfield
  {journal} {\bibinfo  {journal} {Journal of Chemical Physics}\ }\textbf
  {\bibinfo {volume} {144}} (\bibinfo {year} {2016}),\
  10.1063/1.4944935}\BibitemShut {NoStop}%
\bibitem [{\citenamefont {Dasgupta}\ \emph {et~al.}(2021)\citenamefont
  {Dasgupta}, \citenamefont {Lambros}, \citenamefont {Perdew},\ and\
  \citenamefont {Paesani}}]{Dasgupta2021/10.1038/s41467-021-26618-9}%
  \BibitemOpen
  \bibfield  {author} {\bibinfo {author} {\bibfnamefont {S.}~\bibnamefont
  {Dasgupta}}, \bibinfo {author} {\bibfnamefont {E.}~\bibnamefont {Lambros}},
  \bibinfo {author} {\bibfnamefont {J.~P.}\ \bibnamefont {Perdew}},\ and\
  \bibinfo {author} {\bibfnamefont {F.}~\bibnamefont {Paesani}},\ }\bibfield
  {title} {\enquote {\bibinfo {title} {Elevating density functional theory to
  chemical accuracy for water simulations through a density-corrected many-body
  formalism},}\ }\href {https://doi.org/10.1038/s41467-021-26618-9} {\bibfield
  {journal} {\bibinfo  {journal} {Nature Communications}\ }\textbf {\bibinfo
  {volume} {12}},\ \bibinfo {pages} {1--12} (\bibinfo {year}
  {2021})}\BibitemShut {NoStop}%
\bibitem [{\citenamefont {Ehrlich}\ \emph {et~al.}(2011)\citenamefont
  {Ehrlich}, \citenamefont {Moellmann}, \citenamefont {Reckien}, \citenamefont
  {Bredow},\ and\ \citenamefont {Grimme}}]{Ehrlich2011/10.1002/CPHC.201100521}%
  \BibitemOpen
  \bibfield  {author} {\bibinfo {author} {\bibfnamefont {S.}~\bibnamefont
  {Ehrlich}}, \bibinfo {author} {\bibfnamefont {J.}~\bibnamefont {Moellmann}},
  \bibinfo {author} {\bibfnamefont {W.}~\bibnamefont {Reckien}}, \bibinfo
  {author} {\bibfnamefont {T.}~\bibnamefont {Bredow}},\ and\ \bibinfo {author}
  {\bibfnamefont {S.}~\bibnamefont {Grimme}},\ }\bibfield  {title} {\enquote
  {\bibinfo {title} {System-dependent dispersion coefficients for the dft-d3
  treatment of adsorption processes on ionic surfaces},}\ }\href
  {https://doi.org/10.1002/CPHC.201100521} {\bibfield  {journal} {\bibinfo
  {journal} {ChemPhysChem}\ }\textbf {\bibinfo {volume} {12}},\ \bibinfo
  {pages} {3414--3420} (\bibinfo {year} {2011})}\BibitemShut {NoStop}%
\bibitem [{\citenamefont {Caldeweyher}\ \emph {et~al.}(2020)\citenamefont
  {Caldeweyher}, \citenamefont {Mewes}, \citenamefont {Ehlert},\ and\
  \citenamefont {Grimme}}]{Caldeweyher2020/10.1039/D0CP00502A}%
  \BibitemOpen
  \bibfield  {author} {\bibinfo {author} {\bibfnamefont {E.}~\bibnamefont
  {Caldeweyher}}, \bibinfo {author} {\bibfnamefont {J.~M.}\ \bibnamefont
  {Mewes}}, \bibinfo {author} {\bibfnamefont {S.}~\bibnamefont {Ehlert}},\ and\
  \bibinfo {author} {\bibfnamefont {S.}~\bibnamefont {Grimme}},\ }\bibfield
  {title} {\enquote {\bibinfo {title} {Extension and evaluation of the d4
  london-dispersion model for periodic systems},}\ }\href
  {https://doi.org/10.1039/D0CP00502A} {\bibfield  {journal} {\bibinfo
  {journal} {Physical Chemistry Chemical Physics}\ }\textbf {\bibinfo {volume}
  {22}},\ \bibinfo {pages} {8499--8512} (\bibinfo {year} {2020})}\BibitemShut
  {NoStop}%
\bibitem [{\citenamefont {Bučko}\ \emph {et~al.}(2014)\citenamefont {Bučko},
  \citenamefont {Lebègue}, \citenamefont {Ángyán},\ and\ \citenamefont
  {Hafner}}]{Bucko2014/10.1063/1.4890003/194229}%
  \BibitemOpen
  \bibfield  {author} {\bibinfo {author} {\bibfnamefont {T.}~\bibnamefont
  {Bučko}}, \bibinfo {author} {\bibfnamefont {S.}~\bibnamefont {Lebègue}},
  \bibinfo {author} {\bibfnamefont {J.~G.}\ \bibnamefont {Ángyán}},\ and\
  \bibinfo {author} {\bibfnamefont {J.}~\bibnamefont {Hafner}},\ }\bibfield
  {title} {\enquote {\bibinfo {title} {Extending the applicability of the
  tkatchenko-scheffler dispersion correction via iterative hirshfeld
  partitioning},}\ }\href {https://doi.org/10.1063/1.4890003} {\bibfield
  {journal} {\bibinfo  {journal} {Journal of Chemical Physics}\ }\textbf
  {\bibinfo {volume} {141}},\ \bibinfo {pages} {34114} (\bibinfo {year}
  {2014})}\BibitemShut {NoStop}%
\bibitem [{\citenamefont {Kostal}\ \emph {et~al.}(2023)\citenamefont {Kostal},
  \citenamefont {Mason}, \citenamefont {Martinez-Seara},\ and\ \citenamefont
  {Jungwirth}}]{Kostal2023/10.1021/ACS.JPCLETT.3C00856}%
  \BibitemOpen
  \bibfield  {author} {\bibinfo {author} {\bibfnamefont {V.}~\bibnamefont
  {Kostal}}, \bibinfo {author} {\bibfnamefont {P.~E.}\ \bibnamefont {Mason}},
  \bibinfo {author} {\bibfnamefont {H.}~\bibnamefont {Martinez-Seara}},\ and\
  \bibinfo {author} {\bibfnamefont {P.}~\bibnamefont {Jungwirth}},\ }\bibfield
  {title} {\enquote {\bibinfo {title} {Common cations are not polarizable:
  Effects of dispersion correction on hydration structures from ab initio
  molecular dynamics},}\ }\href {https://doi.org/10.1021/ACS.JPCLETT.3C00856}
  {\bibfield  {journal} {\bibinfo  {journal} {Journal of Physical Chemistry
  Letters}\ }\textbf {\bibinfo {volume} {14}},\ \bibinfo {pages} {4403--4408}
  (\bibinfo {year} {2023})}\BibitemShut {NoStop}%
\bibitem [{\citenamefont {Benavides}, \citenamefont {Aragones},\ and\
  \citenamefont {Vega}(2016)}]{Benavides2016/10.1063/1.4943780}%
  \BibitemOpen
  \bibfield  {author} {\bibinfo {author} {\bibfnamefont {A.~L.}\ \bibnamefont
  {Benavides}}, \bibinfo {author} {\bibfnamefont {J.~L.}\ \bibnamefont
  {Aragones}},\ and\ \bibinfo {author} {\bibfnamefont {C.}~\bibnamefont
  {Vega}},\ }\bibfield  {title} {\enquote {\bibinfo {title} {Consensus on the
  solubility of nacl in water from computer simulations using the chemical
  potential route},}\ }\href {https://doi.org/10.1063/1.4943780} {\bibfield
  {journal} {\bibinfo  {journal} {The Journal of Chemical Physics}\ }\textbf
  {\bibinfo {volume} {144}},\ \bibinfo {pages} {124504} (\bibinfo {year}
  {2016})}\BibitemShut {NoStop}%
\bibitem [{\citenamefont {Berendsen}, \citenamefont {Grigera},\ and\
  \citenamefont {Straatsma}(1987)}]{Berendsen1987/10.1021/J100308A038}%
  \BibitemOpen
  \bibfield  {author} {\bibinfo {author} {\bibfnamefont {H.~J.}\ \bibnamefont
  {Berendsen}}, \bibinfo {author} {\bibfnamefont {J.~R.}\ \bibnamefont
  {Grigera}},\ and\ \bibinfo {author} {\bibfnamefont {T.~P.}\ \bibnamefont
  {Straatsma}},\ }\bibfield  {title} {\enquote {\bibinfo {title} {The missing
  term in effective pair potentials},}\ }\href
  {https://doi.org/10.1021/J100308A038} {\bibfield  {journal} {\bibinfo
  {journal} {Journal of Physical Chemistry}\ }\textbf {\bibinfo {volume}
  {91}},\ \bibinfo {pages} {6269--6271} (\bibinfo {year} {1987})}\BibitemShut
  {NoStop}%
\bibitem [{\citenamefont {Yao}\ and\ \citenamefont
  {Kanai}(2018)}]{Yao2018/10.1021/ACS.JCTC.7B00846}%
  \BibitemOpen
  \bibfield  {author} {\bibinfo {author} {\bibfnamefont {Y.}~\bibnamefont
  {Yao}}\ and\ \bibinfo {author} {\bibfnamefont {Y.}~\bibnamefont {Kanai}},\
  }\bibfield  {title} {\enquote {\bibinfo {title} {Free energy profile of nacl
  in water: First-principles molecular dynamics with scan and
  $\mathrm{\omega}$b97x-v exchange-correlation functionals},}\ }\href
  {https://doi.org/10.1021/ACS.JCTC.7B00846} {\bibfield  {journal} {\bibinfo
  {journal} {Journal of Chemical Theory and Computation}\ }\textbf {\bibinfo
  {volume} {14}},\ \bibinfo {pages} {884--893} (\bibinfo {year}
  {2018})}\BibitemShut {NoStop}%
\bibitem [{\citenamefont {Geissler}, \citenamefont {Dellago},\ and\
  \citenamefont {Chandler}(1999)}]{Geissler1999}%
  \BibitemOpen
  \bibfield  {author} {\bibinfo {author} {\bibfnamefont {P.~L.}\ \bibnamefont
  {Geissler}}, \bibinfo {author} {\bibfnamefont {C.}~\bibnamefont {Dellago}},\
  and\ \bibinfo {author} {\bibfnamefont {D.}~\bibnamefont {Chandler}},\
  }\bibfield  {title} {\enquote {\bibinfo {title} {Kinetic pathways of ion pair
  dissociation in water},}\ }\href {https://doi.org/10.1021/JP984837G}
  {\bibfield  {journal} {\bibinfo  {journal} {Journal of Physical Chemistry B}\
  }\textbf {\bibinfo {volume} {103}},\ \bibinfo {pages} {3706--3710} (\bibinfo
  {year} {1999})}\BibitemShut {NoStop}%
\bibitem [{\citenamefont {Timko}, \citenamefont {Bucher},\ and\ \citenamefont
  {Kuyucak}(2010)}]{Timko2010/10.1063/1.3360310/939701}%
  \BibitemOpen
  \bibfield  {author} {\bibinfo {author} {\bibfnamefont {J.}~\bibnamefont
  {Timko}}, \bibinfo {author} {\bibfnamefont {D.}~\bibnamefont {Bucher}},\ and\
  \bibinfo {author} {\bibfnamefont {S.}~\bibnamefont {Kuyucak}},\ }\bibfield
  {title} {\enquote {\bibinfo {title} {Dissociation of nacl in water from ab
  initio molecular dynamics simulations},}\ }\href
  {https://doi.org/10.1063/1.3360310} {\bibfield  {journal} {\bibinfo
  {journal} {Journal of Chemical Physics}\ }\textbf {\bibinfo {volume} {132}},\
  \bibinfo {pages} {114510} (\bibinfo {year} {2010})}\BibitemShut {NoStop}%
\bibitem [{\citenamefont {Finney}\ and\ \citenamefont
  {Salvalaglio}(2022)}]{Finney2022/10.1039/D1FD00089F}%
  \BibitemOpen
  \bibfield  {author} {\bibinfo {author} {\bibfnamefont {A.~R.}\ \bibnamefont
  {Finney}}\ and\ \bibinfo {author} {\bibfnamefont {M.}~\bibnamefont
  {Salvalaglio}},\ }\bibfield  {title} {\enquote {\bibinfo {title} {Multiple
  pathways in nacl homogeneous crystal nucleation},}\ }\href
  {https://doi.org/10.1039/D1FD00089F} {\bibfield  {journal} {\bibinfo
  {journal} {Faraday Discussions}\ }\textbf {\bibinfo {volume} {235}},\
  \bibinfo {pages} {56--80} (\bibinfo {year} {2022})}\BibitemShut {NoStop}%
\bibitem [{\citenamefont {Pluhařová}\ \emph {et~al.}(2013)\citenamefont
  {Pluhařová}, \citenamefont {Marsalek}, \citenamefont {Schmidt},\ and\
  \citenamefont {Jungwirth}}]{Pluharova2013/10.1021/JZ402177Q}%
  \BibitemOpen
  \bibfield  {author} {\bibinfo {author} {\bibfnamefont {E.}~\bibnamefont
  {Pluhařová}}, \bibinfo {author} {\bibfnamefont {O.}~\bibnamefont
  {Marsalek}}, \bibinfo {author} {\bibfnamefont {B.}~\bibnamefont {Schmidt}},\
  and\ \bibinfo {author} {\bibfnamefont {P.}~\bibnamefont {Jungwirth}},\
  }\bibfield  {title} {\enquote {\bibinfo {title} {Ab initio molecular dynamics
  approach to a quantitative description of ion pairing in water},}\ }\href
  {https://doi.org/10.1021/jz402177q} {\bibfield  {journal} {\bibinfo
  {journal} {Journal of Physical Chemistry Letters}\ }\textbf {\bibinfo
  {volume} {4}},\ \bibinfo {pages} {4177--4181} (\bibinfo {year}
  {2013})}\BibitemShut {NoStop}%
\bibitem [{\citenamefont {Duignan}\ \emph
  {et~al.}(2020{\natexlab{b}})\citenamefont {Duignan}, \citenamefont {Mundy},
  \citenamefont {Schenter},\ and\ \citenamefont
  {Zhao}}]{Duignan2020/10.1021/ACS.JCTC.0C00300}%
  \BibitemOpen
  \bibfield  {author} {\bibinfo {author} {\bibfnamefont {T.~T.}\ \bibnamefont
  {Duignan}}, \bibinfo {author} {\bibfnamefont {C.~J.}\ \bibnamefont {Mundy}},
  \bibinfo {author} {\bibfnamefont {G.~K.}\ \bibnamefont {Schenter}},\ and\
  \bibinfo {author} {\bibfnamefont {X.~S.}\ \bibnamefont {Zhao}},\ }\bibfield
  {title} {\enquote {\bibinfo {title} {Method for accurately predicting
  solvation structure},}\ }\href
  {https://doi.org/10.1021/ACS.JCTC.0C00300/ASSET/IMAGES/LARGE/CT0C00300_0006.JPEG}
  {\bibfield  {journal} {\bibinfo  {journal} {Journal of Chemical Theory and
  Computation}\ }\textbf {\bibinfo {volume} {16}},\ \bibinfo {pages}
  {5401--5409} (\bibinfo {year} {2020}{\natexlab{b}})}\BibitemShut {NoStop}%
\bibitem [{\citenamefont {Riley}\ \emph {et~al.}(2012)\citenamefont {Riley},
  \citenamefont {Platts}, \citenamefont {Řezáč}, \citenamefont {Hobza},\
  and\ \citenamefont {Hill}}]{Riley2012/10.1021/JP211997B}%
  \BibitemOpen
  \bibfield  {author} {\bibinfo {author} {\bibfnamefont {K.~E.}\ \bibnamefont
  {Riley}}, \bibinfo {author} {\bibfnamefont {J.~A.}\ \bibnamefont {Platts}},
  \bibinfo {author} {\bibfnamefont {J.}~\bibnamefont {Řezáč}}, \bibinfo
  {author} {\bibfnamefont {P.}~\bibnamefont {Hobza}},\ and\ \bibinfo {author}
  {\bibfnamefont {J.~G.}\ \bibnamefont {Hill}},\ }\bibfield  {title} {\enquote
  {\bibinfo {title} {Assessment of the performance of mp2 and mp2 variants for
  the treatment of noncovalent interactions},}\ }\href
  {https://doi.org/10.1021/JP211997B} {\bibfield  {journal} {\bibinfo
  {journal} {Journal of Physical Chemistry A}\ }\textbf {\bibinfo {volume}
  {116}},\ \bibinfo {pages} {4159--4169} (\bibinfo {year} {2012})}\BibitemShut
  {NoStop}%
\bibitem [{\citenamefont {Ren}\ \emph {et~al.}(2011)\citenamefont {Ren},
  \citenamefont {Tkatchenko}, \citenamefont {Rinke},\ and\ \citenamefont
  {Scheffler}}]{Ren2011/10.1103/PHYSREVLETT.106.153003}%
  \BibitemOpen
  \bibfield  {author} {\bibinfo {author} {\bibfnamefont {X.}~\bibnamefont
  {Ren}}, \bibinfo {author} {\bibfnamefont {A.}~\bibnamefont {Tkatchenko}},
  \bibinfo {author} {\bibfnamefont {P.}~\bibnamefont {Rinke}},\ and\ \bibinfo
  {author} {\bibfnamefont {M.}~\bibnamefont {Scheffler}},\ }\bibfield  {title}
  {\enquote {\bibinfo {title} {Beyond the random-phase approximation for the
  electron correlation energy: The importance of single excitations},}\ }\href
  {https://doi.org/10.1103/PHYSREVLETT.106.153003} {\bibfield  {journal}
  {\bibinfo  {journal} {Physical Review Letters}\ }\textbf {\bibinfo {volume}
  {106}},\ \bibinfo {pages} {153003} (\bibinfo {year} {2011})}\BibitemShut
  {NoStop}%
\bibitem [{\citenamefont {Riera}, \citenamefont {Götz},\ and\ \citenamefont
  {Paesani}(2016)}]{Riera2016/10.1039/C6CP02553F}%
  \BibitemOpen
  \bibfield  {author} {\bibinfo {author} {\bibfnamefont {M.}~\bibnamefont
  {Riera}}, \bibinfo {author} {\bibfnamefont {A.~W.}\ \bibnamefont {Götz}},\
  and\ \bibinfo {author} {\bibfnamefont {F.}~\bibnamefont {Paesani}},\
  }\bibfield  {title} {\enquote {\bibinfo {title} {The i-ttm model for ab
  initio-based ion–water interaction potentials. ii. alkali metal ion–water
  potential energy functions},}\ }\href {https://doi.org/10.1039/C6CP02553F}
  {\bibfield  {journal} {\bibinfo  {journal} {Physical Chemistry Chemical
  Physics}\ }\textbf {\bibinfo {volume} {18}},\ \bibinfo {pages} {30334--30343}
  (\bibinfo {year} {2016})}\BibitemShut {NoStop}%
\bibitem [{\citenamefont {Otero-De-La-Roza}\ and\ \citenamefont
  {Johnson}(2012)}]{Otero2012/10.1063/1.4705760/191680}%
  \BibitemOpen
  \bibfield  {author} {\bibinfo {author} {\bibfnamefont {A.}~\bibnamefont
  {Otero-De-La-Roza}}\ and\ \bibinfo {author} {\bibfnamefont {E.~R.}\
  \bibnamefont {Johnson}},\ }\bibfield  {title} {\enquote {\bibinfo {title}
  {Van der waals interactions in solids using the exchange-hole dipole moment
  model},}\ }\href {https://doi.org/10.1063/1.4705760/191680} {\bibfield
  {journal} {\bibinfo  {journal} {Journal of Chemical Physics}\ }\textbf
  {\bibinfo {volume} {136}},\ \bibinfo {pages} {174109} (\bibinfo {year}
  {2012})}\BibitemShut {NoStop}%
\bibitem [{\citenamefont {Chen}\ \emph {et~al.}(2023)\citenamefont {Chen},
  \citenamefont {Lee}, \citenamefont {Ye}, \citenamefont {Berkelbach},
  \citenamefont {Reichman},\ and\ \citenamefont
  {Markland}}]{Chen2023/10.1021/ACS.JCTC.2C01203}%
  \BibitemOpen
  \bibfield  {author} {\bibinfo {author} {\bibfnamefont {M.~S.}\ \bibnamefont
  {Chen}}, \bibinfo {author} {\bibfnamefont {J.}~\bibnamefont {Lee}}, \bibinfo
  {author} {\bibfnamefont {H.~Z.}\ \bibnamefont {Ye}}, \bibinfo {author}
  {\bibfnamefont {T.~C.}\ \bibnamefont {Berkelbach}}, \bibinfo {author}
  {\bibfnamefont {D.~R.}\ \bibnamefont {Reichman}},\ and\ \bibinfo {author}
  {\bibfnamefont {T.~E.}\ \bibnamefont {Markland}},\ }\bibfield  {title}
  {\enquote {\bibinfo {title} {Data-efficient machine learning potentials from
  transfer learning of periodic correlated electronic structure methods: Liquid
  water at afqmc, ccsd, and ccsd(t) accuracy},}\ }\href
  {https://doi.org/10.1021/ACS.JCTC.2C01203} {\bibfield  {journal} {\bibinfo
  {journal} {Journal of Chemical Theory and Computation}\ }\textbf {\bibinfo
  {volume} {19}},\ \bibinfo {pages} {4510--4519} (\bibinfo {year}
  {2023})}\BibitemShut {NoStop}%
\bibitem [{\citenamefont {Daru}\ \emph {et~al.}(2022)\citenamefont {Daru},
  \citenamefont {Forbert}, \citenamefont {Behler},\ and\ \citenamefont
  {Marx}}]{Daru2022/10.1103/PHYSREVLETT.129.226001}%
  \BibitemOpen
  \bibfield  {author} {\bibinfo {author} {\bibfnamefont {J.}~\bibnamefont
  {Daru}}, \bibinfo {author} {\bibfnamefont {H.}~\bibnamefont {Forbert}},
  \bibinfo {author} {\bibfnamefont {J.}~\bibnamefont {Behler}},\ and\ \bibinfo
  {author} {\bibfnamefont {D.}~\bibnamefont {Marx}},\ }\bibfield  {title}
  {\enquote {\bibinfo {title} {Coupled cluster molecular dynamics of condensed
  phase systems enabled by machine learning potentials: Liquid water
  benchmark},}\ }\href {https://doi.org/10.1103/PHYSREVLETT.129.226001}
  {\bibfield  {journal} {\bibinfo  {journal} {Physical Review Letters}\
  }\textbf {\bibinfo {volume} {129}},\ \bibinfo {pages} {226001} (\bibinfo
  {year} {2022})}\BibitemShut {NoStop}%
\bibitem [{\citenamefont {Fransson}, \citenamefont {Wiktor},\ and\
  \citenamefont {Erhart}(2023)}]{Fransson2023/10.1021/ACS.JPCC.3C01542}%
  \BibitemOpen
  \bibfield  {author} {\bibinfo {author} {\bibfnamefont {E.}~\bibnamefont
  {Fransson}}, \bibinfo {author} {\bibfnamefont {J.}~\bibnamefont {Wiktor}},\
  and\ \bibinfo {author} {\bibfnamefont {P.}~\bibnamefont {Erhart}},\
  }\bibfield  {title} {\enquote {\bibinfo {title} {Phase transitions in
  inorganic halide perovskites from machine-learned potentials},}\ }\href
  {https://doi.org/10.1021/ACS.JPCC.3C01542} {\bibfield  {journal} {\bibinfo
  {journal} {Journal of Physical Chemistry C}\ }\textbf {\bibinfo {volume}
  {127}},\ \bibinfo {pages} {13773--13781} (\bibinfo {year}
  {2023})}\BibitemShut {NoStop}%
\bibitem [{\citenamefont {Fong}\ \emph {et~al.}(2020)\citenamefont {Fong},
  \citenamefont {Bergstrom}, \citenamefont {McCloskey},\ and\ \citenamefont
  {Mandadapu}}]{fong2020transport}%
  \BibitemOpen
  \bibfield  {author} {\bibinfo {author} {\bibfnamefont {K.~D.}\ \bibnamefont
  {Fong}}, \bibinfo {author} {\bibfnamefont {H.~K.}\ \bibnamefont {Bergstrom}},
  \bibinfo {author} {\bibfnamefont {B.~D.}\ \bibnamefont {McCloskey}},\ and\
  \bibinfo {author} {\bibfnamefont {K.~K.}\ \bibnamefont {Mandadapu}},\
  }\bibfield  {title} {\enquote {\bibinfo {title} {Transport phenomena in
  electrolyte solutions: Nonequilibrium thermodynamics and statistical
  mechanics},}\ }\href
  {https://aiche.onlinelibrary.wiley.com/doi/full/10.1002/aic.17091} {\bibfield
   {journal} {\bibinfo  {journal} {AIChE Journal}\ }\textbf {\bibinfo {volume}
  {66}},\ \bibinfo {pages} {e17091} (\bibinfo {year} {2020})}\BibitemShut
  {NoStop}%
\bibitem [{\citenamefont {Fong}\ \emph {et~al.}(2021)\citenamefont {Fong},
  \citenamefont {Self}, \citenamefont {McCloskey},\ and\ \citenamefont
  {Persson}}]{Fong2021/10.1021/ACS.MACROMOL.0C02545}%
  \BibitemOpen
  \bibfield  {author} {\bibinfo {author} {\bibfnamefont {K.~D.}\ \bibnamefont
  {Fong}}, \bibinfo {author} {\bibfnamefont {J.}~\bibnamefont {Self}}, \bibinfo
  {author} {\bibfnamefont {B.~D.}\ \bibnamefont {McCloskey}},\ and\ \bibinfo
  {author} {\bibfnamefont {K.~A.}\ \bibnamefont {Persson}},\ }\bibfield
  {title} {\enquote {\bibinfo {title} {Ion correlations and their impact on
  transport in polymer-based electrolytes},}\ }\href
  {https://doi.org/10.1021/ACS.MACROMOL.0C02545} {\bibfield  {journal}
  {\bibinfo  {journal} {Macromolecules}\ }\textbf {\bibinfo {volume} {54}},\
  \bibinfo {pages} {2575--2591} (\bibinfo {year} {2021})}\BibitemShut {NoStop}%
\bibitem [{\citenamefont {Raiteri}, \citenamefont {Schuitemaker},\ and\
  \citenamefont {Gale}(2020)}]{Raiteri2020/10.1021/ACS.JPCB.0C01582}%
  \BibitemOpen
  \bibfield  {author} {\bibinfo {author} {\bibfnamefont {P.}~\bibnamefont
  {Raiteri}}, \bibinfo {author} {\bibfnamefont {A.}~\bibnamefont
  {Schuitemaker}},\ and\ \bibinfo {author} {\bibfnamefont {J.~D.}\ \bibnamefont
  {Gale}},\ }\bibfield  {title} {\enquote {\bibinfo {title} {Ion pairing and
  multiple ion binding in calcium carbonate solutions based on a polarizable
  amoeba force field and ab initio molecular dynamics},}\ }\href
  {https://doi.org/10.1021/ACS.JPCB.0C01582/ASSET/IMAGES/LARGE/JP0C01582_0001.JPEG}
  {\bibfield  {journal} {\bibinfo  {journal} {Journal of Physical Chemistry B}\
  }\textbf {\bibinfo {volume} {124}},\ \bibinfo {pages} {3568--3582} (\bibinfo
  {year} {2020})}\BibitemShut {NoStop}%
\bibitem [{\citenamefont {Chao}\ \emph {et~al.}(2020)\citenamefont {Chao},
  \citenamefont {Zhou}, \citenamefont {Xie}, \citenamefont {Ye}, \citenamefont
  {Li}, \citenamefont {Jaroniec},\ and\ \citenamefont
  {Qiao}}]{Chao2020/10.1126/SCIADV.ABA4098}%
  \BibitemOpen
  \bibfield  {author} {\bibinfo {author} {\bibfnamefont {D.}~\bibnamefont
  {Chao}}, \bibinfo {author} {\bibfnamefont {W.}~\bibnamefont {Zhou}}, \bibinfo
  {author} {\bibfnamefont {F.}~\bibnamefont {Xie}}, \bibinfo {author}
  {\bibfnamefont {C.}~\bibnamefont {Ye}}, \bibinfo {author} {\bibfnamefont
  {H.}~\bibnamefont {Li}}, \bibinfo {author} {\bibfnamefont {M.}~\bibnamefont
  {Jaroniec}},\ and\ \bibinfo {author} {\bibfnamefont {S.~Z.}\ \bibnamefont
  {Qiao}},\ }\bibfield  {title} {\enquote {\bibinfo {title} {Roadmap for
  advanced aqueous batteries: From design of materials to applications},}\
  }\href {https://doi.org/10.1126/SCIADV.ABA4098} {\bibfield  {journal}
  {\bibinfo  {journal} {Science Advances}\ }\textbf {\bibinfo {volume} {6}}
  (\bibinfo {year} {2020}),\ 10.1126/SCIADV.ABA4098}\BibitemShut {NoStop}%
\bibitem [{\citenamefont {Gao}\ and\ \citenamefont
  {Remsing}(2022)}]{Gao2022/10.1038/s41467-022-29243-2}%
  \BibitemOpen
  \bibfield  {author} {\bibinfo {author} {\bibfnamefont {A.}~\bibnamefont
  {Gao}}\ and\ \bibinfo {author} {\bibfnamefont {R.~C.}\ \bibnamefont
  {Remsing}},\ }\bibfield  {title} {\enquote {\bibinfo {title} {Self-consistent
  determination of long-range electrostatics in neural network potentials},}\
  }\href {https://doi.org/10.1038/s41467-022-29243-2} {\bibfield  {journal}
  {\bibinfo  {journal} {Nature Communications 2022 13:1}\ }\textbf {\bibinfo
  {volume} {13}},\ \bibinfo {pages} {1--11} (\bibinfo {year}
  {2022})}\BibitemShut {NoStop}%
\bibitem [{\citenamefont {Grisafi}\ and\ \citenamefont
  {Ceriotti}(2019)}]{grisafi2019/10.1063/1.5128375/197247}%
  \BibitemOpen
  \bibfield  {author} {\bibinfo {author} {\bibfnamefont {A.}~\bibnamefont
  {Grisafi}}\ and\ \bibinfo {author} {\bibfnamefont {M.}~\bibnamefont
  {Ceriotti}},\ }\bibfield  {title} {\enquote {\bibinfo {title} {Incorporating
  long-range physics in atomic-scale machine learning},}\ }\href
  {https://doi.org/10.1063/1.5128375/197247} {\bibfield  {journal} {\bibinfo
  {journal} {Journal of Chemical Physics}\ }\textbf {\bibinfo {volume} {151}},\
  \bibinfo {pages} {204105} (\bibinfo {year} {2019})}\BibitemShut {NoStop}%
\bibitem [{\citenamefont {Ko}\ \emph {et~al.}(2023)\citenamefont {Ko},
  \citenamefont {Finkler}, \citenamefont {Goedecker},\ and\ \citenamefont
  {Behler}}]{Finkler2023/10.1021/ACS.JCTC.2C01146}%
  \BibitemOpen
  \bibfield  {author} {\bibinfo {author} {\bibfnamefont {T.~W.}\ \bibnamefont
  {Ko}}, \bibinfo {author} {\bibfnamefont {J.~A.}\ \bibnamefont {Finkler}},
  \bibinfo {author} {\bibfnamefont {S.}~\bibnamefont {Goedecker}},\ and\
  \bibinfo {author} {\bibfnamefont {J.}~\bibnamefont {Behler}},\ }\bibfield
  {title} {\enquote {\bibinfo {title} {Accurate fourth-generation machine
  learning potentials by electrostatic embedding},}\ }\href
  {https://doi.org/10.1021/ACS.JCTC.2C01146/ASSET/IMAGES/LARGE/CT2C01146_0007.JPEG}
  {\bibfield  {journal} {\bibinfo  {journal} {Journal of Chemical Theory and
  Computation}\ }\textbf {\bibinfo {volume} {19}},\ \bibinfo {pages}
  {3567--3579} (\bibinfo {year} {2023})}\BibitemShut {NoStop}%
\bibitem [{\citenamefont {Batatia}\ \emph {et~al.}(2022)\citenamefont
  {Batatia}, \citenamefont {Batzner}, \citenamefont {Kovács}, \citenamefont
  {Musaelian}, \citenamefont {Simm}, \citenamefont {Drautz}, \citenamefont
  {Ortner}, \citenamefont {Kozinsky},\ and\ \citenamefont
  {Csányi}}]{Batatia2022/2205.06643}%
  \BibitemOpen
  \bibfield  {author} {\bibinfo {author} {\bibfnamefont {I.}~\bibnamefont
  {Batatia}}, \bibinfo {author} {\bibfnamefont {S.}~\bibnamefont {Batzner}},
  \bibinfo {author} {\bibfnamefont {D.~P.}\ \bibnamefont {Kovács}}, \bibinfo
  {author} {\bibfnamefont {A.}~\bibnamefont {Musaelian}}, \bibinfo {author}
  {\bibfnamefont {G.~N.~C.}\ \bibnamefont {Simm}}, \bibinfo {author}
  {\bibfnamefont {R.}~\bibnamefont {Drautz}}, \bibinfo {author} {\bibfnamefont
  {C.}~\bibnamefont {Ortner}}, \bibinfo {author} {\bibfnamefont
  {B.}~\bibnamefont {Kozinsky}},\ and\ \bibinfo {author} {\bibfnamefont
  {G.}~\bibnamefont {Csányi}},\ }\bibfield  {title} {\enquote {\bibinfo
  {title} {The design space of e(3)-equivariant atom-centered interatomic
  potentials},}\ }\href {https://arxiv.org/abs/2205.06643v2} {\  (\bibinfo
  {year} {2022})}\BibitemShut {NoStop}%
\bibitem [{\citenamefont {Batzner}\ \emph {et~al.}(2021)\citenamefont
  {Batzner}, \citenamefont {Musaelian}, \citenamefont {Sun}, \citenamefont
  {Geiger}, \citenamefont {Mailoa}, \citenamefont {Kornbluth}, \citenamefont
  {Molinari}, \citenamefont {Smidt},\ and\ \citenamefont
  {Kozinsky}}]{Batzner2021/10.1038/s41467-022-29939-5}%
  \BibitemOpen
  \bibfield  {author} {\bibinfo {author} {\bibfnamefont {S.}~\bibnamefont
  {Batzner}}, \bibinfo {author} {\bibfnamefont {A.}~\bibnamefont {Musaelian}},
  \bibinfo {author} {\bibfnamefont {L.}~\bibnamefont {Sun}}, \bibinfo {author}
  {\bibfnamefont {M.}~\bibnamefont {Geiger}}, \bibinfo {author} {\bibfnamefont
  {J.~P.}\ \bibnamefont {Mailoa}}, \bibinfo {author} {\bibfnamefont
  {M.}~\bibnamefont {Kornbluth}}, \bibinfo {author} {\bibfnamefont
  {N.}~\bibnamefont {Molinari}}, \bibinfo {author} {\bibfnamefont {T.~E.}\
  \bibnamefont {Smidt}},\ and\ \bibinfo {author} {\bibfnamefont
  {B.}~\bibnamefont {Kozinsky}},\ }\bibfield  {title} {\enquote {\bibinfo
  {title} {E(3)-equivariant graph neural networks for data-efficient and
  accurate interatomic potentials},}\ }\href
  {https://doi.org/10.1038/s41467-022-29939-5} {\bibfield  {journal} {\bibinfo
  {journal} {Nature Communications 2022 13:1}\ }\textbf {\bibinfo {volume}
  {13}},\ \bibinfo {pages} {1--11} (\bibinfo {year} {2021})}\BibitemShut
  {NoStop}%
\bibitem [{\citenamefont {Bochkarev}\ \emph {et~al.}(2022)\citenamefont
  {Bochkarev}, \citenamefont {Lysogorskiy}, \citenamefont {Ortner},
  \citenamefont {Csányi},\ and\ \citenamefont
  {Drautz}}]{Bochkarev2022/10.1103/PHYSREVRESEARCH.4.L042019}%
  \BibitemOpen
  \bibfield  {author} {\bibinfo {author} {\bibfnamefont {A.}~\bibnamefont
  {Bochkarev}}, \bibinfo {author} {\bibfnamefont {Y.}~\bibnamefont
  {Lysogorskiy}}, \bibinfo {author} {\bibfnamefont {C.}~\bibnamefont {Ortner}},
  \bibinfo {author} {\bibfnamefont {G.}~\bibnamefont {Csányi}},\ and\ \bibinfo
  {author} {\bibfnamefont {R.}~\bibnamefont {Drautz}},\ }\bibfield  {title}
  {\enquote {\bibinfo {title} {Multilayer atomic cluster expansion for
  semilocal interactions},}\ }\href
  {https://doi.org/10.1103/PHYSREVRESEARCH.4.L042019/FIGURES/5/MEDIUM}
  {\bibfield  {journal} {\bibinfo  {journal} {Physical Review Research}\
  }\textbf {\bibinfo {volume} {4}},\ \bibinfo {pages} {L042019} (\bibinfo
  {year} {2022})}\BibitemShut {NoStop}%
\bibitem [{\citenamefont {Siria}, \citenamefont {Bocquet},\ and\ \citenamefont
  {Bocquet}(2017)}]{Siria201710.1038/s41570-017-0091}%
  \BibitemOpen
  \bibfield  {author} {\bibinfo {author} {\bibfnamefont {A.}~\bibnamefont
  {Siria}}, \bibinfo {author} {\bibfnamefont {M.~L.}\ \bibnamefont {Bocquet}},\
  and\ \bibinfo {author} {\bibfnamefont {L.}~\bibnamefont {Bocquet}},\
  }\bibfield  {title} {\enquote {\bibinfo {title} {New avenues for the
  large-scale harvesting of blue energy},}\ }\href
  {https://doi.org/10.1038/s41570-017-0091} {\bibfield  {journal} {\bibinfo
  {journal} {Nature Reviews Chemistry 2017 1:11}\ }\textbf {\bibinfo {volume}
  {1}},\ \bibinfo {pages} {1--10} (\bibinfo {year} {2017})}\BibitemShut
  {NoStop}%
\bibitem [{\citenamefont {Marbach}\ and\ \citenamefont
  {Bocquet}(2019)}]{Marbach2019/10.1039/C8CS00420J}%
  \BibitemOpen
  \bibfield  {author} {\bibinfo {author} {\bibfnamefont {S.}~\bibnamefont
  {Marbach}}\ and\ \bibinfo {author} {\bibfnamefont {L.}~\bibnamefont
  {Bocquet}},\ }\bibfield  {title} {\enquote {\bibinfo {title} {Osmosis, from
  molecular insights to large-scale applications},}\ }\href
  {https://doi.org/10.1039/C8CS00420J} {\bibfield  {journal} {\bibinfo
  {journal} {Chemical Society Reviews}\ }\textbf {\bibinfo {volume} {48}},\
  \bibinfo {pages} {3102--3144} (\bibinfo {year} {2019})}\BibitemShut {NoStop}%
\bibitem [{\citenamefont {Tunuguntla}\ \emph {et~al.}(2017)\citenamefont
  {Tunuguntla}, \citenamefont {Henley}, \citenamefont {Yao}, \citenamefont
  {Pham}, \citenamefont {Wanunu},\ and\ \citenamefont
  {Noy}}]{Tunuguntla2017/10.1126/SCIENCE.AAN2438}%
  \BibitemOpen
  \bibfield  {author} {\bibinfo {author} {\bibfnamefont {R.~H.}\ \bibnamefont
  {Tunuguntla}}, \bibinfo {author} {\bibfnamefont {R.~Y.}\ \bibnamefont
  {Henley}}, \bibinfo {author} {\bibfnamefont {Y.~C.}\ \bibnamefont {Yao}},
  \bibinfo {author} {\bibfnamefont {T.~A.}\ \bibnamefont {Pham}}, \bibinfo
  {author} {\bibfnamefont {M.}~\bibnamefont {Wanunu}},\ and\ \bibinfo {author}
  {\bibfnamefont {A.}~\bibnamefont {Noy}},\ }\bibfield  {title} {\enquote
  {\bibinfo {title} {Enhanced water permeability and tunable ion selectivity in
  subnanometer carbon nanotube porins},}\ }\href
  {https://doi.org/10.1126/SCIENCE.AAN2438} {\bibfield  {journal} {\bibinfo
  {journal} {Science}\ }\textbf {\bibinfo {volume} {357}},\ \bibinfo {pages}
  {792--796} (\bibinfo {year} {2017})}\BibitemShut {NoStop}%
\bibitem [{\citenamefont {Robin}, \citenamefont {Kavokine},\ and\ \citenamefont
  {Bocquet}(2021)}]{Robin2021/10.1126/SCIENCE.ABF7923}%
  \BibitemOpen
  \bibfield  {author} {\bibinfo {author} {\bibfnamefont {P.}~\bibnamefont
  {Robin}}, \bibinfo {author} {\bibfnamefont {N.}~\bibnamefont {Kavokine}},\
  and\ \bibinfo {author} {\bibfnamefont {L.}~\bibnamefont {Bocquet}},\
  }\bibfield  {title} {\enquote {\bibinfo {title} {Modeling of emergent memory
  and voltage spiking in ionic transport through angstrom-scale slits},}\
  }\href {https://doi.org/10.1126/SCIENCE.ABF7923} {\bibfield  {journal}
  {\bibinfo  {journal} {Science}\ }\textbf {\bibinfo {volume} {373}},\ \bibinfo
  {pages} {687--691} (\bibinfo {year} {2021})}\BibitemShut {NoStop}%
\bibitem [{\citenamefont {Fong}\ \emph {et~al.}(2024)\citenamefont {Fong},
  \citenamefont {Sumic}, \citenamefont {O'Neill}, \citenamefont {Schran},
  \citenamefont {Grey},\ and\ \citenamefont
  {Michaelides}}]{Fong2024/10.26434/CHEMRXIV-2024-R67MX}%
  \BibitemOpen
  \bibfield  {author} {\bibinfo {author} {\bibfnamefont {K.~D.}\ \bibnamefont
  {Fong}}, \bibinfo {author} {\bibfnamefont {B.}~\bibnamefont {Sumic}},
  \bibinfo {author} {\bibfnamefont {N.}~\bibnamefont {O'Neill}}, \bibinfo
  {author} {\bibfnamefont {C.}~\bibnamefont {Schran}}, \bibinfo {author}
  {\bibfnamefont {C.~P.}\ \bibnamefont {Grey}},\ and\ \bibinfo {author}
  {\bibfnamefont {A.}~\bibnamefont {Michaelides}},\ }\bibfield  {title}
  {\enquote {\bibinfo {title} {The interplay of solvation and polarization
  effects on ion pairing in nanoconfined electrolytes},}\ }\href
  {https://doi.org/10.26434/CHEMRXIV-2024-R67MX} {\  (\bibinfo {year} {2024}),\
  10.26434/CHEMRXIV-2024-R67MX}\BibitemShut {NoStop}%
\bibitem [{\citenamefont {Behler}\ and\ \citenamefont
  {Parrinello}(2007)}]{Behler2007/10.1103/PHYSREVLETT.98.146401}%
  \BibitemOpen
  \bibfield  {author} {\bibinfo {author} {\bibfnamefont {J.}~\bibnamefont
  {Behler}}\ and\ \bibinfo {author} {\bibfnamefont {M.}~\bibnamefont
  {Parrinello}},\ }\bibfield  {title} {\enquote {\bibinfo {title} {Generalized
  neural-network representation of high-dimensional potential-energy
  surfaces},}\ }\href {https://doi.org/10.1103/PHYSREVLETT.98.146401}
  {\bibfield  {journal} {\bibinfo  {journal} {Physical Review Letters}\
  }\textbf {\bibinfo {volume} {98}},\ \bibinfo {pages} {146401} (\bibinfo
  {year} {2007})}\BibitemShut {NoStop}%
\bibitem [{\citenamefont {Bussi}, \citenamefont {Donadio},\ and\ \citenamefont
  {Parrinello}(2007)}]{Bussi2007/10.1063/1.2408420}%
  \BibitemOpen
  \bibfield  {author} {\bibinfo {author} {\bibfnamefont {G.}~\bibnamefont
  {Bussi}}, \bibinfo {author} {\bibfnamefont {D.}~\bibnamefont {Donadio}},\
  and\ \bibinfo {author} {\bibfnamefont {M.}~\bibnamefont {Parrinello}},\
  }\bibfield  {title} {\enquote {\bibinfo {title} {Canonical sampling through
  velocity rescaling},}\ }\href {https://doi.org/10.1063/1.2408420} {\bibfield
  {journal} {\bibinfo  {journal} {Journal of Chemical Physics}\ }\textbf
  {\bibinfo {volume} {126}},\ \bibinfo {pages} {14101} (\bibinfo {year}
  {2007})}\BibitemShut {NoStop}%
\bibitem [{\citenamefont {Ceriotti}\ \emph {et~al.}(2010)\citenamefont
  {Ceriotti}, \citenamefont {Parrinello}, \citenamefont {Markland},\ and\
  \citenamefont {Manolopoulos}}]{Ceriotti2010/10.1063/1.3489925}%
  \BibitemOpen
  \bibfield  {author} {\bibinfo {author} {\bibfnamefont {M.}~\bibnamefont
  {Ceriotti}}, \bibinfo {author} {\bibfnamefont {M.}~\bibnamefont
  {Parrinello}}, \bibinfo {author} {\bibfnamefont {T.~E.}\ \bibnamefont
  {Markland}},\ and\ \bibinfo {author} {\bibfnamefont {D.~E.}\ \bibnamefont
  {Manolopoulos}},\ }\bibfield  {title} {\enquote {\bibinfo {title} {Efficient
  stochastic thermostatting of path integral molecular dynamics},}\ }\href
  {https://doi.org/10.1063/1.3489925} {\bibfield  {journal} {\bibinfo
  {journal} {Journal of Chemical Physics}\ }\textbf {\bibinfo {volume} {133}},\
  \bibinfo {pages} {124104} (\bibinfo {year} {2010})}\BibitemShut {NoStop}%
\end{thebibliography}
%aipnum4-2.bst 2019-01-14 (MD) hand-edited version of apsrev4-1.bst
%Control: production of eprint (1) enabled
%

\end{document}

\def\mytitle{%
        To Pair or not to Pair? Machine-Learned Explicitly-Correlated Electronic Structure for NaCl in Water}

\title{Supporting Information for: \mytitle}
\author{Niamh O'Neill}%
\affiliation{%
Yusuf Hamied Department of Chemistry, University of Cambridge, Lensfield Road, Cambridge, CB2 1EW, UK
}
\affiliation{%
Cavendish Laboratory, Department of Physics, University of Cambridge, Cambridge, CB3 0HE, UK
}
\affiliation{%
Lennard-Jones Centre, University of Cambridge, Trinity Ln, Cambridge, CB2 1TN, UK
}

\author{Benjamin X. Shi}%
\affiliation{%
Yusuf Hamied Department of Chemistry, University of Cambridge, Lensfield Road, Cambridge, CB2 1EW, UK
}
\affiliation{%
Lennard-Jones Centre, University of Cambridge, Trinity Ln, Cambridge, CB2 1TN, UK
}

\author{Kara Fong}%
\affiliation{%
Yusuf Hamied Department of Chemistry, University of Cambridge, Lensfield Road, Cambridge, CB2 1EW, UK
}
\affiliation{%
Lennard-Jones Centre, University of Cambridge, Trinity Ln, Cambridge, CB2 1TN, UK
}

\author{Angelos Michaelides}%
\email{am452@cam.ac.uk}
\affiliation{%
Yusuf Hamied Department of Chemistry, University of Cambridge, Lensfield Road, Cambridge, CB2 1EW, UK
}
\affiliation{%
Lennard-Jones Centre, University of Cambridge, Trinity Ln, Cambridge, CB2 1TN, UK
}

\author{Christoph Schran}%
\email{cs2121@cam.ac.uk}
\affiliation{%
Cavendish Laboratory, Department of Physics, University of Cambridge, Cambridge, CB3 0HE, UK
}
\affiliation{%
Lennard-Jones Centre, University of Cambridge, Trinity Ln, Cambridge, CB2 1TN, UK
}
\title{Supporting Information for: To Pair or not to Pair? Machine-Learned Explicitly-Correlated Electronic Structure for NaCl in Water}

{\maketitle}

\tableofcontents

\onecolumngrid

\section{Electronic Structure}\label{sec:elec-struc}
\subsection{DFT settings}
All DFT calculations were performed in CP2K to generate both total energies and nuclear gradients \cite{Kuhne2020/10.1063/5.0007045}.
The electronic density was partitioned into core and valence contributions, with core electrons described using the norm-conserving Goedecker, Teter and Hutter (GTH) pseudopotentials \cite{Goedecker1996/10.1103/PhysRevB.54.1703}.
Na 2s and 2p electrons were also treated explicitly given the well-known issue of non-linear core-valence exchange/correlation.
Valence electrons were described using the MOLOPT TZV2P basis set \cite{VandeVondele2007/10.1063/1.2770708}.
To cover the range of DFT XC functional approximations, functionals were chosen from ascending rungs of Jacob's Ladder.
In short we used the generalised gradient approximation, revPBE \cite{Perdew1996/10.1103/PhysRevLett.77.3865,Zhang1998/10.1103/PhysRevLett.80.890}, meta GGA r$^2$SCAN \cite{Furness2020/10.1021/ACS.JPCLETT.0C02405} which improves on the numerical instabilities of SCAN \cite{Sun2015/10.1103/PHYSREVLETT.115.036402}, van der Waals inclusive optB88-vdW \cite{Klimes2009/10.1088/0953-8984/22/2/022201} and hybrid revPBE0 \cite{Adamo1999/10.1063/1.478522}. revPBE and revPBE0 were used with with the zero-damping variant of Grimme's D3 dispersion correction \cite{Grimme2010/10.1063/1.3382344}.
The revPBE0 calculations were performed using the auxiliary density matrix method \cite{Giodon2010/10.1021/CT1002225}, to reduce the cost of computing the exact exchange component.

\begin{figure}[ht]
	    \centering
                \includegraphics[width=1.0\textwidth]{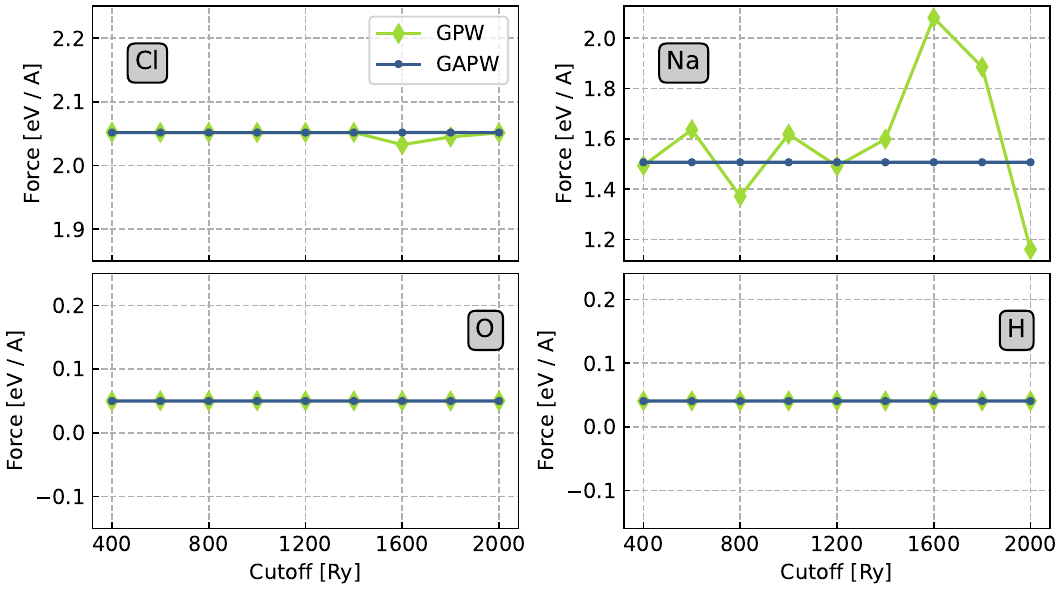}
		        
				\caption{Convergence plots of forces on each atom type (Na, Cl, O, H) vs plane wave cutoff for GAPW (blue) and GPW (green) methods.
				 \label{fig:force-converge}}
\end{figure}

\begin{figure}[ht]
	    \centering
	        \includegraphics[width=0.7\textwidth]{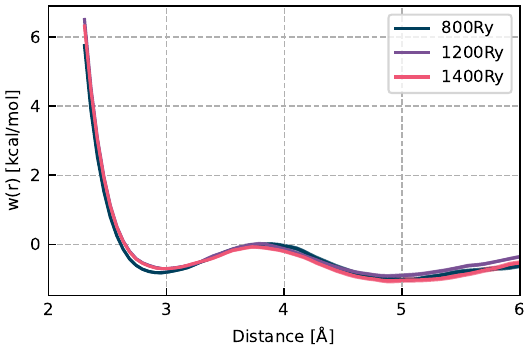}
		        
				\caption{PMF for revPBE-D3 MLP with PW cutoffs at 800, 1200 and 1400 Ry.
				 \label{fig:pmf-pw}}
\end{figure}

\begin{figure}[ht]
	    \centering
	        \includegraphics[width=0.7\textwidth]{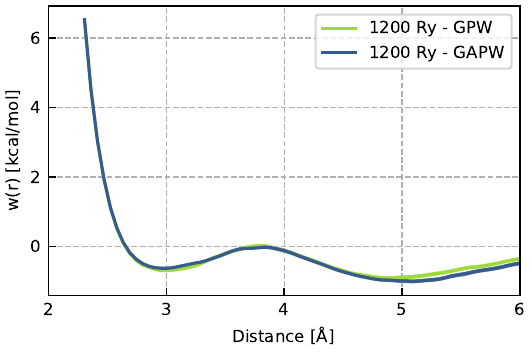}
		        
				\caption{PMF for revPBE-D3 MLP with 1200 Ry PW cutoff for both GPW and GAPW methods.
				 \label{fig:pmf-gapw}}
\end{figure}

Figure \ref{fig:force-converge} shows the convergence of the forces on the four atom types with respect to plane-wave cutoff.
A large cutoff is required to converge the forces on Na.
Using the Gaussian and Augmented Plane Wave (GAPW) method \cite{Lippert1999/10.1007/S002140050523} resolves this issue, with forces converging after 400 Ry.
However, the GAPW method is not available for wavefunction methods RPA and MP2.
We therefore consider the effect of these stochastic forces on the sodium atoms with respect to the property of interest in this work - the potential of mean force of Na and Cl in water.
Figure \ref{fig:pmf-pw} shows the PMF for 3 machine learning potential (MLP) models at revPBE-D3 level of theory using the GPW method with increasing plane-wave cutoff.
This property is well-converged with PW cutoff, with an acceptable error of approximately 0.1 kcal/mol.
Additionally, Figure \ref{fig:pmf-gapw} shows that again the target property, the PMF is not affected by using the GPW method over the GAPW method.
Therefore in summary, all DFT calculations were performed using the GPW method with 1200 Ry plane-wave cutoff.

\subsection{Correlated wave-function theory}
\begin{figure}[ht]
	    \centering
	        \includegraphics[width=1.0\textwidth]{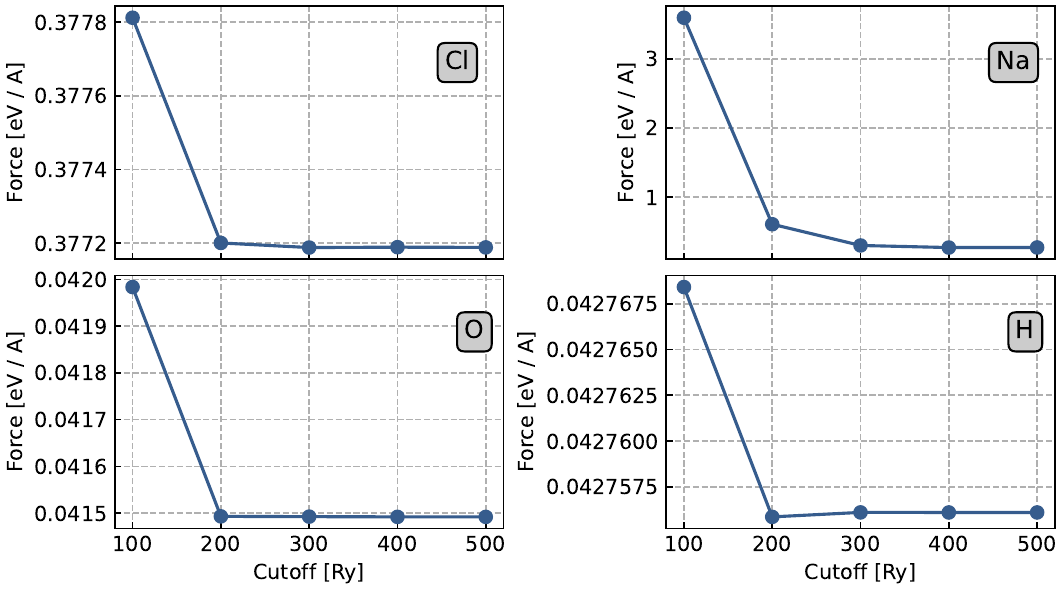}
		        
				\caption{Convergance plots of forces on each atom type (Na, Cl, O, H) vs integral plane wave cutoff.
				 \label{fig:RPA-integral}}
\end{figure}

Random-phase approximation (RPA) and second-order M$\o$ller Plesset perturbation theory (MP2) were performed in CP2K \cite{Kuhne2020/10.1063/5.0007045} to generate both total energies and nuclear gradients.
We used DFT with the PBE functional as the starting point for the RPA correlation energy calculations.
The resolution-of-identity (RI) techniques was used for these methods~\cite{Bussy2023/10.1063/5.0144493/2886896}. 
We used triple-zeta (TZ) quality correlation consistent basis sets for H and O (taken from CP2K) as well as Na and Cl.
Auxiliary basis sets for the RI integral operations were generated using the automatic auxiliary basis of Stoychev et al. \cite{Stoychev2017/10.1021/ACS.JCTC.6B01041, Lehtola2021/10.1021/ACS.JCTC.1C00607} for Na and Cl, with the defaults (from CP2K) used for H and O.
A planewave cutoff of 1200 Ry was again used.
We used the GTH-HF pseudopotentials from Goedecker-Teter-Hutter \cite{Goedecker1996/10.1103/PhysRevB.54.1703} for all of the atoms.
Figure \ref{fig:RPA-integral} shows the force convergence on the atoms with respect to the GPW integral cutoff.
A cutoff of 300 Ry was shown to be well converged within 0.00001 meV/\AA.
The calculation is also sensitive to the number of quadrature points  and so 20 were used, which has an error of \textless \, 1 \,mHartree according to literature \cite{DelBen2015/10.1016/J.CPC.2014.10.021}.


\section{Development and validation of machine learning potential}\label{si:development}
\subsection{Automated work flow}
\begin{figure}[ht]
	    \centering
	        \includegraphics[width=1.0\textwidth]{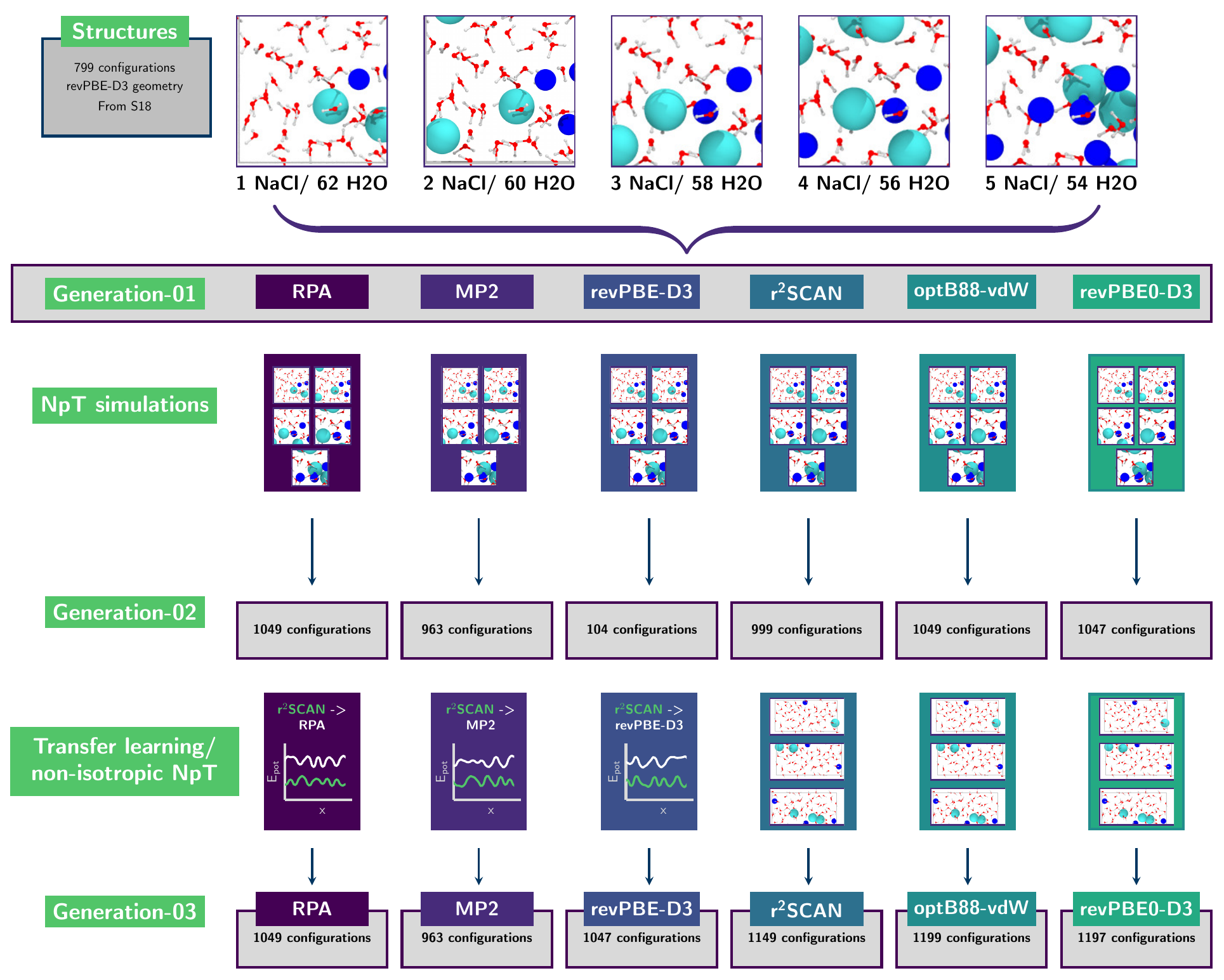}
		        
				\caption[Schematic of training procedure for the 6 ML models in this work.  ]{Schematic of training procedure for the 6 ML models in this work.  
				 \label{fig:mlp-dev}}
\end{figure}
The procedure for developing the committee neural network potentials (C-NNP) used in this work was followed as described in Ref~\citenum{Schran2021/10.1073/pnas.2110077118}.
Overall 6 models were trained to describe NaCl ions in water at different levels of electronic structure theory (revPBE-D3, optB88-vdW, r$^2$SCAN, revPBE0-D3, MP2 and RPA).
The training of the individual models was divided into
\textcolor{black}{three} generations, and is graphically depicted in Figure \ref{fig:mlp-dev}. 
The first generation comprised a common training set for all models, adapted from previous work \cite{ONeill2022/2211.04345}.
Specifically, only configurations corresponding to ions in solution were used, since this previous model also contained solid NaCl in water, which was used to explore dissolution.
These configurations comprised increasing concentrations of NaCl ions in water (See Figure \ref{fig:mlp-dev} for specifics).
The forces and energies of this common training set were then computed for the different levels of theory (See Section \ref{sec:elec-struc} for the electronic structure details) used in this paper, and individual models then trained as described below.
In order to ensure the relevant configuration space for a given level of theory was suitably covered by the models, NpT simulations were then performed with each model for all of the solution concentrations used in Generation 1.
An active learning procedure \cite{Schran2021/10.1073/pnas.2110077118} was then employed to select relevant configurations from each concentration to add to the model training set.
For a given active learning iteration, 20 random structures from a reference trajectory were used to initialise the model. %
After training 8 NNP members, forces and energies of 2000 randomly selected structures from the reference trajectory were predicted to ascertain the force and energy committee disagreements.
20 structures with the largest mean force disagreement were added to the training set for the next round of active learning.
Convergence was reached when new structures added to the training set did not improve the committee disagreement between points already in the training set, indicating the training set was sufficiently diverse.
\textcolor{black}{
Overall, approximately 200 additional structures were added per model.
In order to ensure the models could treat large ion-ion separations (beyond the 6~\AA ~\ cutoff of the model), a third iteration of training was performed (Generation 3 in Figure \ref{fig:mlp-dev}). 
This generation entailed two different protocols depending on the electronic structure method.
For revPBE0-D3, r$^2$SCAN and optB88-vdW, additional NpT simulations were performed on non-isotropic simulation cells of initial dimensions 10 x 10 x 20  \AA \ containing 1, 2 and 3 ion pairs in 64, 62 and 60 waters respectively.
Another active learning procedure as described above was performed, to add approximately 150 structures in total per model,  leading to the training set of each model comprising $\sim$ 1100 structures (See Figure \ref{fig:mlp-dev}) for model-specific details.
For RPA, MP2 and revPBE-D3, for reasons of computational cost (RPA and MP2) and computational complexity (revPBE-D3), a transfer learning protocol was employed to ensure that the second generation models could treat the large ion-ion separations.
Transfer learning is a well-established machine learning technique, and has been successfully used in recent work for example in training a CCSD(T) water model with sparse training data \cite{Chen2023/10.1021/ACS.JCTC.2C01203}.
In this approach, the third generation r$^2$SCAN model was used as a baseline for the training of the RPA, MP2 and revPBE-D3 models.
In this way, the training was initialised from the pre-optimised r$^2$SCAN weights (rather than randomly seeded weights as is conventionally done) using the relevant training set from the second generation.
The specific benefits of this approach for our work are two-fold, typically requiring less training data and also easing the fitting procedure \cite{Pan2010/10.1109/TKDE.2009.191}.
}

\subsection{Details of Model}
The chemical environment around each atom was described using a general set of atom-centered symmetry functions~\cite{Behler2011/10.1063/1.3553717}.
There are 10 radial and 4 angular functions for each pair and triple of atoms, following Ref.~\citenum{Schran2021/10.1073/pnas.2110077118}.
All symmetry functions used a cutoff function of angular cosine form with a cutoff radius of 12 Bohr.
The committee was comprised of 8 NNP members, of identical architecture with 2 hidden layers and 25 neurons in each layer.
In all cases, random sub-sampling was performed to introduce variability between the committee members, where 10\% of the total set of structures were discarded.
The weights and biases of the NNPs were optimised using the n2p2 code \cite{Singraber2019/10.1021/acs.jctc.8b01092}.
Individual models during active learning were optimised for 15 epochs, while the final C-NNP model used in simulations was optimised for 50 epochs.

We have explicitly incorporated long-range effects beyond the cutoff of the symmetry functions (12 Bohr) of the machine learning potential.
The predicted energy can in general be written as a sum of short range and long range contributions ($E_{sr}$ and $E_{coul}$ respectively):	$E_{\mathrm{tot}} = E_{\mathrm{sr}}+E_{\mathrm{coul}}$.
The long-range model was thus trained on the difference between the standard short-ranged model and the Coulomb contribution, calculated using point charges of +/- 1 respectively for Na and Cl and using TIP3P model parameters for water \cite{Jorgensen1983/10.1063/1.445869}.
We used this model in all production simulations, where the  Coulomb contributions were explicitly included via particle mesh ewald summation.
Details on the validation of the final models are presented in the next Section.

\subsection{Validation}
\begin{figure}[ht]
	    \centering
	        \includegraphics[width=0.7\textwidth]{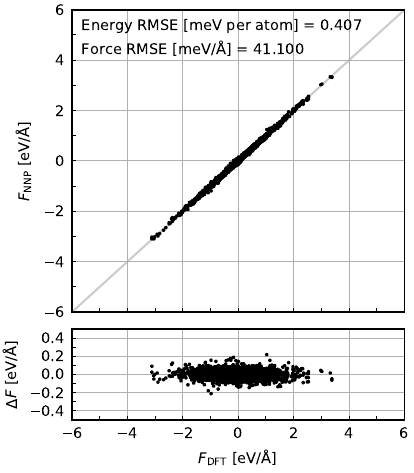}
		        
				\caption[Correlation plot for r$^2$SCAN C-NNP model predicted forces and corresponding reference DFT forces.]{Correlation plot for r$^2$SCAN C-NNP predicted forces and corresponding reference DFT forces, with light grey line showing a perfect correlation coefficient of 1. 
				 \label{fig:rmse}}
\end{figure}

We validate the ability of the model to reproduce its underlying reference method based on a validation set of 100 configurations.
These comprise a scan along the inter-ion separation coordinate up to 6 \AA{} (giving a computationally reasonable box size for RPA and MP2 reference calculations).
The force and energy RMSE for each model are shown in Table \ref{tab:rmse} for each model, and a representative correlation plot for r$^2$SCAN in Figure \ref{fig:rmse}.
These errors compare favorably with our previous model \cite{ONeill2022/2211.04345} with RMSE values for both forces (37.0 meV/\AA{}) and energies (0.3 meV/atom) as well as similar reactive systems, which have been studied using machine learning potentials, such as the work by Behler et al. in Ref.~\citenum{Quaranta2017/10.1021/ACS.JPCLETT.7B00358} who quote a force and energy RMSE for a model describing proton transport at ZnO/\ch{H2O} interfaces of 140.4 meV/\AA{} and 1.0 meV/atom respectively.
 
\textbf{Training errors, forces}

\begin{table}[h]
\caption{Summary of force and energy RMSE for each ML model.\label{tab:rmse}}
\begin{tabular}{ l c c }
\toprule
\multicolumn{1}{l}{Model} & {\begin{tabular}[c]{@{}c@{}}Energy RMSE \\ {[}meV/atom{]}\end{tabular}} & {\begin{tabular}[c]{@{}c@{}}Force RMSE\\  {[}meV/Å{]}\end{tabular}} \\ \midrule
revPBE-D3                 & 2.366  & 27.280 \\
optB88-vdW                & 0.503  & 37.435 \\
r$^2$SCAN                    & 0.407  & 41.100 \\
revPBE0-D3                & 0.780  & 39.430 \\
MP2                       & 0.763  & 40.160 \\
RPA                       & 0.819  & 38.327 \\ \bottomrule
\end{tabular}
\end{table}

\section{Simulation Details}
\subsection{System setup}
The NaCl PMF was obtained from molecular dynamics simulation using the machine learning potentials from above.
\textcolor{black}{
The system comprised 6 NaCl ion pairs in 332 waters giving a concentration of 1 mol/kg in a cubic simulation cell with periodic boundary conditions in the $x,y$ and $z$ directions.}
For each model, a 1 ns NpT simulation was first performed to obtain the equillibrium density.
From the NpT simulations, 10 configurations were sampled to give uncorrelated starting configurations for production simulations run at the model density.
Production simulations were performed in the NVT ensemble, with a timestep of 1 fs at 300 K.
The PMF was then obtained from the average of the rdfs from the 10 independant simulations, using the relation $w(\mathbf{r}) = -k_BT\,\mathrm{ln}\,g(\mathbf{r_{\mathrm{Na-Cl}}})$ where ${k_B}$ and ${T}$ are the Boltzman constant and temperature (300 K), with $g_{\mathrm{Na-Cl}}(r)$ being the radial distribution function between Na and Cl ion pairs.
The standard deviation between replicates was used to quantify the statistical uncertainty.
Overall, over 200 ns of \textit{ab initio} quality simulations were performed, highlighting the major advantage of the machine learning approach.
\textcolor{black}{
The density and diffusion coefficients were computed as a function of concentration for MP2 and RPA.
For a fixed number of 270 waters, increasing numbers of ion pairs were added to the box up to a final concentration of roughly 4 mol/kg (corresponding to 20 ion pairs in 270 waters.)
The density was computed from an NpT simulation at each concentration, with a block averaging procedure to obtain statistical error bars.
The diffusion coefficients were computed from NVT simulations at the equillibrium density obtained from the previous step, and production simulations were at least 4 ns long.
Diffusion coefficients ($D$) were computed from the Einstein relation
\begin{equation}
    D = \lim_{t\to\infty} \frac{1}{6} \frac{d \langle |\mathbf{r}(t)-\mathbf{r}(0) |^2 \rangle}{dt}
\end{equation}
The standard Yuh and Hummer correction for finite-box size \cite{Yeh2004/10.1021/JP0477147} was added, using the SPC/E viscosity $\eta = 0.66$ cP.}

\subsection{Convergence tests}
\begin{figure}[ht]
	    \centering
	        \includegraphics[width=0.7\textwidth]{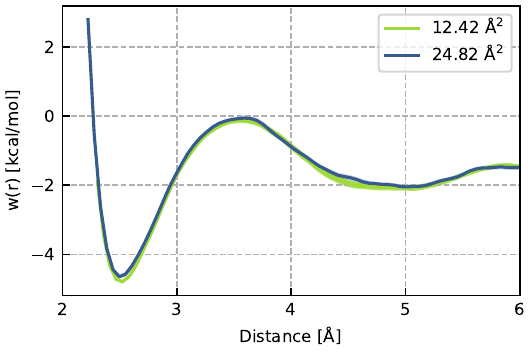}
		        
				\caption{Comparison of finite size effects for small and large box sizes comprising a  single ion pair and 1 M solution respectively.
				 \label{fig:box-size}}
\end{figure}

This section describes several tests to ensure our simulation protocol was statistically converged and did not suffer from finite-size effects.
\newline
\textbf{Simulation box size}
Figure \ref{fig:box-size} shows the PMF of the large 24.82 \AA{} cubic box comprising 6 NaCl ion pairs and 332 waters used in this work and the 12.42 \AA{} cubic box of one ion pair and 62 waters used in validation tests.

\begin{figure*}[ht]
	    \centering
	        \includegraphics[width=0.7\textwidth]{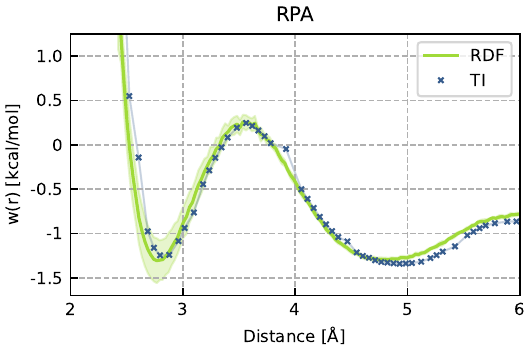}
		        
				\caption{Comparison of thermodynamic integration and RDF approaches for the RPA model.
				 \label{fig:TI-rdf}}
\end{figure*}

\textbf{Sampling - thermodynamic integration vs RDF}
To ensure that all regions along the inter-ion separation coordinate were sufficintly sampled, we compare a thermodynamic integration scheme as described in \cite{Spirik1998/10.1063/1.477419} to the RDF approach used in this work.
Figure \ref{fig:TI-rdf} shows the PMF obtained from RDF and thermodynamic integration (TI) for the RPA model, showing the equivalence of the RDF approach for a 1 M system and thermodynamic integration for 1 ion pair within their own statistical errors.

\textbf{Simulation time}

\begin{figure}[ht]
	    \centering
	        \includegraphics[width=0.7\textwidth]{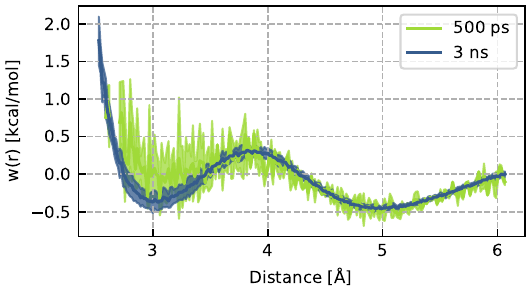}
		        
				\caption{Comparison of PMF obtained from MLP simulations for 500 ps vs fully converged MLP simulations of 10 replicates of 3 ns each.
				 \label{fig:aimd}}
\end{figure}

One of the major advantages of machine learning potentials is the much greater timescales accessible during simulations than for standard \textit{ab initio} methods.
Here we compute the PMF using one of the MLPs over a time period of 500 ps, for 2 replicates, a time period at the extreme upper end of that accessible for \textit{ab initio} simulations (Note that for the hybrid and wavefunction methods even this timescale is unfeasible).
In figure \ref{fig:aimd}, we compare this with the ease in which the PMF can be converged over 10 replicates each of over 3 ns with the MLP.

\newpage
\section{Additional Results}
\textcolor{black}{
\subsection{Diffusion Coefficients}
The water self-diffusion coefficients were computed as a function of concentration for the MP2 and RPA models. Figure \ref{fig:diffusion} shows the computed diffusion coefficients for MP2 and RPA compared with various values from literature. In all cases the self-diffusion coefficient $D$ has been normalised with respect to the self-diffusion coefficient in bulk water with no ions ($D_0$).
\begin{figure}[h!]
    \centering
\includegraphics[width=0.7\textwidth]{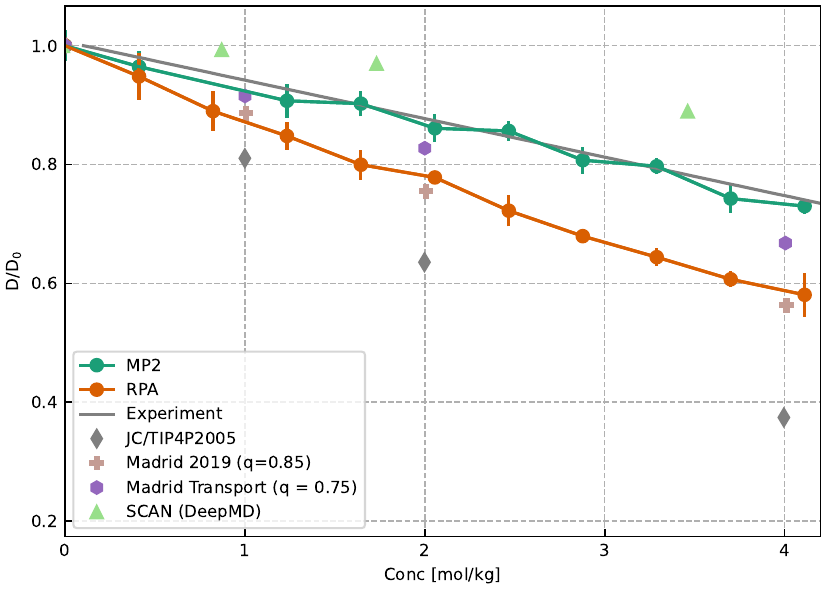}
    \caption{Normalised self-diffusion coefficient of water as a function of concentration for the MP2 and RPA models trained in this work and compared experiment (298 K) to various classical force-fields (298.15) \cite{Blazquez2023/10.1063/5.0136498/2871744} and a DeepMD potential based on SCAN (330 K). \cite{Panagiotopoulos2023/10.1021/ACS.JPCB.2C07477} }
    \label{fig:diffusion}
\end{figure}
}
\newpage
\textcolor{black}{%
\textbf{PMF Insights}\\
\textbf{Ion-ion binding energy:}
In order to explore the effect of ion-ion interactions on the PMF, we computed the binding curve of a Na Cl ion pair in vacuum using the reference electronic structure method for each model. 
The energy was computed for 100 ion separation distances.
The binding energy was obtained from the minimum in the binding curve after alignment of the curves at 3.5 \AA \ (inside the transition state for all models), and the curves are shown in Figure \ref{fig:binding-curve}.\\
\begin{figure}
    \centering\includegraphics[width=0.7\textwidth]{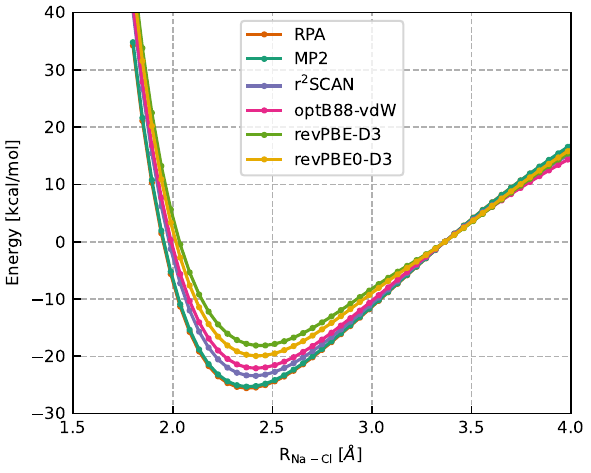}
    \caption{\ce{Na^+} \ce{Cl^-} ion-ion interaction curve aligned at 3.5 \AA~ for the various electronic structure methods used in this paper.}
    \label{fig:binding-curve}
\end{figure}
\begin{figure}
    \centering
    \includegraphics[width=0.7\textwidth]{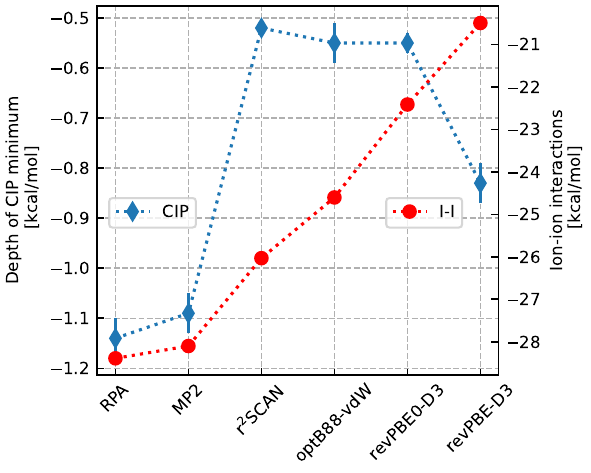}
    \caption{Relationship between CIP well depth and strength of ion-ion interactions.}
    \label{fig:ion-ion}
\end{figure}
}

\textcolor{black}{%
\textbf{Ion-water interaction energy}
To obtain a proxy for the ion-water interactions we computed the energy of interaction for an ion pair in water at the 3 regions of interest -- CIP, TS and SSIP.
This interaction energy of the ion pair with water $E_{IW}$ is given by
\begin{equation}
    E_{IW} = E_{sys} - E_W - E_G
\end{equation} where $E_{sys}$ is the complete interacting system, $E_W$ is the energy of the water with the ion-pair removed and $E_G$ is the energy of the gas-phase ion-pair.
Column 3 of Table \ref{tab:iw} shows the ion-water interaction energies for each functional.
Moreover, the interaction energy of a single ion in water was also computed in a similar manner, where the interaction energy of the ion of identity $X$ with water $E_{XW}$ is given by:
\begin{equation}
    E_{XW} = E_{sys} - E_W - E_GX
\end{equation}
where again $E_{sys}$ is the complete interacting system, $E_W$ is the energy of the water with the ion removed and $E_G$ is the energy of the gas-phase ion.
Table \ref{tab:iw} shows the interaction energy of an individual \ce{Na^+}/\ce{Cl^-} ion with water.
revPBE-D3 has the strongest \ce{Na^+} - water interaction since it is a GGA, with both delocalisation error and the D3 correction resulting in overbinding \cite{Bryenton2023/10.1002/WCMS.1631}.
Inclusion of Hartree-Fock exchange in revPBE0-D3 reduces this overbinding.
RPA and MP2 show very similar behaviour with moderate \ce{Na^+} - water interaction but strong \ce{Cl-} - water interaction.
Meanwhile most DFT functionals find it hard to agree with RPA or MP2 on the \ce{Cl^-} - water interactions. This is a  well known problem since it is difficult to localise the extra electron. 
This issue is worst for GGAs (revPBE-D3), but is improved with meta-GGA (r$^2$SCAN) and vdW-inclusive (optB88-vdW) functionals \cite{Dellostritto2020/10.1039/C9CP06821J}
}

\begin{table}
\caption{Individual ion/water (\ce{Na^+}-W/\ce{Cl^-}-W) interaction energies and total ion-pair-water interaction energies ([\ce{Na^+} - \ce{Cl^-}] - W)}
\label{tab:iw}
\begin{tabular}{@{}llll@{}}
\toprule
Model      & \multicolumn{1}{c}{\begin{tabular}[c]{@{}c@{}}\ce{Na^+}-W\\ (kcal/mol)\end{tabular}} & \multicolumn{1}{c}{\begin{tabular}[c]{@{}c@{}}\ce{Cl^-}-W\\ (kcal/mol)\end{tabular}} & \multicolumn{1}{c}{\begin{tabular}[c]{@{}c@{}}[\ce{Na^+} - \ce{Cl^-}] - W\\ (kcal/mol)\end{tabular}} \\ \midrule

RPA        & -173.78         & 4.79  &   -158.81       \\
MP2        & -172.31         & 3.21  &    -159.77      \\
r$^2$SCAN     & -169.29         & 8.83   &   -151.81    \\
optB88-vdW & -170.95         & 5.05   &    -152.78     \\
revPBE-D3  & -175.65         & 12.68   &   -144.97      \\
revPBE0-D3 & -172.45         & 13.28    &    -146.72   \\ \bottomrule
\end{tabular}
\end{table}

\begin{figure}[ht]
	    \centering
	        \includegraphics[width=1.0\textwidth]{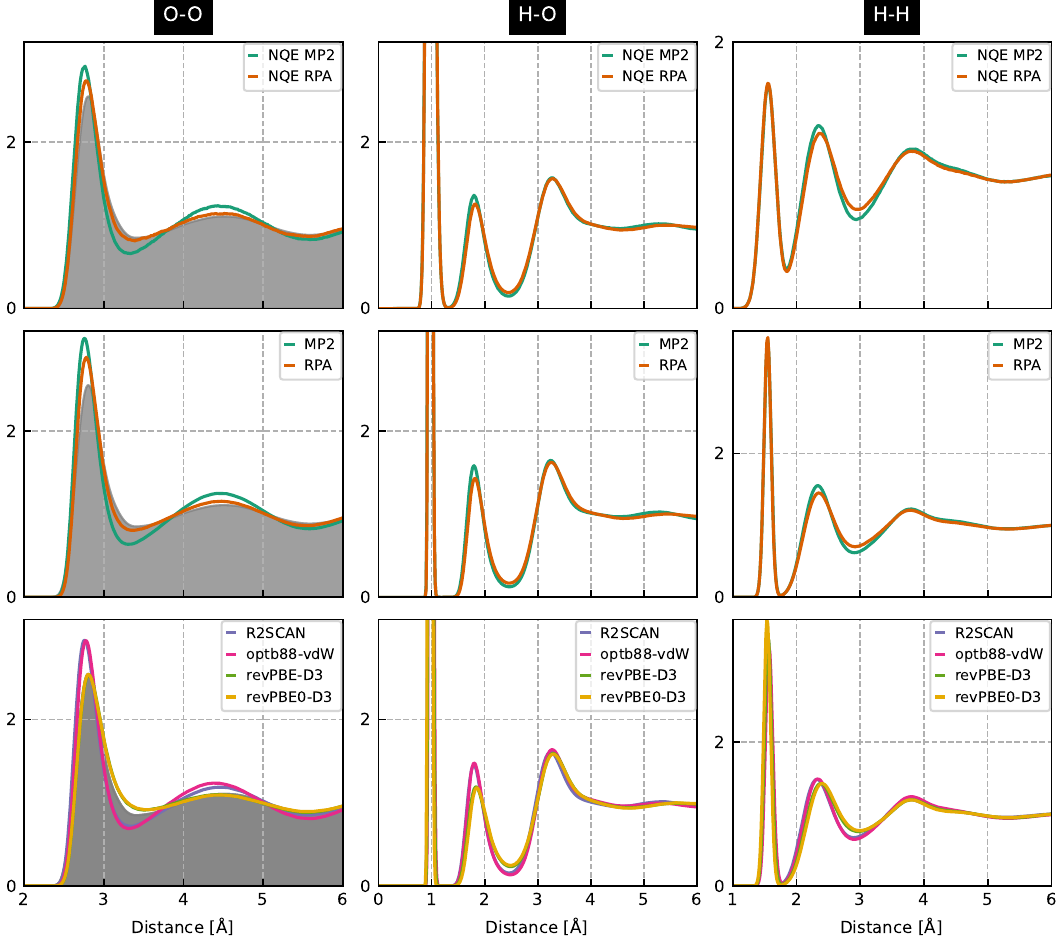}
		        
				\caption{Water - water RDFs for bulk water.
				 \label{fig:water-rdfs}}
\end{figure}

\begin{figure}[ht]
	    \centering
	        \includegraphics[width=1.0\textwidth]{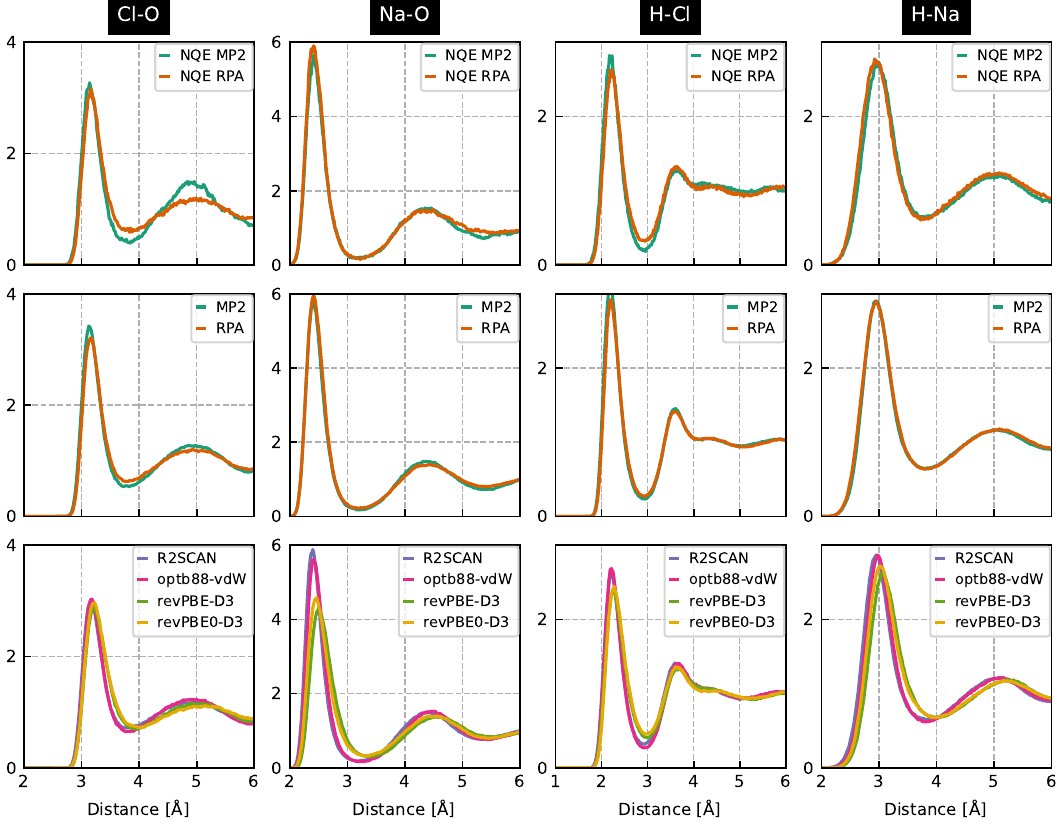}
		        
				\caption{Ion - water RDFs for one NaCl in a 95 water box.
				 \label{fig:ions-rdfs}}
\end{figure}

\begin{figure}[ht]
	    \centering
	        \includegraphics[width=1.0\textwidth]{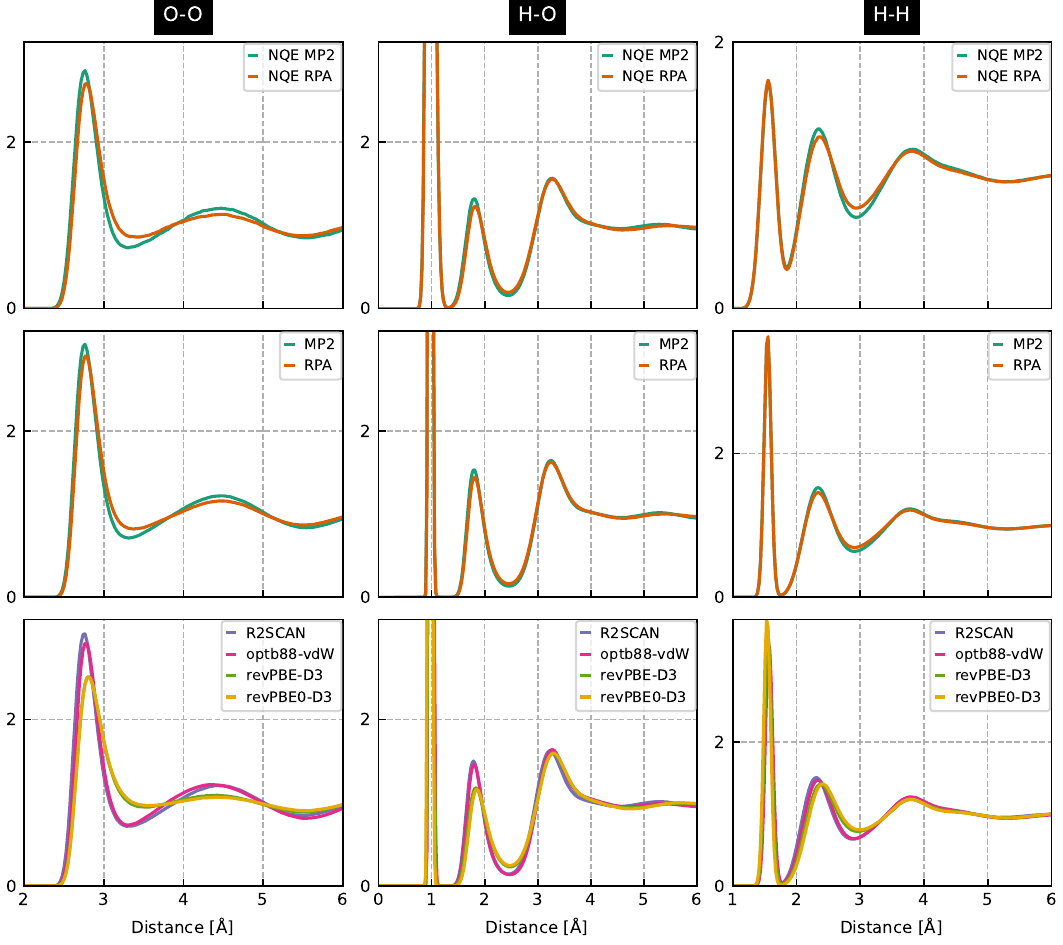}
		        
				\caption{Water - water RDFs for one NaCl in a 95 water box.
				 \label{fig:water-ions-rdfs}}
\end{figure}

\textbf{Additional RDFs}
Figures \ref{fig:water-rdfs}, \ref{fig:ions-rdfs} and \ref{fig:water-ions-rdfs} summarise all of the RDFs computed in this work.
Figure \ref{fig:water-rdfs} shows the bulk water RDFs.
Figures \ref{fig:water-ions-rdfs} and \ref{fig:ions-rdfs} show the ion-water and water-water RDFs respectively for the system of one NaCl ion pair in 95 waters.

\FloatBarrier
%\bibliography{si.bib}
%Control: production of eprint (-1) disabled
%